\documentclass[twocolumn]{aastex631}

\usepackage{amsmath}
\usepackage{xspace}

\newcommand\HI{H{\sc i}}

\newcommand{\highel}{${e}_{\text{high}}$\xspace}
\newcommand{\lowel}{${e}_{\text{low}}$\xspace}
\newcommand{\HII}{\ensuremath{\textrm{\ion{H}{2}}}}

\shorttitle{DRAGONS Faraday Depth Survey}
\shortauthors{A. Ordog et al.}
\graphicspath{{./}{figures/}}
\begin{document}

\title{GMIMS-DRAGONS: A Faraday Depth Survey of the Northern Sky Covering 350--1030 MHz}

\correspondingauthor{Anna Ordog}
\email{aordog@uwo.ca}

\newcommand{\UBCO}{Department of Computer Science, Math, Physics, \& Statistics, University of British Columbia, Okanagan Campus, Kelowna, BC, V1V 1V7, Canada}

\newcommand{\UCalgary}{Department of Physics and Astronomy, University of Calgary, 2500 University Drive NW, Calgary, Alberta, T2N 1N4, Canada}

\newcommand{\DRAO}{Dominion Radio Astrophysical Observatory, Herzberg Research Centre for Astronomy and Astrophysics, National Research Council Canada, PO Box 248, Penticton, BC, V2A 6J9, Canada}

\newcommand{\UWO}{Department of Physics \& Astronomy, University of Western Ontario, 1151 Richmond Street, London, ON, N6A 3K7, Canada}

\newcommand{\UTas}{School of Natural Sciences, University of Tasmania, Hobart, Tas 7000 Australia}

\newcommand{\INAF}{INAF-Istituto di Radioastronomia, Via Gobetti 101, 40129 Bologna, Italy}

\newcommand{\INAFOAA}{INAF – Osservatorio Astrofisico di Arcetri, Largo E. Fermi 5, 50125 Firenze, Italy}

\newcommand{\LPENS}{Laboratoire de Physique de l’École Normale Supérieure, ENS, Université PSL, CNRS, Sorbonne Université, Université de Paris, F-75005 Paris, France}

\newcommand{\RU}{Department of Astrophysics/IMAPP, Radboud University, PO Box 9010, 6500 GL Nijmegen, The Netherlands}

\newcommand{\CSIROBentley}{ATNF, CSIRO Space \& Astronomy, Bentley, WA, Australia}

\newcommand{\Caltech}{Caltech Owen's Valley Radio Observatory, Big Pine 93513, CA, USA}

\newcommand{\Onsala}{Department of Earth and Space Sciences, Chalmers University of Technology, Onsala Space Observatory, 439 92 Onsala, Sweden}

\newcommand{\Waterloo}{Department of Physics and Astronomy, University of Waterloo, 200 University Avenue West Waterloo, ON, N2L 3G1, Canada}

\newcommand{\USC}{Department of Physics \& Astronomy, University of South Carolina, Columbia, SC 29208, USA}
\author[0000-0002-2465-8937]{Anna Ordog}
\affiliation{\UBCO }
\affiliation{\DRAO }
\affiliation{\UWO }

\author[0000-0001-5181-6673]{Rebecca A. Booth}
\affiliation{\UCalgary }

\author[0000-0003-1455-2546]{T.L. Landecker}
\affiliation{\DRAO}

\author[0000-0002-3973-8403]{Ettore Carretti}
\affiliation{\INAF }

\author[0000-0001-7301-5666]{Alex S. Hill}
\affiliation{\UBCO }
\affiliation{\DRAO }

\author[0000-0003-4781-5701]{Jo-Anne C. Brown}
\affiliation{\UCalgary }

\author{Artem Davydov}
\affiliation{\UCalgary }

\author{Leonardo Moutinho Caffarello}
\affiliation{\UBCO }

\author{Luca B. Galler}
\affiliation{\UCalgary }
\affiliation{\Waterloo }

\author[0000-0002-0204-2891]{Jonas Flygare}
\affiliation{\Caltech }
\affiliation{\Onsala }

\author[0000-0001-7722-8458]{Jennifer L. West}
\affiliation{\DRAO }

\author{A.G. Willis}
\affiliation{\DRAO }

\author[0000-0001-8749-1436]{Mehrnoosh Tahani}
\affiliation{\USC }

\author{G.J. Hovey}
\affiliation{\DRAO }
\affiliation{\Onsala }

\author[0009-0001-8007-1890]{Dustin Lagoy}
\affiliation{\DRAO }

\author{Stephen Harrison}
\affiliation{\DRAO }

\author{Michael A. Smith}
\affiliation{\DRAO }

\author{Charl Baard}
\affiliation{\DRAO }

\author{Rob H. Messing}
\affiliation{\DRAO }

\author{D.~A. Del Rizzo}
\affiliation{\DRAO }

\author{Benoit Robert}
\affiliation{\DRAO }

\author[0000-0002-4217-5138]{Timothy Robishaw}
\affiliation{\DRAO }

\author[0000-0002-6300-7459]{John M. Dickey}
\affiliation{\UTas }

\author[0009-0007-0974-0645]{George Morgan}
\affiliation{\UBCO }

\author{Ian R. Kennedy}
\affiliation{\UBCO }

\author[0000-0002-5288-312X]{Marijke Haverkorn}
\affiliation{\RU }

\author[0000-0003-0932-3140]{Andrea Bracco}
\affiliation{\INAFOAA }
\affiliation{\LPENS }

\author{John Conway}
\affiliation{\Onsala }



\begin{abstract}
Polarized synchrotron emission at meter to centimeter wavelengths provides an effective tracer of the Galactic magnetic field. Calculating Faraday depth, the most useful parameter for mapping the line-of-sight magnetic field, requires observations covering wide frequency bands with many channels. As part of the Global Magneto-Ionic Medium Survey (GMIMS), we have observed polarized emission spanning 350--1030~MHz over the northern sky, in the declination range ${-20\arcdeg}\leq{\delta}\leq{90\arcdeg}$. We used the 15~m telescope at the Dominion Radio Astrophysical Observatory (DRAO), equipped to receive orthogonal circular polarizations, with the Onsala Space Observatory band~1 feed developed for the SKA Project. Angular resolution varies across the band from $1.3\arcdeg$ to $3.6\arcdeg$. A digital spectrometer provided 42~kHz frequency resolution. Data were taken with the telescope moving rapidly in azimuth and are absolutely calibrated in intensity. Approximately 25\% of the data were lost due to radio-frequency interference. The resolution in Faraday depth is $\sim6$~rad~m$^{-2}$, and features as wide as $\sim38$~rad~m$^{-2}$ are represented. The median sensitivity of the Faraday depth cube is 11~mK. Approximately 55\% of sight-lines in this survey show Faraday complexity. This dataset, called ``DRAO GMIMS of the Northern Sky'' (DRAGONS), is the first to probe Faraday depth of the northern sky in its frequency range and will support many scientific investigations. The data will be used to calibrate surveys with higher angular resolution, particularly Galactic foreground maps from the Canadian Hydrogen Intensity Mapping Experiment, and to provide information on large structures for aperture-synthesis telescopes, particularly the DRAO Synthesis Telescope. The data are available through the Canadian Astronomy Data Centre.
\end{abstract}

\keywords{ISM: magnetic fields --- polarization --- radio continuum: ISM ---
surveys --- techniques: polarimetric}

\section{Introduction}
\label{sec:intro}

Magnetic fields play a significant role in the physics of the Galaxy, contributing to the formation of clouds and stars \citep[e.g.,][]{Federrath_2012, Tahani_2022a, Tahani_2022b}, and influencing Galactic evolution \citep{Kim_1996}. The magnetic field shares the energy of the interstellar medium (ISM) in approximately equal parts (on sufficiently large scales) with the energetic charged particles and the motion of the gas \citep{ferr01}. The interplay between magnetic fields, cosmic rays, gas, dust, and stars is responsible for the structures appearing in the ISM on a wide range of spatial scales. Mapping the magnetic field across the sky, and along the line of sight (LOS), will provide fundamental information on the Galaxy and its future development. This 3D mapping is challenging, but important, and high quality, all-sky data are necessary for constraining global, large-scale models \citep[e.g.,][]{Unger2024}. Recent studies have shown that mapping 3D magnetic field vectors of Galactic objects such as molecular clouds reveals their formation and evolution history \citep[e.g.,][]{Tahani_2022a,Tahani_2022b}.

One of the best means of studying the LOS component of the Galactic magnetic field (GMF) is to map the Faraday rotation of polarized synchrotron emission observed with radio telescopes. The frequency dependence of the polarization angle rotation encodes information about the magnetic field and electron density in the volume through which the emission passes. This is quantified by the Faraday depth, $\phi$, defined as
\begin{equation}
    \frac{\phi}{\text{rad m}^{-2}} = 0.81\int_{r=0}^{r=d} \frac{n_e}{\text{cm}^{-3}}\frac{B_{||}}{\mu \text{G}}\frac{dr}{\text{pc}},
\end{equation}
where $n_e$ is the electron number density, and $B_{||}$ is the LOS component of the magnetic field, which is positive when directed toward the observer. We follow the convention in Equation~17 of \cite{Ferr2021}, with the integral calculated along the LOS path length increments, $dr$, from the observer located at $r=0$ to the polarized emission source located at a distance $r=d$. 

In the simplest case of an emission source located at a discrete distance experiencing Faraday rotation through a magnetized, ionized medium between the source and observer, the LOS is characterized by a single Faraday depth, known as a rotation measure (RM), which can be calculated from a linear relationship between the measured polarization angle and the square of the wavelength, $\lambda$. However, in the case of the diffuse polarized synchrotron emission in the interstellar medium (ISM), emission and Faraday rotation can be mixed, with sources located over a range of different distances, $d$, all undergoing the Faraday rotation caused by varying pathlengths through the intervening medium. The resulting effects on the observed, complex polarization can, in part, be disentangled by applying `Faraday synthesis' \citep[also known as `RM synthesis';][]{burn_1966, brentjens_de_bruyn_2005}, which recovers a spectrum of Faraday depths describing the LOS instead of a single value. The complex Faraday depth spectrum, $\tilde{P}(\phi)$, can be calculated as 
\begin{equation}
    \tilde{P}(\phi)=\int_{-\infty}^{\infty}W(\lambda^2)\tilde{P}(\lambda^2)e^{-2i\lambda^2\phi}d\lambda^2,
\end{equation}
where $\tilde{P}(\lambda^2)$ is the measured complex polarization across the available observing frequencies (wavelengths), and $W(\lambda^2)$ is a weighting function representing the finite $\lambda^2$ coverage. The range of scales in Faraday depth space that can be recovered using this method depends on the observing bandwidth and the width and sampling of the frequency channels. Broad frequency coverage (octave bandwidth) and narrow channel widths ($\sim1$~MHz or less at frequencies below 2~GHz) are best suited to maximizing the information that can be gained from measuring Faraday rotation. 

Faraday rotation is a tracer of both the magnetic field and the ionized gas density. Recent studies have shown that Faraday rotation is sensitive to relatively low electron densities, meaning that it is able to probe not only the warm ionized medium, but also the warm partially-ionized medium \citep{heiles2012} and even the low ionization of the warm neutral medium \citep{foster2013,vaneck2017,bracco2022}, particularly at low frequencies ($<1$~GHz).

The Global Magneto-Ionic Medium Survey (GMIMS) is paving the way for full-sky, broadband, single-antenna polarization maps that are ideal for the application of the Faraday synthesis technique, enabling the study of complex effects of mixed synchrotron emission and Faraday rotation in the ISM. The GMIMS initiative is loosely comprised of six component surveys, divided into three bands in each of the northern and southern hemispheres. Three of these surveys have been published: (i) GMIMS-Low-Band-South (GMIMS-LBS), covering 300--480~MHz \citep{Wolleben_2019}, (ii) the Southern Twenty Centimeter All-sky Polarization Survey (STAPS), contributing GMIMS-High-Band-South (GMIMS-HBS) and covering 1324--1770~MHz \citep{Sun_2025}, both using Murriyang, CSIRO's Parkes 64 m radio telescope, and (iii) GMIMS-High-Band-North (GMIMS-HBN), covering 1280--1750~MHz \citep{woll21} using the Dominion Radio Astrophysical Observatory (DRAO) 26~m John A. Galt telescope. Of the remaining planned surveys, observations are close to completion for the POSSUM (Polarisation Sky Survey of the Universe's Magnetism) EMU (Evolutionary Map of the Universe) GMIMS All-Stokes UWL (Ultra Wideband Low) Survey (PEGASUS; E.~Carretti et al. 2026, in preparation), covering 704--1440~MHz and contributing GMIMS-Mid-Band-South, using the Murriyang telescope at Parkes. 

\begin{figure}
    \centering
    \includegraphics[width=1\hsize]{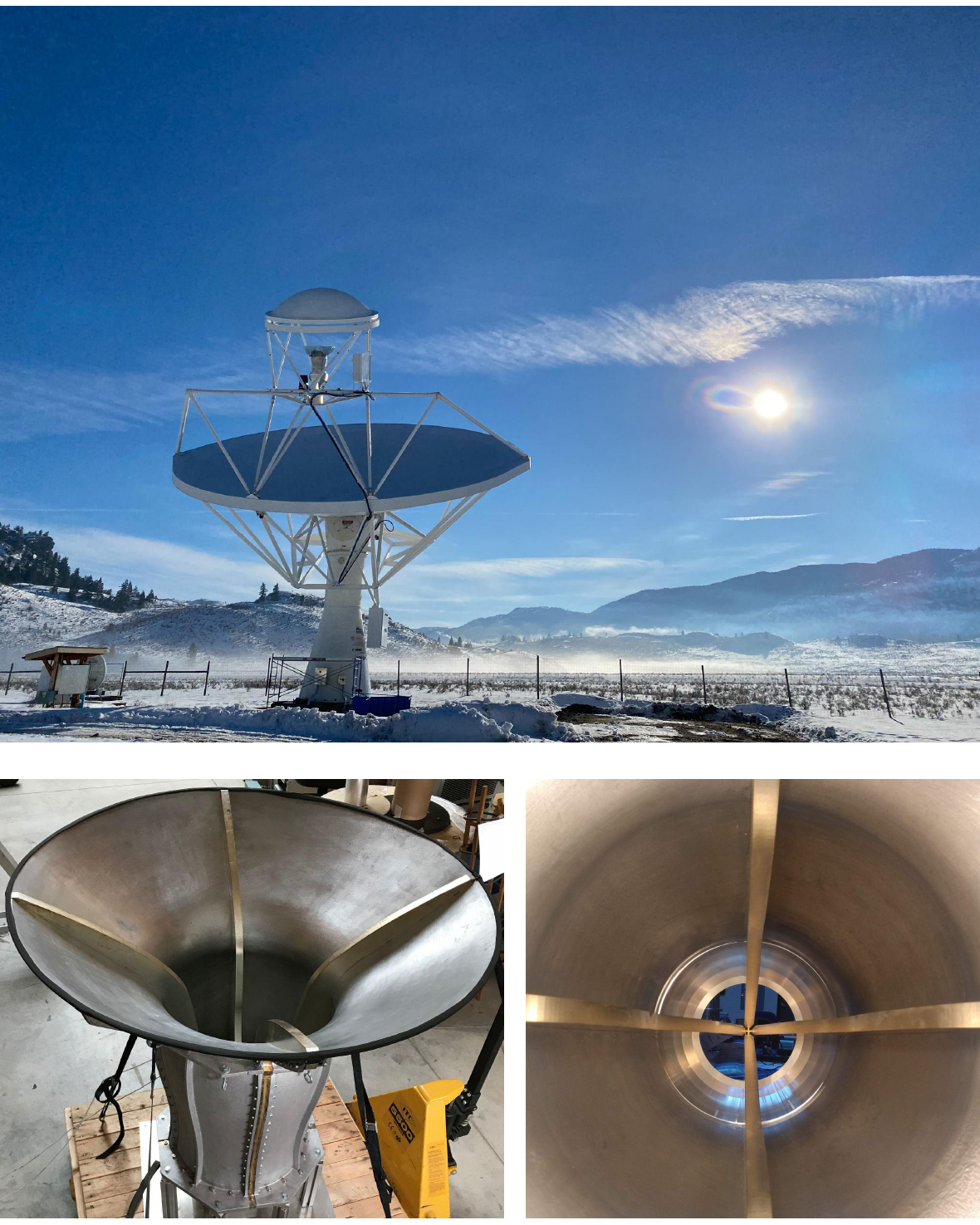}
    \caption{The DRAO-15 telescope, a ${15}\times{18}$\,m offset Gregorian reflector (\textit{top}) and the Onsala Space Observatory quad-ridged, flared horn feed (\textit{bottom}).}
    \label{fig:DRAO15m}
\end{figure}

Interferometric data from several other instruments and surveys are complementary to the GMIMS single-antenna datasets. A polarization study using data from the Canadian Hydrogen Intensity Mapping Experiment \citep[CHIME;][]{chime_overview} has been published \citep{Mohammed_2024}, demonstrating great promise for incorporating CHIME maps into GMIMS as a high-resolution component in the northern hemisphere, covering 400--800~MHz. Separately, the LOFAR Two-meter Sky Survey (LoTSS-DR2) has demonstrated a low-frequency (120--167~MHz), interferometric approach to diffuse emission Faraday synthesis maps over a significant portion of the northern sky \citep{Erceg_2022,Erceg_2024}. POSSUM is also producing high-resolution diffuse emission polarization maps \citep{Gaensler2025}, for which PEGASUS will ultimately provide the short baselines.

Here we present the DRAO GMIMS of the Northern Sky (DRAGONS) survey, covering 350--1030~MHz, which contributes GMIMS-Low-Band-North. The frequency range corresponds to a $\lambda^2$ range of 0.085~m$^2$ to 0.73~m$^2$, the widest $\lambda^2$ coverage of all GMIMS surveys, allowing for unprecedented sensitivity to Faraday complexity. The overlap in sky coverage ($-20\arcdeg\leq\delta\leq20\arcdeg$) and frequency (350--480~MHz) with GMIMS-LBS allows for a useful comparison and data validation between the two low-frequency GMIMS surveys. The complete sky-coverage overlap with CHIME data, and 400--800~MHz frequency overlap, along with complementary sensitivity to different ranges of angular scales, will allow DRAGONS to be used to calibrate CHIME polarization maps and provide information on missing short baselines.

This paper is organized as follows. In Section~\ref{sec:telescope} we describe the DRAO 15~m telescope, its signal chain, and the science commissioning steps. In Section~\ref{sec:obs_cal} we outline the observing strategy and describe the data collected for the survey and calibration. In Section~\ref{sec:process} we describe the data processing pipeline and map-making steps. In Section~\ref{sec:quality} we present Stokes $I$, $Q$, and $U$ maps, and a data quality assessment with comparisons to previously published data sets. In Section~\ref{sec:faraday} we describe the Faraday synthesis procedures and present the resulting Faraday depth cubes. In Section~\ref{sec:results} we discuss some preliminary science results, with comparisons to other GMIMS data products. We conclude in Section~\ref{sec:concl} with a summary of DRAGONS science papers currently underway, and a look toward future work with the broadband capabilities of this survey.

\section{Telescope and Receiver}\label{sec:telescope}
The telescope we use was originally proposed as an SKA Project design. Although not selected for the SKA, the prototype instrument, installed at the south-west corner of the DRAO site, has proven to be useful as a single-antenna telescope with its rapid survey capability. We describe the components comprising the telescope below, including the unique reflector configuration, the feed and receiver, and the signal chain. We also describe the characterization of the telescope completed during the science commissioning phase preceding this survey.

\subsection{The 15~m Telescope}\label{subsec:dish}
We conducted survey observations with the ${15}\times{18}$\,m offset Gregorian reflector at the DRAO, pictured in Figure \ref{fig:DRAO15m}. The telescope was designed to obtain an unblocked aperture of 15\,m diameter \citep{Lacy2012, knee16}. We refer to it as the DRAO-15. Both reflecting surfaces are built from carbon fiber, with a surface rms ${\sim}250~{\mu}{\rm{m}}$. We equipped the telescope to receive left- and right-hand circular polarization (denoted here by $L$ and $R$ respectively). Offset reflectors have poor polarization performance, with a position offset between $L$ and $R$ beams \citep{chu73}, but the problem is overcome in this telescope by \cite{Baker_2014,Baker_2020} through shaping of the reflector surfaces following the technique described by \citet[][hereafter `Mizigutch technique']{mizu76}. Further shaping of the reflector surfaces is used to enhance aperture efficiency. 

\begin{figure}
    \centering
    \includegraphics[width=1\hsize]{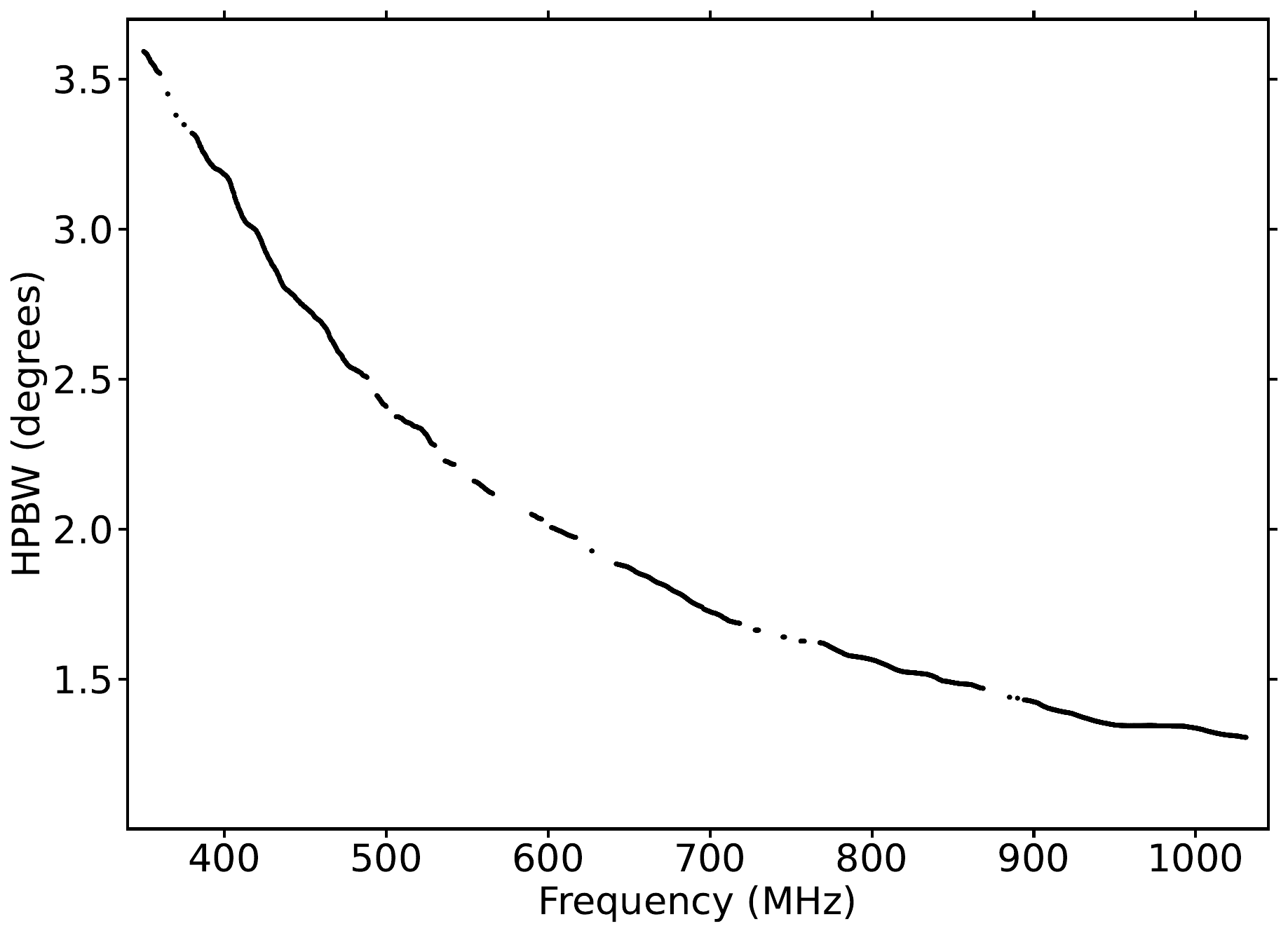}
    \caption{The measured HPBW of the telescope as a function of frequency, determined from the FWHM of Gaussians fitted to raster scans of compact sources (see Section~\ref{subsubsec:raster}). Gaps in the measured beamwidth data are channels with too much RFI for a successful Gaussian fit. A boxcar median smoothing kernel of width 49 channels ($\sim4$~MHz) was applied to the measured FWHM values.
    }
    \label{fig:beamw}
\end{figure}

\begin{figure}
    \centering
    \includegraphics[width=1\hsize]{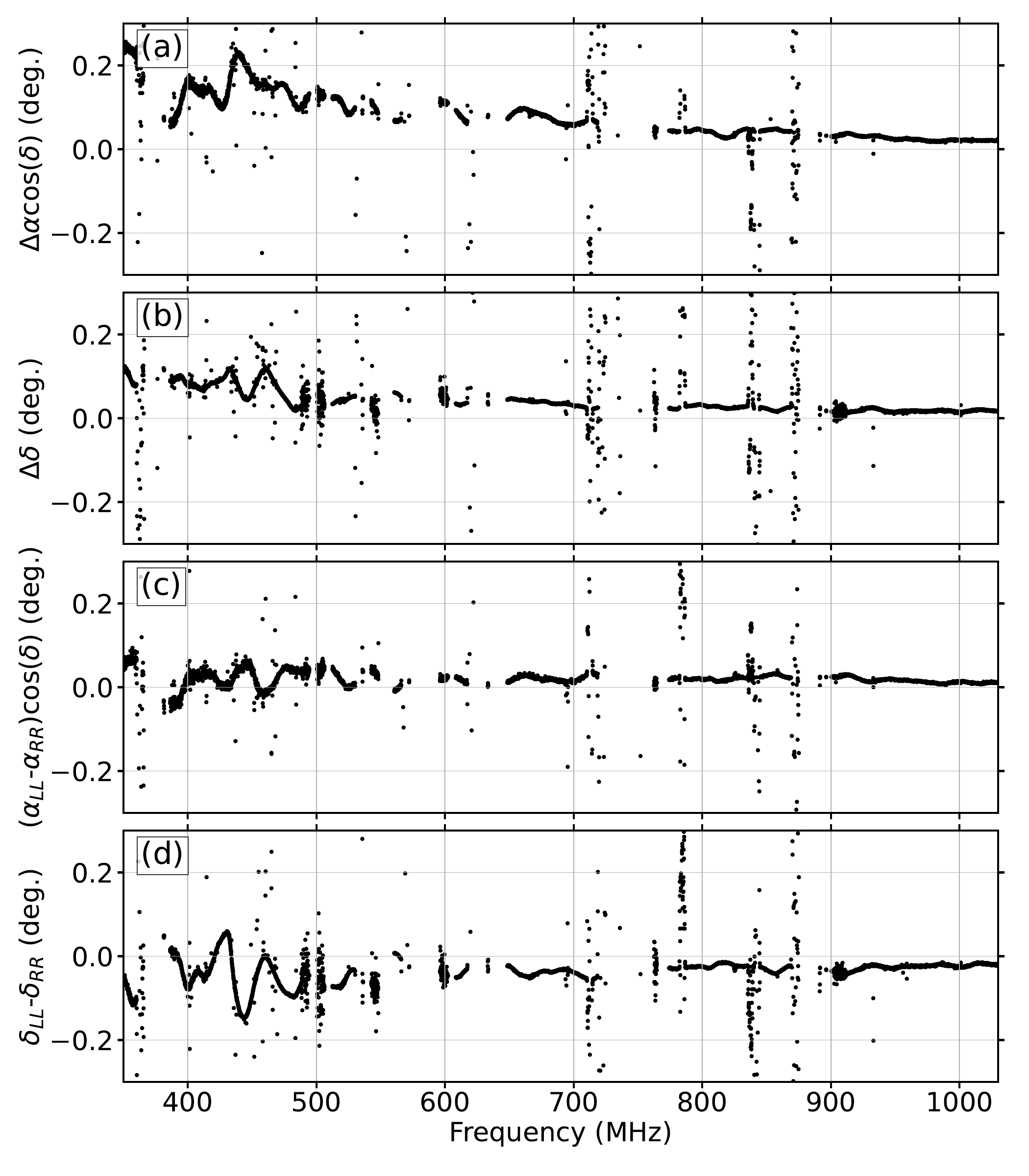}
    \caption{Frequency dependence of pointing measurements using the average of seven Cyg~A raster scans. (\textit{a}) Right ascension offset, (\textit{b}) declination offset, (\textit{c}) difference in right ascension measured in LL* and RR*, and (\textit{d}) difference in declination measured in LL* and RR*. The consistently positive offsets in (\textit{a}) and (\textit{b}) are the residual, systematic errors in the pointing accuracy following the correction described in Section~\ref{subsec:pointing}. The `persistent RFI' mask described in Section~\ref{subsec:rfi} was applied to these data to eliminate most RFI channels. Remaining outliers are due to poor-quality fits to the raster scans in channels contaminated by intermittent RFI.}
    \label{fig:pointingspectr}
\end{figure}

The reflector was fed with a quad-ridged, flared horn \citep{flyg18} with an integrated calibration noise coupler \citep{flyg17}, the prototype feed developed by Chalmers University's Onsala Space Observatory (OSO) for the SKA-mid Band\,1. The feed accepts two orthogonal linearly polarized signals. To measure the linearly polarized Galactic emission, circular polarization is preferred over linear because Stokes parameters $Q$ and $U$ can then both be derived using correlation techniques, as described in \cite{mcco06} and \cite{robi21}. This method is critically important for single-antenna systems, as it avoids the inherent instability of deriving Stokes $Q$ from the difference of two large, noisy total-power signals. By converting to a circular basis, both Stokes $Q$ and $U$ are instead derived from the more stable cross-correlation product, which is largely immune to uncorrelated receiver noise and gain fluctuations. To achieve this, the linearly polarized outputs from the feed were connected to a transmission-line hybrid coupler{\footnote{R{\&}D Microwaves model HD-A01}} before amplification so that the telescope outputs became $L$ and $R$. Losses in transmission lines, in connectors, and in the hybrid amount to ${\sim}12$\,K at 350\,MHz and ${\sim}35$\,K at 1030\,MHz. 

Figure~\ref{fig:beamw} shows the measured half-power beamwidth (HPBW) across the band. The HPBW varies from $3.6^{\circ}$ at 350\,MHz to $1.3^{\circ}$ at 1030\,MHz (all frequency channel maps are convolved to a common $3.6\arcdeg$ resolution prior to Faraday synthesis; Section~\ref{subsec:rmsynth}). Figure~\ref{fig:pointingspectr} shows the positions of Cyg~A derived from raster scans (described in Section~\ref{subsubsec:raster}), averaged over seven scans. The top two panels show the difference between the expected and measured positions. There is some variation of beam position with frequency, but the offsets are typically less than 5\% of the beamwidth. The bottom two panels of Figure~\ref{fig:pointingspectr} show the offset between the RR* and LL* beams. These measurements show the extent to which the offset reflector, corrected by the Mizigutch technique, provides coincident beams in the two hands of circular polarization. This small offset has a negligible effect on our data.

\subsection{The 350--1030~MHz Receiver}\label{subsec:rec}
In the frequency band of this survey, commercially available devices were able to provide adequate receiver sensitivity and bandwidth. Low-noise amplifiers were followed by conventional amplification and band-equalizing stages placed on the feed platform. The receiver passband was defined by the cut-off frequency of the feed at ${\sim}350$\,MHz, and by a low-pass filter, with 3~dB frequency ${\sim}1030$\,MHz, placed at the end of the feed-platform amplifier chain. The amplified outputs were transported to the control building by a radio frequency (RF)-over-fiber (RFoF) device. Signals were further amplified to $-20$\,dBm, the specified input level to the analog-to-digital converters. The calibration signal was provided by a solid-state noise source heated to a constant temperature near $30\arcdeg$~C.\footnote{Outside nighttime temperatures did not exceed $30\arcdeg$C during the survey, and it is simpler to only heat the noise source rather than also implementing cooling in warmer weather.}

\subsection{The Digital Spectrometer}
\label{subsec:spectrum}
Since the key data product from this survey is a Faraday depth cube, the requirements for the spectral channelization are primarily driven by bandwidth depolarization effects \citep{Raja2014,Schnitzeler2015,Pratley2020}, which could, in principle, be met by a spectrometer that analyzes incoming signals to a frequency resolution of ${\sim}1$\,MHz (see Section~\ref{subsec:moment} for details). Faraday synthesis on such data produces a reliable Faraday depth cube (and we ultimately bin the channels to 1~MHz in the map-making stage). Also, digitization to four bits would satisfactorily limit any additional noise from the quantization process. However, in the real world, our observing frequencies carry heavy communications traffic, and we need much finer digitization in order to have the dynamic range to detect the astronomical signal in the presence of the much stronger radio frequency interference (RFI) signals. Channels and samples spoiled by RFI must be excised from the data, and to minimize data loss it is useful to have a sampling that is much finer than required by the demands of Faraday synthesis. Fortunately, modern digital electronics permits the design of spectrometers with many bits and many narrow frequency channels.

The digital spectrometer \citep{lago22} analyzes the receiver output signals in five steps:
\begin{enumerate}
\item {it digitizes time-domain signals into 10-bit samples at a rate of $F_s=2.2$\,Gigasamples per second;}
\item {it produces a coarse spectrum, with channel width ${F_s}/{16} = 137.5$\,MHz, of each input in a polyphase filterbank (PFB) implemented with a field programmable gate array, and attaches a time stamp to each data sample (GPS-based time);}
\item {it computes a fine spectrum of each of the 137.5\,MHz-wide channels in a second PFB, producing 3300 channels with an individual resolution of 41.67\,kHz;}
\item {it produces the correlated data products from the left- and right-circular channels;}
\item {it integrates the correlated data into 0.6 second samples.}
\end{enumerate}
Steps 1 and 2 are executed on an ICEboard \citep{band16}. Steps 3 and 5 are implemented on a graphics processing unit (GPU). The ICEboard and GPU are connected via two 40 Gbit/s network interfaces. The data rate after fine channelization is 25 terabytes per day. During the integration of 0.6\, second, the raw data are discarded, and the integrated data are sent via a single 40 Gbit/s interface to a storage cluster. The cluster provides low-cost redundancy and efficient time-series data management. The stored data are processed by the backend pipeline into an HDF-based data format \citep{hobb20}.

Two of the 137.5~MHz-wide bands lie below the cutoff frequency of the feed and therefore do not contain useful data. The number of spectrometer channels used is about 16,300. Inter-channel isolation (between 41.6~kHz channels) is 70\,dB, and the passband ripple is 0.1~dB. Aliasing at the 1030~MHz roll-off of the low-pass filter is ${\sim}40$~dB. The four correlated outputs produced by the spectrometer are the total power from the left- and right-circular channels, RR* and LL*, and the linear polarization products, RL* and LR*.

\subsection{Telescope Control System}\label{subsec:software}
Survey observations were made as azimuth scans at two fixed elevations (Section~\ref{subsec:obs}). With its azimuth-elevation configuration, the DRAO-15 telescope has the capability to scan quickly in azimuth (up to 180 deg/minute) at a fixed elevation. The azimuth axis uses two drives to prevent backlash.\footnote{Discrepancy in the motion due to the clearance between the gear teeth.} The elevation drive has a maximum speed of 60~deg/minute, allowing the telescope to change between the two survey elevations ($49\arcdeg$ and $20\arcdeg$; Section~\ref{subsec:obs}) in less than 4~minutes (slow-down and settling time included), permitting efficient scheduling of scans at the two elevations used for the survey within an observing night. 

The low-level control software was written in \texttt{C++}, while the observing schedule for the survey (including azimuth scans and calibration scans) consisted of a set of \texttt{python} scripts. For the cross-scans for the pointing model (Section~\ref{subsec:pointing}) and the raster scans of bright calibrators (Section~\ref{subsubsec:raster}) \texttt{python} scripts provided the telescope with a series of pointings in right ascension (RA) and declination, updated every 0.5 seconds, tracing out the pattern of the scan by commanding the telescope to `chase' the changing target pointing.

\begin{figure}
    \centering
    \includegraphics[width=1\hsize]{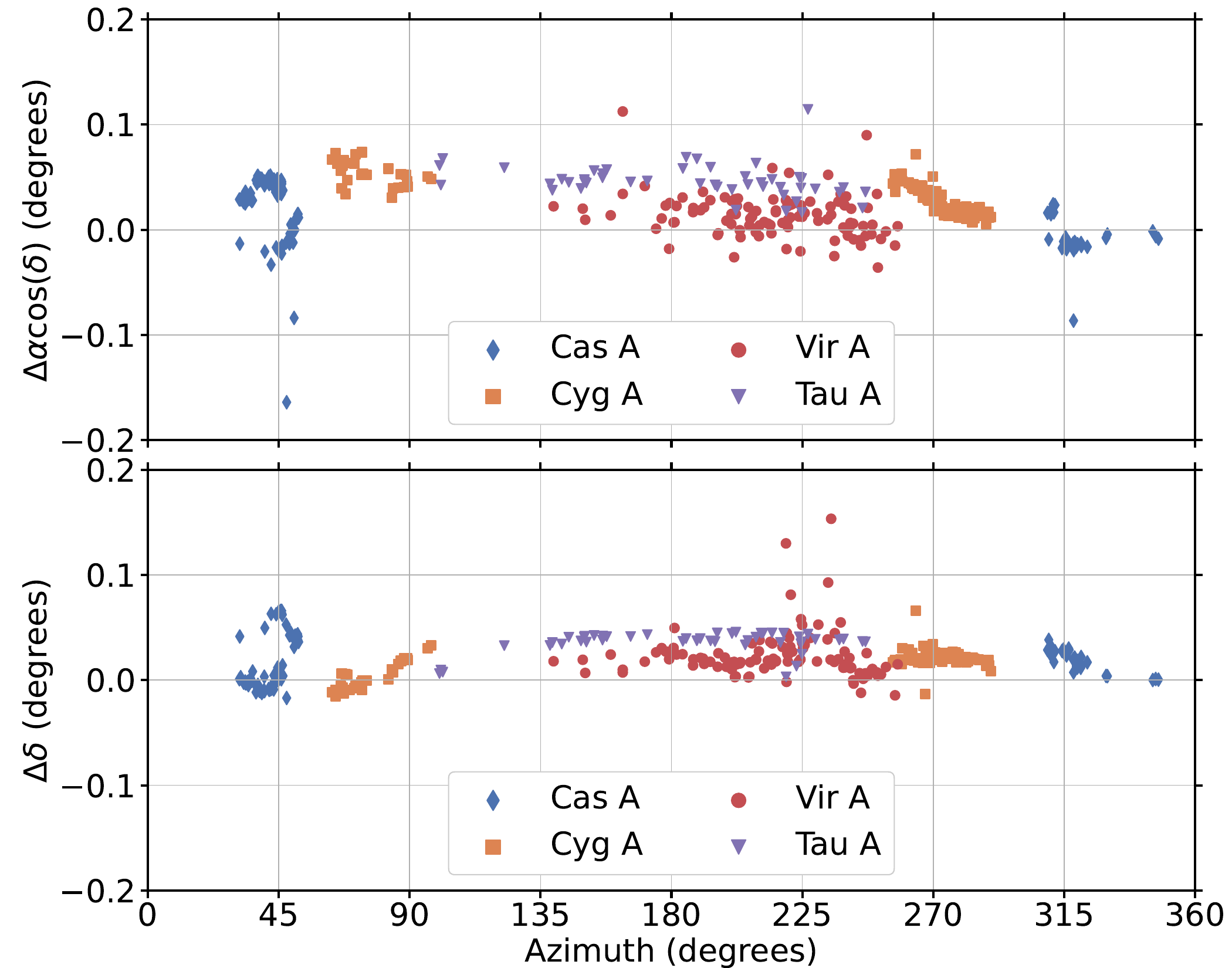}
    \caption{Right ascension (\textit{top}) and declination (\textit{bottom}) offsets between the expected and observed pointing positions as a function of azimuth. These measurements are derived from Gaussian fits to the calibration raster scans of Cyg~A, Cas~A, Tau~A, and Vir~A (Section~\ref{subsubsec:raster}) produced over the course of the survey.}
    \label{fig:pointing820}
\end{figure}

\subsection{Pointing Model}\label{subsec:pointing}
Commissioning the DRAO-15 included an assessment and correction of the pointing accuracy. We used the pointing model software package \texttt{TPoint}\footnote{Available from Software Bisque: \\ \url{https://www.bisque.com/product/tpoint/}} \citep{Wallace94} to compare position measurements of bright radio sources to their expected sky positions and to calculate model parameters to feed into the telescope control system. We observed eight radio sources, with a minimum flux density of 60~Jy at 1~GHz, approximately regularly spaced in declination from $-12\arcdeg$ to $59\arcdeg$ for this stage. Each source was scheduled for observation several times over the course of a few days in order to maximize hour-angle coverage. A total of 110 observations provided sufficient coverage in azimuth and elevation (the native coordinates of the telescope) for \texttt{TPoint} to produce a suitable model. 

Each observation consisted of a `cross-scan': one $8\arcdeg$-long scan in RA and one $8\arcdeg$-long scan in declination across the expected source position. Peak positions of 1D Gaussian functions fitted to each scan in the LL* correlation product at frequencies above 900~MHz determined the measured positions. The six model parameters fitted to the data with \texttt{TPoint} were: azimuth offset,  elevation offset, non-perpendicularity between the azimuth and elevation axes, elevation-dependent azimuth shift, azimuth-axis east-west misalignment and azimuth-axis north-south misalignment. After applying the corrections we re-observed a smaller subset of sources in order to assess the new pointing accuracy. The combined corrections produced a substantial improvement, reducing the systematic offsets between the expected and observed source positions by $\sim80\%$ in both elevation and azimuth, and reducing the rms errors by $6\%$ in elevation and by $50\%$ in azimuth. The overall residual uncertainty in pointing is $\sim5\%$ of the beamwidth, making this a sufficiently accurate correction for mapping large-scale structures. 

We applied the pointing model at the beginning of the survey and no further corrections were made throughout. However the daily calibration raster scans using the four brightest sources (Cyg~A, Cas~A, Tau~A, and Vir~A; Section~\ref{subsubsec:raster}) provided continuous checks on the pointing accuracy. We show the RA and declination offsets in Figure~\ref{fig:pointingspectr} as a function of frequency based on an averaged set of Cyg~A raster scans, and in Figure~\ref{fig:pointing820} at 820~MHz as a function of telescope azimuth from the full set of raster scans. While there is a slight residual pointing accuracy dependence on both frequency and azimuth, the pointing accuracy remains within 5\% of the beamwidth over the course of the survey, indicating sufficient pointing stability.

\subsection{System and Noise-Source Temperature}\label{subsec:HCTF}
Separate from the calibration measurements and procedures that we ultimately applied to the survey data (described in Section~\ref{subsec:calobs}), we measured the noise temperature of the receiver and the noise equivalent of the injected calibration signal prior to installing the receiver on the telescope. This was accomplished using the DRAO Hot-Cold Test Facility \citep[HCTF;][]{hove18}. The HCTF is a large metal funnel, with its wide mouth open at the top. The floor of the funnel is a flat metal plate, 1~m square, and the top opening is 3~m square, 3~m above the floor. The feed and receiver were placed in it, pointing upwards. Noise measurements were put on an absolute scale by exposing the feed and receiver to two terminations at known temperatures. A hot termination consists of a sliding roof lined with microwave absorber, and a cold termination was provided by sliding the roof open, exposing feed and receiver to the sky. We measured the physical temperature of the hot absorber directly, and derived the brightness temperature of the sky from the \texttt{PyGDSM} sky model \citep{price16,PGSM17}. Measurements were made with laboratory spectrum analyzers with a frequency resolution of 3\,MHz. 

\begin{figure}
    \centering
    \includegraphics[width=1\hsize]{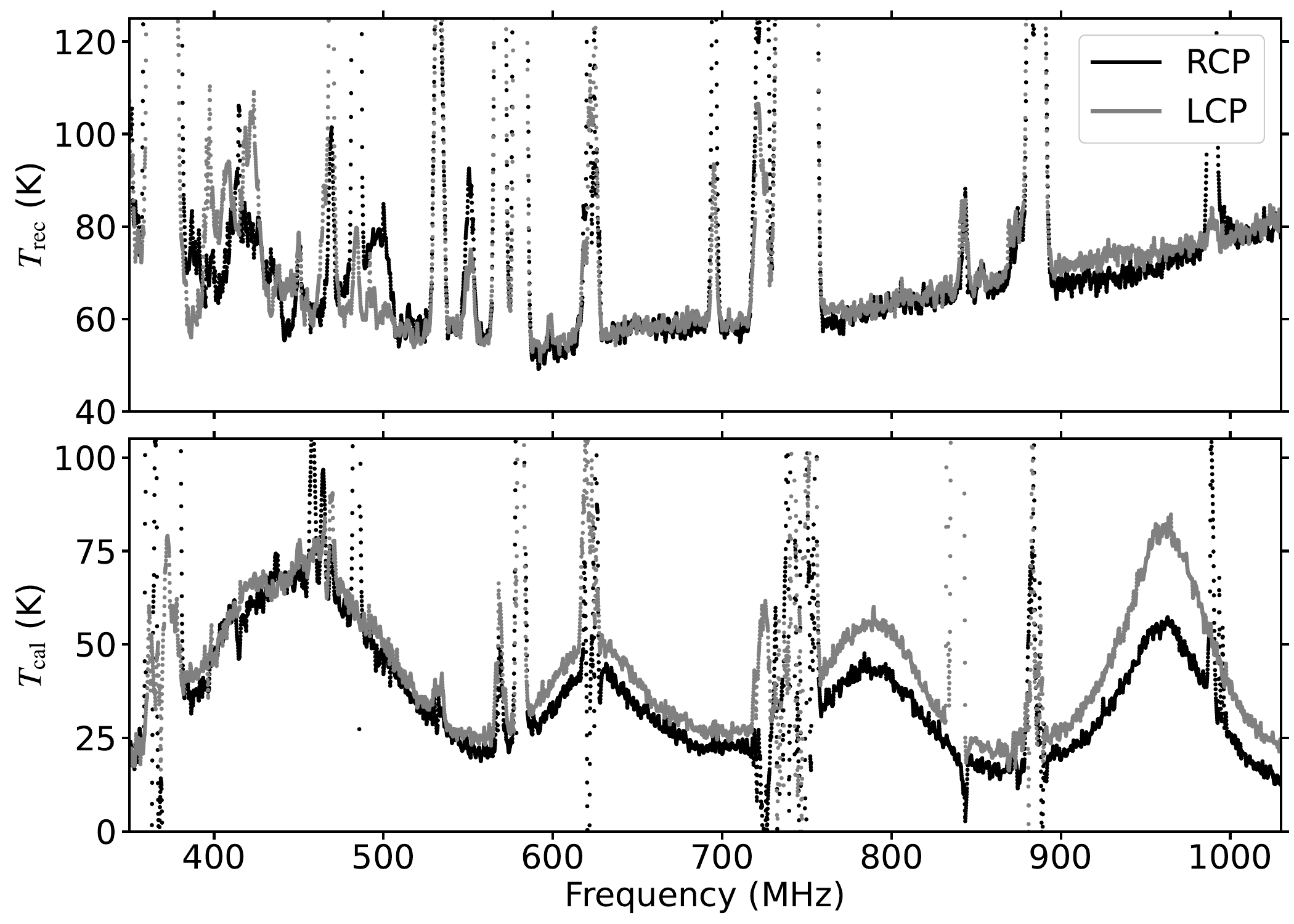}
    \caption{The receiver noise temperature in kelvins (\textit{top}) and the kelvin equivalent of the calibration noise source (\textit{bottom}) as determined from the HCTF measurements. Results are shown for both right- and left-hand circular polarization (RCP; black and LCP; gray).}
    \label{fig:hctftemps}
\end{figure}

\begin{figure}
    \centering
    \includegraphics[width=1\hsize]{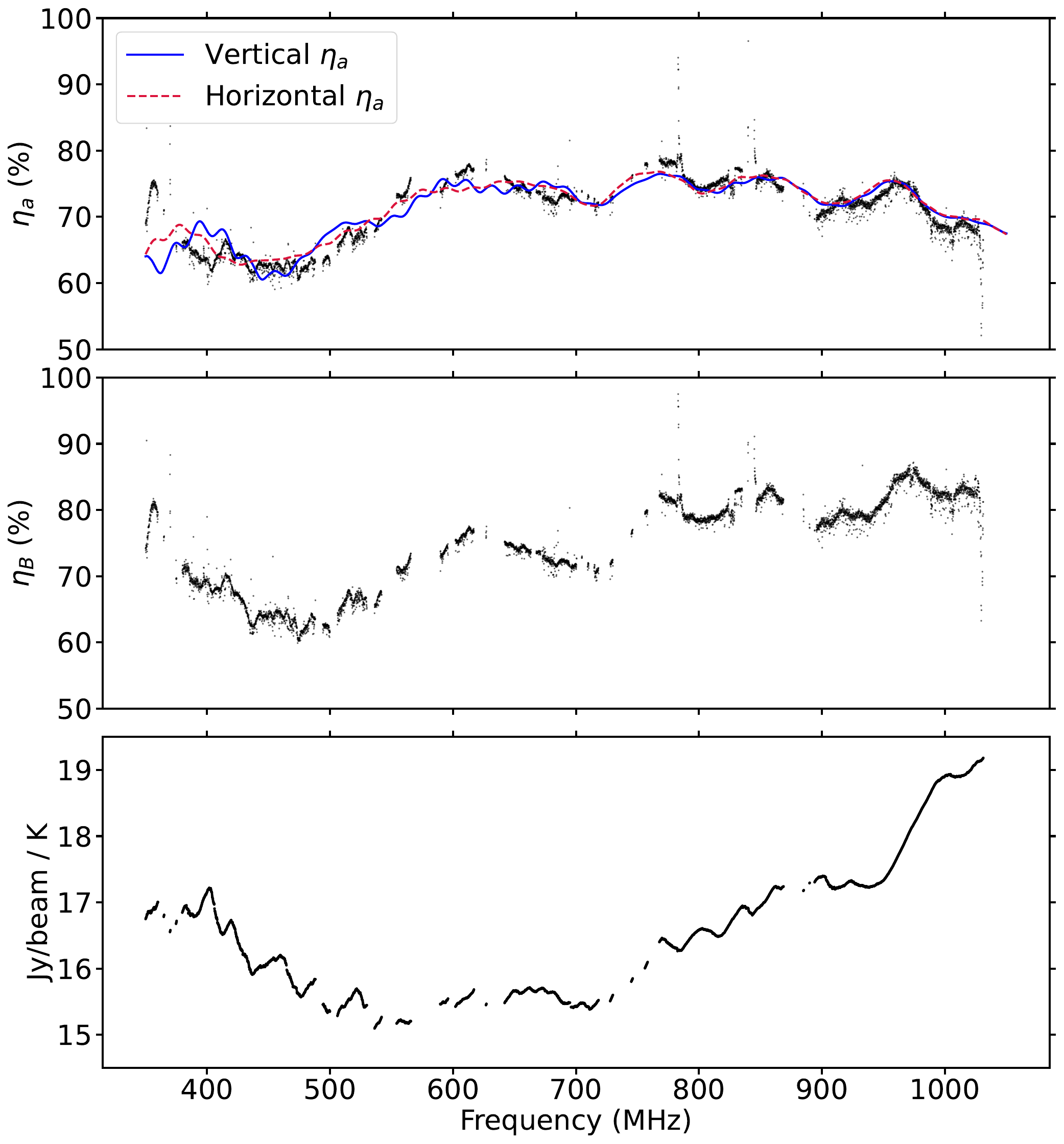}
    \caption{\textit{Top:} Comparison of measured aperture efficiency (black dots) with aperture efficiency calculated with electromagnetic simulation software. The blue and red curves shows the calculated results for vertical and horizontal linear polarization respectively. The measurements were made with circular polarization. \textit{Middle:} The measured beam efficiency across the band. \textit{Bottom:} The Jy~beam$^{-1}$ to K ratio across the band.}
    \label{fig:kconversion}
\end{figure}

The receiver temperature and calibration noise-source temperature are shown in Figure~\ref{fig:hctftemps}. The receiver temperature varies smoothly between $\sim80$~K at the high and low ends of the band and $\sim55$~K at 600~MHz. The calibration noise-source temperature fluctuates between $\sim15$ and $\sim80$~K. The large-scale, sinusoidal fluctuations in the noise-source temperature are likely caused by reflections between the noise-source coupler (Section~\ref{subsubsec:noise}) and the metal plate to which it was mounted. We estimate the accuracy of these measurements as ${\pm}3$\,K above 500\,MHz. Accuracy below 500\,MHz is limited by the size of the ${\sim}45$~\,cm-long pyramidal cones of the absorber that line the sliding roof. To investigate the effects of aliasing into the band we repeated the HCTF measurements out to 4~GHz, revealing that a negligible 0.05 to 0.3~K would be added to the system temperature by aliasing.

We calculate the expected noise in the final 1~MHz channel widths used for published data cubes using the radiometer equation applied to the receiver temperature values in Figure~\ref{fig:hctftemps} and approximate sky brightness temperatures from the \texttt{PyGDSM} sky model \citep{price16,PGSM17}. An integration time of $2\times0.6=1.2$~s\footnote{The effective integration time is twice the spectrometer integration time since each pointing is sampled twice in the scan strategy (Section~\ref{subsec:obs}).} and 1~MHz channel-width yields an expected noise between 55 and 100 mK.

\subsection{Aperture Efficiency}\label{subsec:aperture_eff}
As described in Section~\ref{subsec:abscal}, we achieved the conversion to brightness temperature (K) by calculating the main beam solid angle, as a function of frequency, from the measured beamwidths, assuming that the main beam is Gaussian. Nevertheless, we carried out an absolute calibration of the telescope, with the aim of improving our understanding of the instrument. We compared the antenna temperature produced by Cyg~A with absolute noise standards, resistors at the temperature of liquid nitrogen, at ambient temperature, and in a temperature controlled oven at approximately $100^{\circ}$\,C. We made this measurement using the RR* beam only. In the top panel of Figure~\ref{fig:kconversion} we show the measured aperture efficiency, ${\eta}_A$, compared with that computed using the General Reflector antenna Analysis Software Package by TICRA (GRASP)\footnote{\url{https://www.ticra.com/software/grasp/}} electromagnetic simulator. The excellent agreement, generally within 1\%, gives us great confidence in the GRASP computations of the radiation pattern of the antenna, which will be useful for future improvements to the ground emission calculations described in Section~\ref{subsec:ground}.

\section{Observations}\label{sec:obs_cal}
Here we describe the observing strategy used to collect the survey data, the calibration observations, and the basic properties of the resulting dataset.

\subsection{Observing Strategy}\label{subsec:obs}

\begin{figure}
    \centering
    \includegraphics[width=1\hsize]{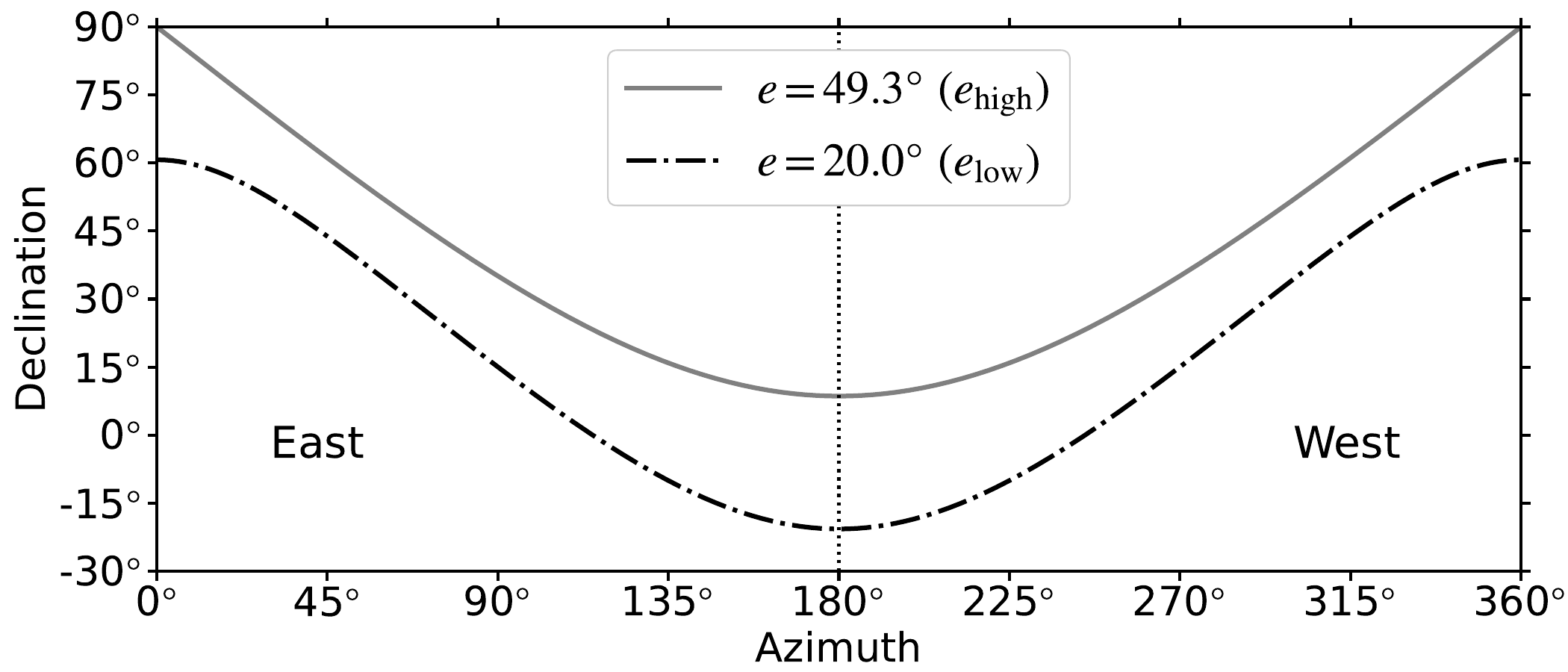}
    \caption{The declination coverage as a function of azimuth for the two elevations used in the survey: \highel and \lowel. This pattern produces four independent sets of maps: east and west at each elevation, with a $49\arcdeg$ declination overlap in $11\arcdeg\leq\delta\leq\ 60\arcdeg$. between \highel and \lowel.}
    \label{fig:decvsaz}
\end{figure}

\begin{figure}
    \centering
    \includegraphics[width=1\hsize]{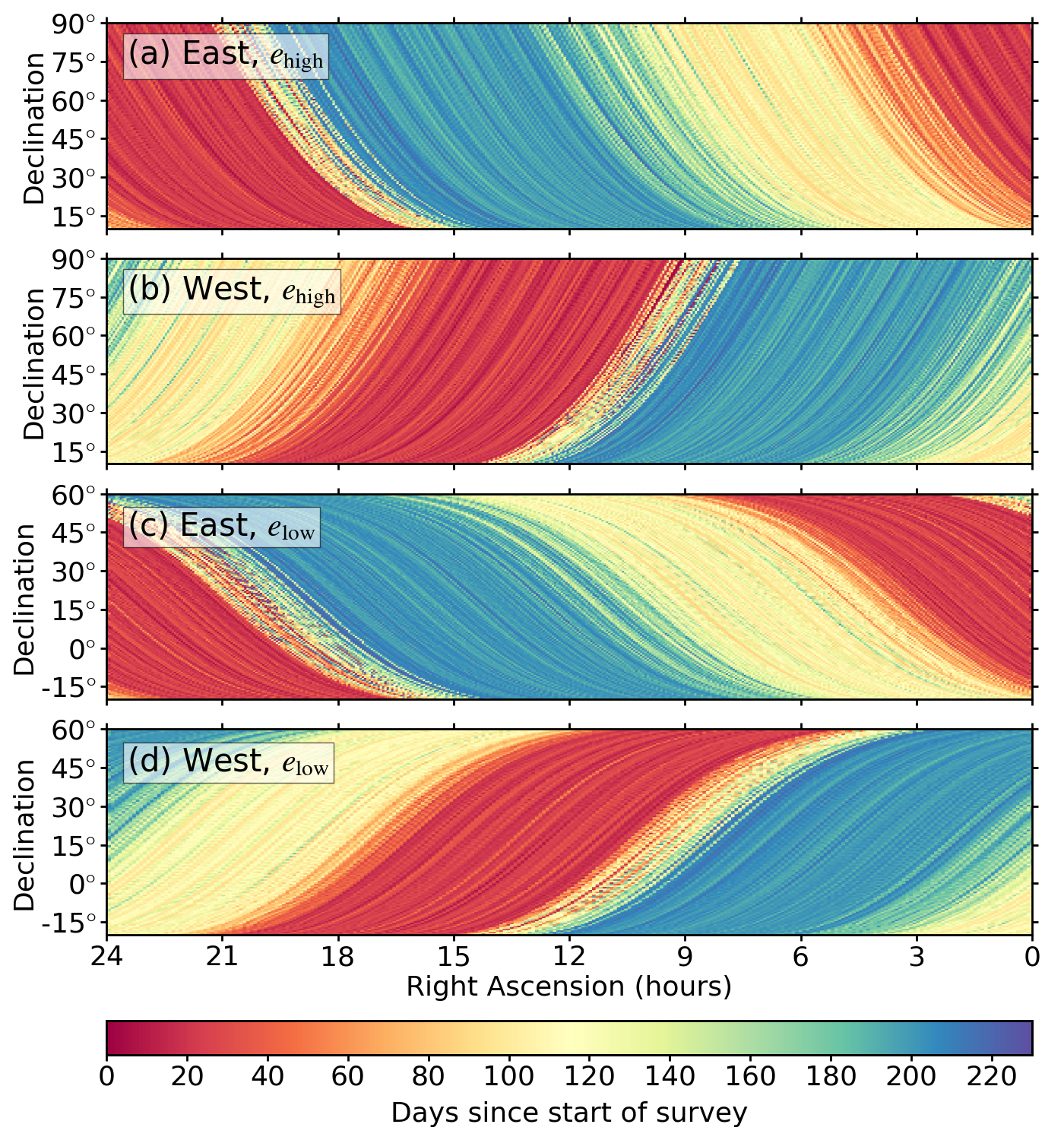}
    \caption{The completed east and west scans in equatorial coordinates for the two elevations: \highel (\textit{a},\textit{b}) and \lowel (\textit{c},\textit{d}). The colors represent the days since the start of the survey when each scan was observed: Days 0--49 (P1; 2022, June--July), Days 65--135 (P2; 2022, August--October) and Days 170--235 (P3; 2022, December--2023, January).}
    \label{fig:scans1}
\end{figure}

\begin{figure}
    \centering
    \includegraphics[width=1\hsize]{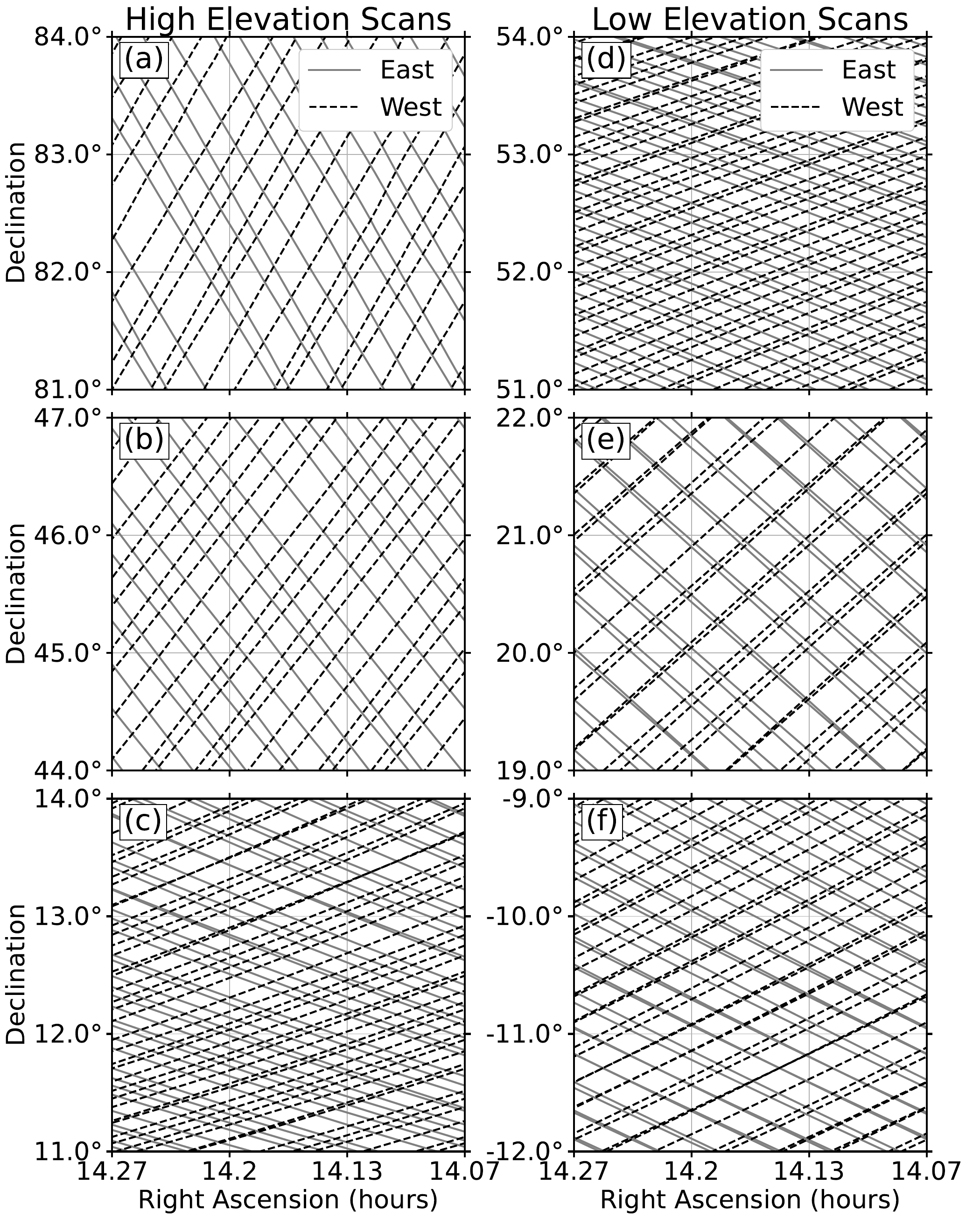}
    \caption{Examples of the intersecting east and west scans in six $3\arcdeg\times3\arcdeg$ regions in equatorial coordinates for the \highel (\textit{a}--\textit{c}) and \lowel (\textit{d}--\textit{f}) scans, showing the variation in sampling with declination.}
    \label{fig:scans2}
\end{figure}

We used the observing technique developed by \cite{carr19}, scanning in azimuth at fixed elevation. The timing was selected such that scans adjacent to each other on the sky provided close to uniform spatial coverage, and the length of the scans was chosen such that scans would also cross one another, allowing for the possibility of basket-weaving to reconcile their base levels (see Section~\ref{subsec:basket}). A scan consists of all the data collected for one sweep in azimuth with the telescope moving either clockwise or counterclockwise over $360^{\circ}$ at $20^{\circ}$~minute$^{-1}$, starting and ending at azimuth ${a}={0^{\circ}}$ (north). Scans were made at two elevations, ${e}={49.3^{\circ}}$, the elevation of the North Celestial Pole (NCP) at DRAO, and at ${e}={20^{\circ}}$, just above the lower elevation limit for the telescope. We denote these as \highel and \lowel scans respectively. The \highel scans covered the declination range $11\arcdeg\leq\delta\leq90\arcdeg$ and \lowel scans covered $-20\arcdeg\leq\delta\leq60\arcdeg$ (Figure~\ref{fig:decvsaz}). In the map-making stage we split each scan into `east' ($0\arcdeg\leq{a}\leq180\arcdeg$) and `west' ($180\arcdeg\leq{a}\leq\ 360\arcdeg$) halves (Figure~\ref{fig:scans1}) which produced the intersecting scan patterns (Figure~\ref{fig:scans2}). We produced separate maps from the \highel and \lowel datasets. We then combined these using a weighted average in the overlap region $11\arcdeg\leq\delta\leq\ 60\arcdeg$, with the \highel (\lowel) map weights changing linearly from 0 to 1 (1 to 0) across $\delta=11\arcdeg$ to $\delta=60\arcdeg$, thereby producing the final maps covering $-20\arcdeg\leq\delta\leq90\arcdeg$.

Scans started at sidereal times evenly spaced by an interval equivalent to 20$\arcmin$ on the celestial sphere. This guaranteed full Nyquist sampling of the sky at the highest survey frequency (1030 MHz) where the HPBW is $1.3\arcdeg$. Examples of the spacing between scans for different declination ranges in the \highel and \lowel scans are shown in Figure~\ref{fig:scans2}. Apart from a few instances of missed scans, scan separations are typically at most 20\arcmin. The survey was completed in three phases over the course of seven months: phase 1 (P1; June--July, 2022), phase 2 (P2; August--October 2022) and phase 3 (P3; December 2022--January 2023). Observations were made at night in order to avoid contamination from solar emission received through the sidelobes. Although the offset reflector has low sidelobes, the Sun in the sidelobes can contribute spurious polarized signals as bright as the brightest Galactic emission, which we observed when running a short test version of the survey that included daytime data.

In the survey planning process we generated a `library' of scans, following \cite{woll21}, with each scan having a unique 4-digit scan ID. This consisted of 1440 \lowel and 1440 \highel scans. For each phase of the survey we generated a daily scan schedule, drawing from the scans with sidereal start times occurring after sunset and sidereal end times occurring before sunrise. These schedules were generated with two objectives: (i) to achieve survey efficiency, and (ii) to add some randomness to the scan sequence. These objectives were, obviously, in conflict, so we adopted a reasonable compromise, aiming to avoid fixed patterns in the observing sequence. Sets of both \highel and \lowel scans were completed on each night. After each survey night, if a scan was deemed unsuccessful (e.g. a result of excessive RFI or equipment malfunction) the scan ID was returned to the master scan library to be rescheduled on a subsequent night. In total, 1401 of the 1440 \highel and 1406 of the 1440 \lowel scans were completed for inclusion in the mapping stage. We made efforts to reschedule failed scans, but were in some cases limited by a lack of available nighttime coverage of the required sidereal observing times. In the processing steps we also discarded a small number of scans from individual frequency channels based on an rms filtering threshold (Section~\ref{subsec:filter}).

\subsection{Calibration Observations}\label{subsec:calobs}
Intensity calibration proceeded in three stages: (1) we used a calibration noise source to correct gain fluctuations, (2) we observed raster scans of bright radio sources at the beginning and end of each survey night to calibrate the data in Jy~beam$^{-1}$, and (3) we used beamwidth measurements to convert the units from Jy~beam$^{-1}$ into brightness temperature (K). These procedures were similar to those used for the GMIMS-HBN survey \citep{woll21}. We did not use any polarized on-sky sources for polarization angle calibration, but found good agreement between the DRAGONS polarization angles and previous datasets without the need for angle corrections (Section~\ref{subsec:dwingeloo}).

\subsubsection{Injected Noise Signal}\label{subsubsec:noise}
The noise-source signal was injected into the feed with an integrated calibration noise coupler \citep{flyg17}. The noise source was switched on for 6 seconds every 3 minutes while the telescope was scanning, synchronized with the recorded integrations. We randomized the start time of the first noise-source instance for each scan in order to avoid coherent patterns of flagged data in the maps. We used the recorded noise-source signals to correct outside-temperature-dependent gain variations in the signal chain over the course of each observing night. The azimuth scans as well as the calibration raster scans (see Section~\ref{subsubsec:raster}) were initially converted into noise-source units, as described in Section~\ref{subsubsec:apply_NS_cal}. For this purpose, the reliance on the stability of the noise source was minimal, requiring only that its output should be stable through each night. However, for the survey-wide calibrations that were ultimately applied, the noise source was required to be stable over the course of the entire survey, and we found this criterion to be adequately met (Section~\ref{subsubsec:apply_Jy_cal}). 

\subsubsection{Raster Scans}\label{subsubsec:raster}
At the beginning and end of each night of the survey we observed one or two (timing permitting) of the four strong calibrators, Cyg~A, Cas~A, Tau~A, and Vir~A. These observations consisted of a raster scan of the source in equatorial coordinates, scanning up and down in declination, with tracks spaced by 0.5$\arcdeg$ in $\alpha$/cos($\delta_{\text{source}}$). The raster scans covered an area $8\arcdeg \times 8\arcdeg$ centered on each source. The noise source was fired during the raster scans of the calibrators (at the turning points in the declination tracks), allowing for the raster scan data to also be converted to noise-source units. 

\begin{figure}
    \centering
    \includegraphics[width=1\hsize]{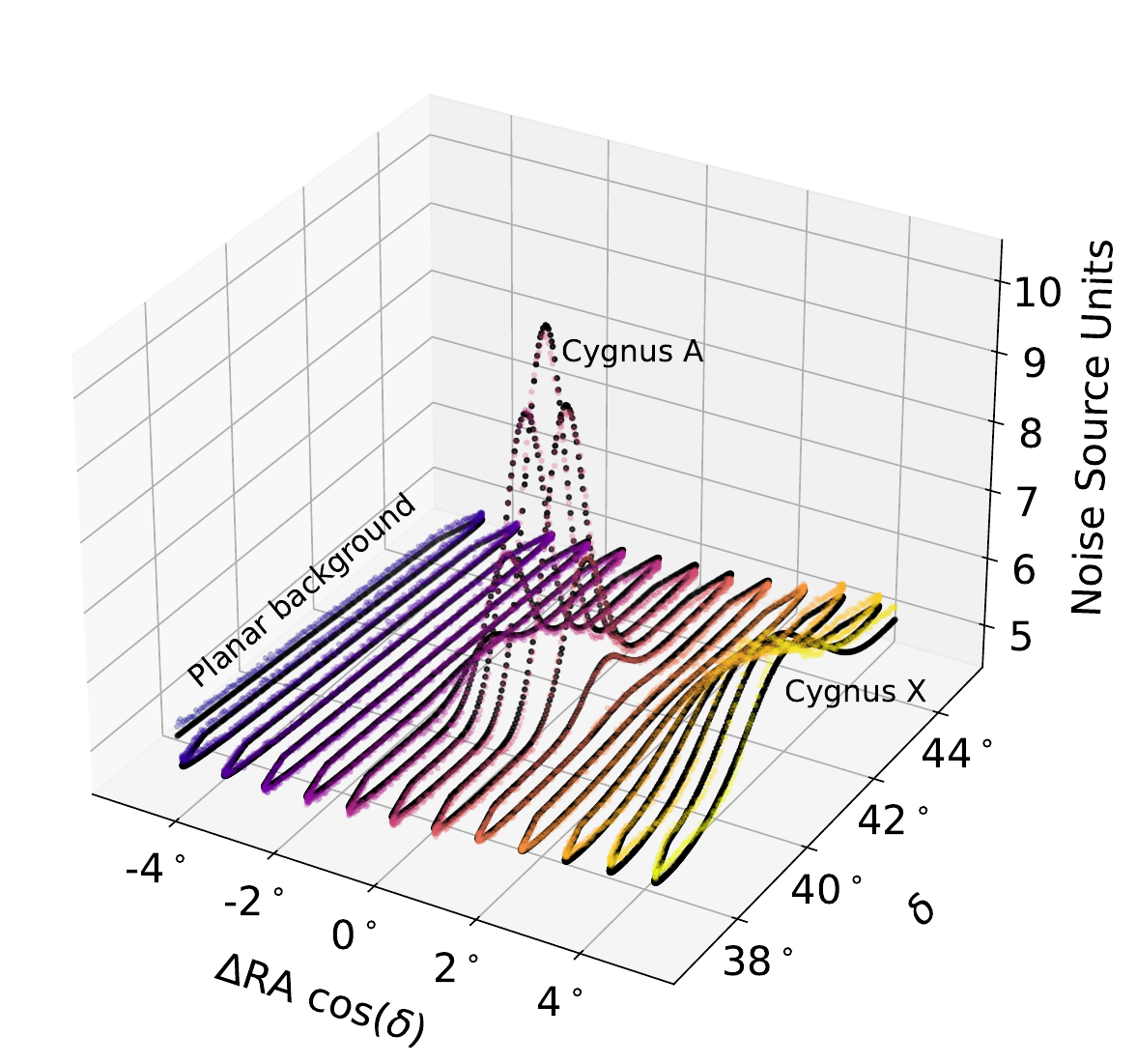}
    \caption{An example of a Cyg~A raster scan at 1010 MHz. The colored points are the raster scan data converted into noise-source units. The black dots show the fitted model, which consists of a 2D sloped planar background, a Gaussian centered at the position of Cyg~A, and a second Gaussian to approximate the effect of nearby Cyg~X on the planar background.}
    \label{fig:raster3D}
\end{figure}

\begin{table}[t]
\begin{center}
\caption{Parameters of the  polarization survey}\label{specs}
\begin{tabular}{ll}
\hline \hline
Antenna dimensions            & $15{\times}18$~m \\
Reflector optics      & Shaped offset Gregorian with \\
& Mizigutch condition \\
Aperture diameter            & 15\,m \\
Feed                         & dual circular polarization \\
Frequency coverage           & 350 to 1030\,MHz \\
Receiver noise temperature   & 55 to 80\,K \\
Angular resolution           & $3.6^{\circ}$ to $1.3^{\circ}$ \\
Frequency resolution         & 83.3\,kHz \\
Coverage (declination)  & ${-20^{\circ}} < {\delta} < {+90^{\circ}}$ (J2000) \\
Coverage (right ascension) & ${0^{h} < \alpha} < {24^{h}}$ (J2000) \\
Spatial sampling             & Full Nyquist (2807 of 2880  \\
   & planned scans) \\
Observation dates            & 2022 June to 2023 January \\
Data loss to RFI{\thinspace}$^{\dagger}$ & $\sim{25}$\% of frequency band\\
Intensity calibration        & absolute \\
\hline
\end{tabular}
\end{center}
~~~~~~~~Note: $^{\dagger}$ RFI = radio frequency interference
\end{table}

As the four calibrators are unresolved in the $1.3\arcdeg$ to $3.6\arcdeg$ beam of the DRAO-15, in the raster scans the flux density of each is convolved across the Gaussian of the beam, resulting in a 2D Gaussian representing the source. For each raster scan, we used a Levenberg-Marquardt least-squares algorithm, included in the \texttt{astropy.modeling} package \citep{Astropy}, to fit a 2D Gaussian plus a sloped planar background to each frequency channel in both LL* and RR*. Cyg~A required a second Gaussian to be added to the model to account for nearby Cyg~X in order to accurately characterize the background. An example of the resulting fit for a Cyg~A scan at 1010 MHz is shown in Figure \ref{fig:raster3D}. We compared the amplitude of the fitted Gaussian for each calibrator in noise-source units to the known flux density and spectral index of the calibrators \citep{perl17} to determine the noise-source-to-Jy~beam$^{-1}$ conversion factor (see Section~\ref{subsubsec:apply_Jy_cal}). 

In addition to the amplitudes, the outputs from the fitting process included the HPBW of the Gaussian and its position. The HPBW values from such a scan are shown in Figure~\ref{fig:beamw}, and are used in the absolute calibration step (Section~\ref{subsec:abscal}) to convert the data into units of brightness temperature (K). The positions obtained served as a daily check of telescope pointing (Figure~\ref{fig:pointing820}); no significant changes in pointing were detected in the course of the survey.

\subsection{The Observed Raw Dataset}\label{subsec:dataet}
We list observational details of the survey in Table~\ref{specs}. The total intensity (RR* and LL*) and cross-polarization (RL$^*$ and LR$^*$) data products were recorded for all $\sim8000$ 83.33~kHz-wide channels (down-sampled from the raw 41.67~kHz) for a total data volume of $\sim2$~TB of azimuth scans and $\sim0.5$~TB of calibration raster scans. In addition to the survey data collected, metadata were also recorded pertaining to environmental conditions (i) outside, (ii) in the control building housing the spectrometer and (iii) in the focus box containing the front-end electronics.

\begin{figure}
    \centering
    \includegraphics[width=1\hsize]{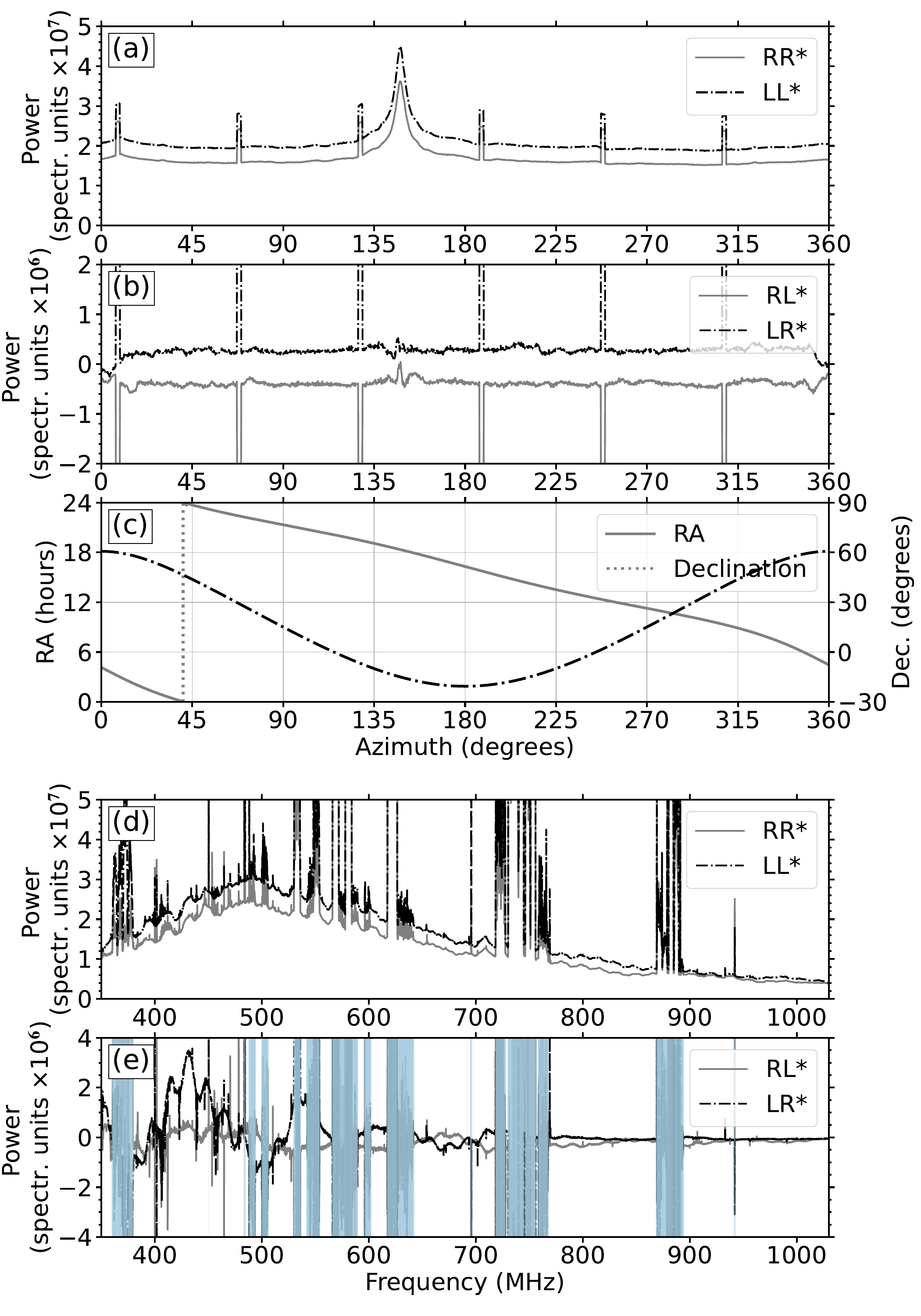}
    \caption{An example of the raw data in a \lowel scan from Phase 1 of the survey. (\textit{a}) The total power, RR* and LL*, and (\textit{b}) the polarized signal, RL* and LR*, as a function of azimuth at 610 MHz in spectrometer units (binned into a 1~MHz wide channel). (\textit{c}) The RA and declination coverage as a function of azimuth for this particular \lowel scan. (\textit{d}) The total power and (\textit{e}) polarized signal frequency spectrum in spectrometer units at ${a}=350\arcdeg$. Please note the factor of 10 difference in displayed scale between the total (panels \textit{a}, \textit{d}) and polarized (panels \textit{b}, \textit{e}) intensities. The regularly spaced, rectangular-shaped signals in the azimuth plots are the noise-source firing (note this signal exceeds the power scale in panel \textit{b}). In panel \textit{e} we show the persistent RFI mask in blue to mitigate confusion between  frequency-dependent fluctuations and the RFI spikes in the polarization data. }
    \label{fig:examplescan}
\end{figure}

\begin{figure}
    \centering
    \includegraphics[width=1\hsize]{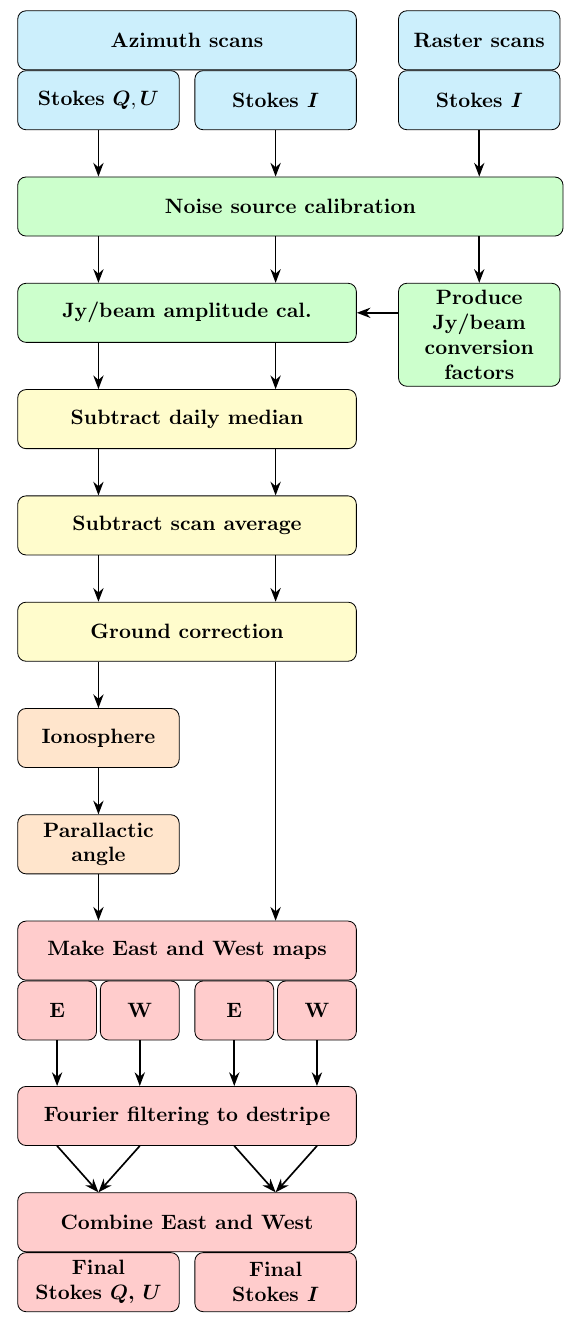}
    \caption{Flowchart depicting the main data processing steps. Blue boxes (rows 1 and 2) represent the raw data in spectrometer units (Section~\ref{subsec:dataet}). Green boxes (rows 3 and 4) correspond to the calibration steps (Sections~\ref{subsubsec:apply_NS_cal}--\ref{subsubsec:apply_Jy_cal}). Yellow (rows 5 -- 7) and orange boxes (rows 8 and 9) represent various amplitude (Sections~\ref{subsec:filter}--\ref{subsec:ground}) and angle (Section~\ref{subsec:iono}) correction steps. Red boxes (rows 10 -- 14) show the map-making steps (Section~\ref{subsec:basket}).}
    \label{fig:flowchart}
\end{figure}

An example of a raw azimuth scan is shown in Figure~\ref{fig:examplescan}, and we note a few key features here. The noise-source signal appears as regularly spaced, rectangular-shaped pulses in the azimuth plots (panels a and b). There is an offset of $\sim20\%$ between the RR* and LL* total power (panel a), which is corrected once the calibration to Jy~beam$^{-1}$ is applied to each of RR* and LL* independently (Sections~\ref{subsubsec:raster},~\ref{subsubsec:apply_Jy_cal}). Instrumental polarization causes an offset in both the RL$^*$ and LR$^*$ products (panel b), which is corrected in the initial calibration stage (Section~\ref{subsec:filter}). The strong source near ${a}=150\arcdeg$ (panel a) is the Galactic midplane, which produces instrumental leakage in the polarized signal (panel b). We quantify this leakage using the calibrated maps in Section~\ref{subsec:leakage}. The increase in polarized signal between ${a}=345\arcdeg$ and ${a}=15\arcdeg$ (the ends of the scan; panel e) corresponds to the highly polarized Fan Region, centered on Galactic coordinates $\ell\approx130\arcdeg$, $b\approx5\arcdeg$ \citep{Hill2017}. 

\section{Data Processing}\label{sec:process}
In this section we describe the steps in processing the survey data, from applying the calibrations described in Section~\ref{subsec:calobs} through to the final maps. These steps are outlined in Figure~\ref{fig:flowchart}. We processed the raw scans, which were stored as \texttt{hdf5} files, using \texttt{python} for RFI flagging (Section~\ref{subsec:rfi}) and calibration into Jy~beam$^{-1}$ (Section~\ref{subsec:apply_cal}), after which we converted them to \texttt{SDFITS} files for the map-making steps. The latter included ground (Section~\ref{subsec:ground}) and ionosphere (Section~\ref{subsec:iono}) correction, followed by destriping and combining east and west maps (Section~\ref{subsec:destripe}), for which we used \texttt{fortran} and \texttt{IDL} routines developed by \cite{carr19} for the S-band Polarization All Sky Survey (S-PASS). We refer to this set of map-making software as `the S-PASS software'.

\subsection{Radio Frequency Interference}\label{subsec:rfi}

Although the DRAO site is largely protected from RFI by the surrounding mountains as well as government regulation, contamination from wireless communications still affects a significant portion of the DRAGONS frequency band. We applied two stages of RFI excision to the 83~kHz channelized survey data prior to feeding into the S-PASS software: (i) identifying persistent contaminated channels (e.g. TV and mobile phone bands) and (ii) flagging each scan for instances of intermittent RFI. We excluded channels with persistent RFI from use in further processing stages. If intermittent RFI contaminated only a small fraction of the scan, we calibrated the remaining data in the scan and included the scan in the map-making. The S-PASS map-making software also includes a step for removing any remaining single-pixel RFI instances, as well as discarding scans with high rms variability, which can be indicative of remaining, low-level RFI (Section~\ref{subsec:filter}).

\begin{figure*}[t]
    \centering
    \includegraphics[width=1\hsize]{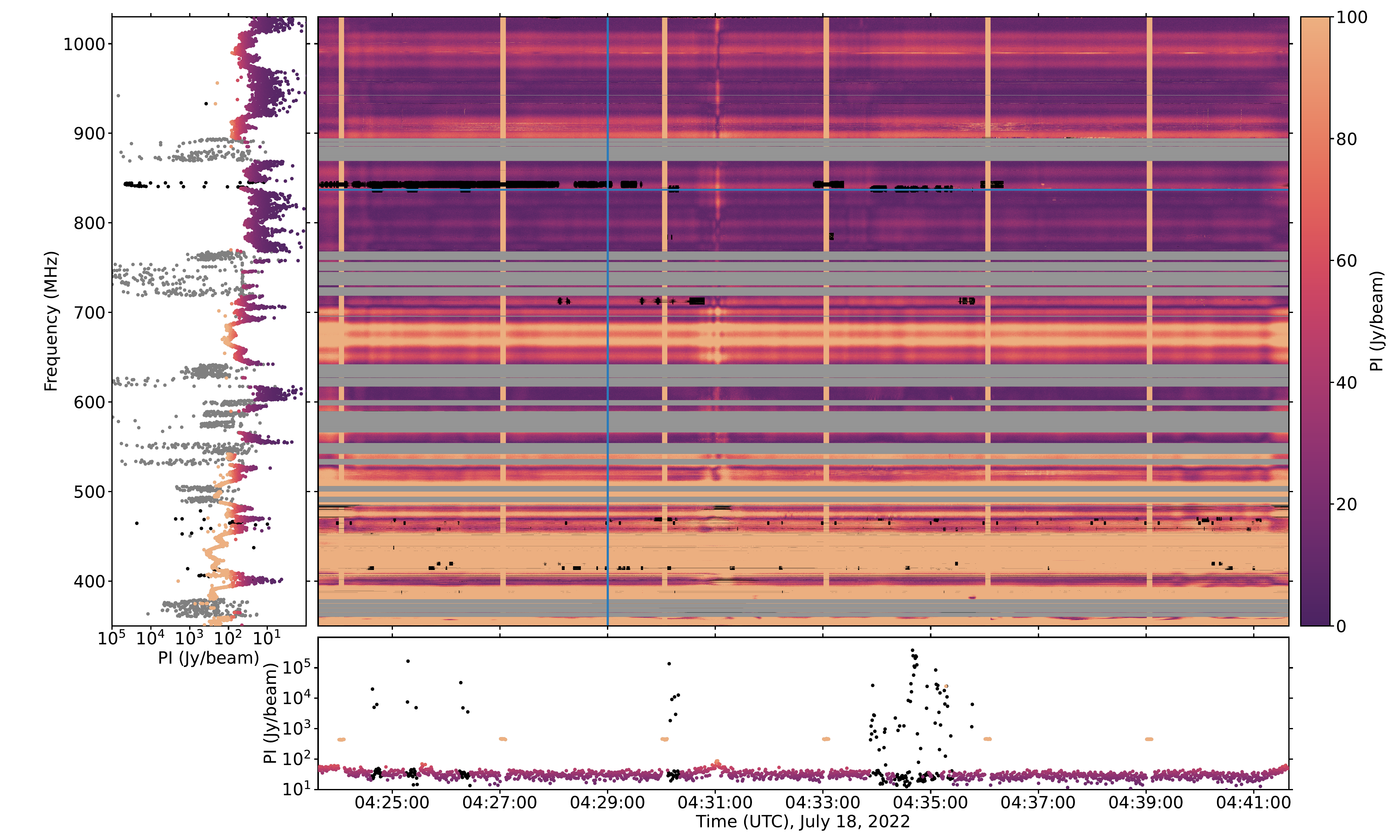}
    \caption{An example of the RFI masking for a \lowel scan from Phase 1 of the survey. The waterfall plot shows the linearly polarized intensity, PI, converted into units of Jy~beam$^{-1}$. The persistent RFI mask is shown in gray, and the intermittent RFI mask for this scan is shown in black. The six vertical stripes of high PI are the calibration noise source. The plots to the left and at the bottom show one-dimensional cuts indicated by the blue lines on the waterfall plot: the vertical cut shows the frequency spectrum (\textit{left}) at a particular timestamp, and the horizontal cut shows the time series data (\textit{bottom}) at 837~MHz}
    \label{fig:rfimask}
\end{figure*}

We identified channels with persistent RFI from the Gaussian fits to the calibration raster scans. For each channel of a raster scan, we determined a $\chi^2$ value from the residuals of the fitted model to the scan data. If the $\chi^2$ for a channel was consistently high across all phases of the survey, the channel was flagged as containing persistent RFI. This eliminated 25\% of the channels, and those remaining were subject to intermittent RFI detection in the individual scans. This persistent RFI mask is shown in gray for one example azimuth scan in Figure~\ref{fig:rfimask}.

In each azimuth scan we identified intermittent RFI as segments of data characterized by a sharp change in signal level as a function of time or frequency in total or linearly polarized intensity (PI). In each frequency channel in the scan, we identified candidate RFI events by flagging integrations with a PI signal 3 times the median PI along the scan and grouping together contiguous instances into a candidate event. For each candidate RFI event, we then verified that it was indeed RFI by finding the edge of the event in the frequency domain, scanning within 5~MHz on either side of the central frequency to find where the slope of PI versus frequency exceeded a set threshold. This avoided the misidentification of brightly polarized regions on the sky as RFI. Independently, we checked for large changes in total intensity along the frequency axis for each integration to pick up RFI instances missed by the time-domain scanning approach. Examples of intermittent RFI and the results of the flagging algorithm are shown in black for the example azimuth scan in Figure~\ref{fig:rfimask}, demonstrating their extent in time and frequency. 

From the persistent and intermittent RFI masks, we produced a combined mask for each azimuth scan. We ultimately binned the data into 1~MHz bins for map-making (Section~\ref{subsec:filter}) and the subsequent Faraday synthesis. Any 83.3 kHz channels identified as containing persistent RFI were excluded from the binning, while channels with only intermittent RFI were included, with the contaminated integrations masked. With 12 83.3~kHz channels contributing to each 1~MHz bin, it was possible to retain useful data from these narrow channels even in cases where more than 50\% of integrations were masked for intermittent RFI, provided the adjacent channels were not heavily contaminated. In some cases, a 1~MHz bin is on the edge of a wider band of persistent RFI, resulting in only a few channels contributing to the bin. In these cases, the 1~MHz channels suffer increased noise (Section~\ref{subsec:errors}) and a few were discarded following visual inspection of the resulting maps (Section~\ref{subsec:rmsynth}). 

\subsection{Calibrating the Scans}\label{subsec:apply_cal}
After the RFI mask for each scan was produced, we applied it to the data, which we then converted first into noise-source units and then into Jy~beam$^{-1}$. 

\subsubsection{Conversion to Noise-source Units}\label{subsubsec:apply_NS_cal}
For all noise-source instances in all raster scans and azimuth scans, we determined the amplitude of the noise source for each of RR*, LL*, RL* and LR* from the median of the 10 integrations during which the noise source was on (6~s) compared to adjacent `off' integrations in the scan. For RL* and LR* the sign information was retained. These amplitudes revealed a time-dependent gain variation over the course of each observing night that correlated strongly with outside temperature fluctuations. The temperature-dependent component was the RFoF connection between the focus and the spectrometer.

We fitted a 9th degree polynomial\footnote{The high order of the polynomial was selected to capture the varying rates of outside temperature fluctuations. This was particularly important for the summer months of observing when the temperatures were least stable.} to the noise-source amplitudes as a function of time for each observing night (including both raster scans and azimuth scans, for each frequency channel and correlation product), and applied these polynomials to all data as a conversion factor into noise-source units. For the total intensity data products, this is simply an amplitude correction, modifying the raw data $\text{RR*}_{\text{raw}}$ and $\text{LL*}_{\text{raw}}$ as
\begin{eqnarray}
    \text{RR*}&=&\text{RR*}_{\text{raw}}/\text{RR*}_{\text{noise}}\\
    \text{LL*}&=&\text{LL*}_{\text{raw}}/\text{LL*}_{\text{noise}},
\end{eqnarray}
where $\text{RR*}_{\text{noise}}$ and $\text{LL*}_{\text{noise}}$ are the fitted polynomial noise-source values. For RL* and LR* the phase must also be taken into account in order to convert the raw polarization data, $\text{RL*}_{\text{raw}}$ and $\text{LR*}_{\text{raw}}$, into noise-source units as
\begin{equation}
    \text{RL*}+i\text{LR*} = e^{2i\tau_{\text{noise}}}(\text{RL*}_{\text{raw}}+i\text{LR*}_{\text{raw}})/\text{PI}_{\text{noise}}.
\end{equation}
Here $\text{PI}_{\text{noise}}$ is the `polarized intensity' of the noise polynomial, $\text{PI}_{\text{noise}}=\sqrt{(\text{RL*}_{\text{noise}})^2+(\text{LR*}_{\text{noise}})^2}$ and $\tau_{\text{noise}}=\frac{1}{2}\tan^{-1}(\text{RL*}_{\text{noise}},\text{LR*}_{\text{noise}})$ is the corresponding phase. We chose the approach of converting the data into noise-source units in this manner rather than deriving a temperature-dependent gain correction because (i) the temperature measurements were taken at longer time intervals of once per 15 minutes compared to the 3-minute cadence of the noise source, and (ii) the temperature measurements were not recorded specifically along the RFoF connection. Following the conversion of the data into noise-source units, we masked the integrations that included the noise-source signal for subsequent processing steps. The Gaussian fits to the raster scans (Section~\ref{subsubsec:raster}) were then done in noise-source units (Figure~\ref{fig:raster3D}).
\newpage
\subsubsection{Conversion to Jy~beam$^{-1}$}\label{subsubsec:apply_Jy_cal}
We determined the noise-source-units-to-Jy~beam$^{-1}$ conversion factor as a function of frequency by comparing the Gaussian-fitted amplitudes of the raster scans in RR* and LL* to the flux density of the corresponding calibrator sources in \citet{perl17}. 

Two of the calibrators, Cas~A and Tau~A, are supernova remnants, and are known to change over time. We tested their suitability as calibrators using the Cyg~A conversion factors to calibrate the Cas~A and Tau~A raster amplitudes for days when both Cyg~A and one of these were observed. Cyg~A is a bright radio galaxy and its flux is not expected to vary in time. From the resulting calibrated Cas~A and Tau~A spectra, we found that these sources have changed substantially since the observations for \citet{perl17} and we rejected them for calibration purposes. When the fourth calibrator, the radio galaxy Vir~A, was calibrated using the conversion factors developed for Cyg~A, the resulting spectrum agreed with \citet{perl17}. However, the lower flux density of Vir~A led to a noisy spectrum, and using conversion factors developed from Vir~A raster scans would have introduced additional noise to the scans. Ultimately, we only used Cyg~A raster scans to calibrate all azimuth scans.

\begin{figure}
    \centering
    \includegraphics[width=1\hsize]{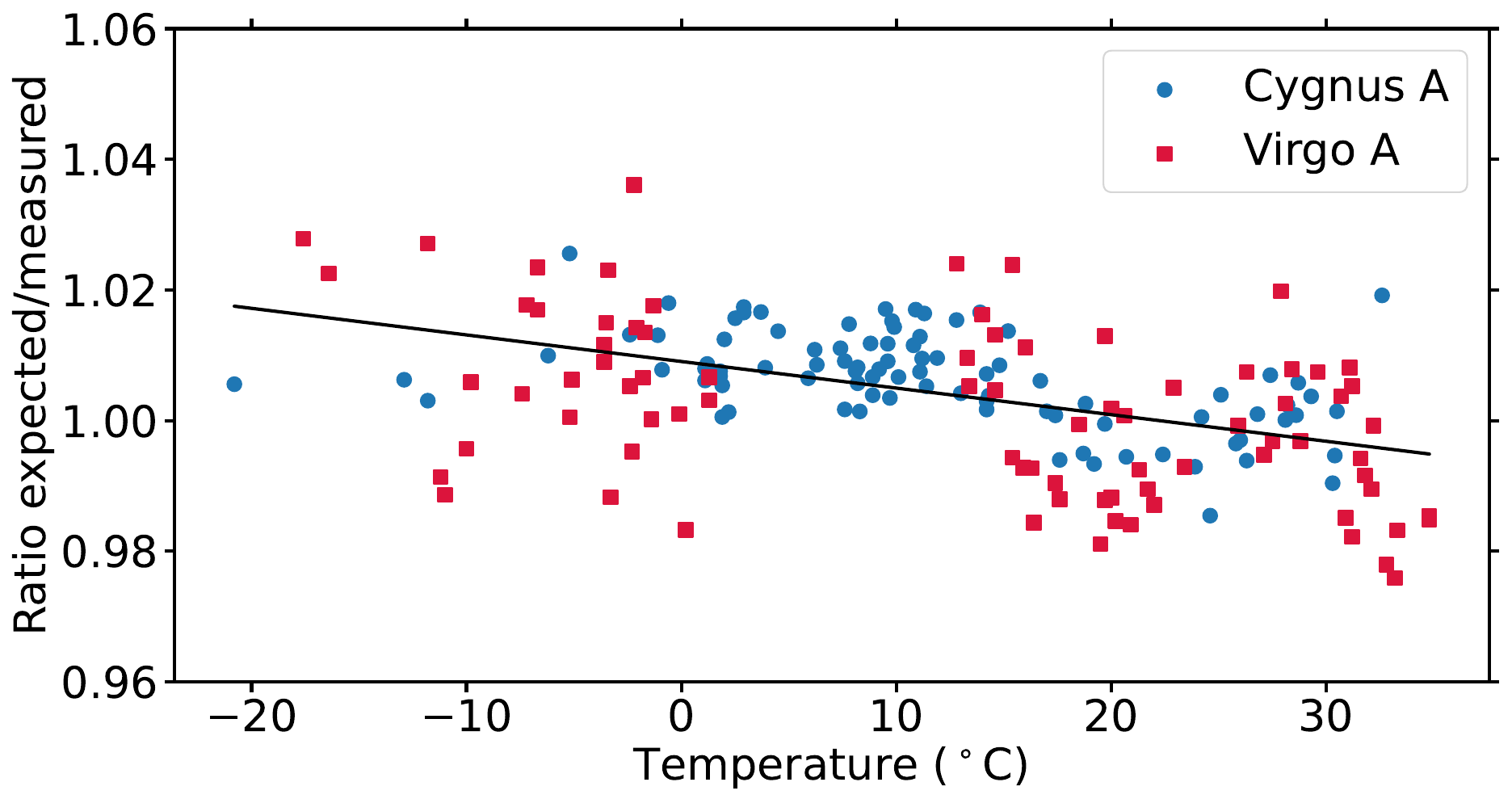}
    \caption{The ratio of the expected \citet{perl17} flux density to the raster scan amplitudes calibrated using the survey-wide calibration factors averaged across the spectrum for each scan as a function of ambient temperature. Vir~A scans are plotted with red squares and Cyg~A scans are plotted with blue circles. A linear fit to the ratio versus temperature plot is shown as a black line.}
    \label{fig:raster_temp}
\end{figure}

We produced survey-wide conversion factors by combining the fitted amplitudes for all Cyg~A raster scans to produce a single set of RR* and LL* amplitude spectra. For the conversion spectra to be valid for all scans, the noise source was required to remain constant throughout the survey. Microwave noise sources are sensitive to temperature, and so the noise source was maintained at a controlled temperature above ambient (at $30\arcdeg$~C) in a well-insulated box. We tested the effectiveness of this in achieving the required stability by applying the master conversion factor spectra to all Cyg~A and Vir~A raster scans and finding the average ratio between the expected spectrum for the source from \citet{perl17} and the measured source spectrum for each scan. Figure \ref{fig:raster_temp} shows the average ratio as a function of ambient temperature for each Cyg~A (blue dots) and Vir~A (red dots) scan. We note that although there is a slight correlation of this ratio with ambient temperature, the ratio remains within 3.5\% of unity across the full range of temperatures. To account for the slight variation in the noise source across the survey, we applied a temperature-based scale factor to each scan. We determined the temperature scale factor from the linear fit to the average expected-to-measured ratio as a function of temperature values, shown as a solid black line in Figure \ref{fig:raster_temp}. 

We applied the noise-source-units-to-Jy~beam$^{-1}$ conversion factors to the RR* and LL* components of all azimuth scans such that each of RR* and LL* represent the total intensity. After this calibration process the median percent difference between RR* and LL* for all scans was less than 5\% across most of the frequency band, peaking at 10\% near 450~MHz. We then calculated the Stokes $I$ total intensity values for the azimuth scans as $I=\frac{1}{2}(\text{RR*}+\text{LL*})$, following \cite{Wolleben_2019,woll21}.\footnote{The correct definition of Stokes~$I$ is $I=\text{RR*}+\text{LL*}$, \textit{without} the factor of 1/2 \citep{robi21}. However, our method of calibrating RR* and LL* intensities independently produces maps of total and polarized intensity that agree in brightness with previous radio surveys.}

At this stage, no further angle corrections were applied to RL* and LR*, and it was assumed that $Q=\text{RL*}$ and $U=\text{LR*}$ \citep[similar to][]{Wolleben_2019,woll21}. The noise-source-units-to-Jy~beam$^{-1}$ conversion factor, $R_{QU}$, used for Stokes $Q$ and $U$ is
\begin{equation}
    R_{QU}=\sqrt{R_{\text{RR}}R_{\text{LL}}},
\end{equation}
where $R_{\text{RR}}$ and $R_{\text{LL}}$ are the conversion factors for the RR* and LL* total intensity components, respectively. 

Finally, we applied a correction for atmospheric attenuation to the azimuth scans, calculated using Equation~1 of \citet{Gibbins1986}. The attenuation factor varies with frequency and the distance through the atmosphere, which we determined as,
\begin{equation}
    d = \frac{6 - 0.55}{\cos(\theta)}\, \textrm{km},
\end{equation}
where 6 km is the atmospheric scale height from sea level \citep{Allnutt1989}, 0.55 km is the elevation of the DRAO, and $\theta$ is the zenith angle of the scan. For the low elevation scans, the attenuation adjustment increases the scan flux density by 0.8\% at 350 MHz to 1.9\% at 1030 MHz. For the high elevation scans, the attenuation adjustment ranges from 0.4\% to 0.8\%. 

\subsection{Base level Adjustments and Filtering}\label{subsec:filter}
Following RFI excision and calibration into Jy~beam$^{-1}$, we binned the Stokes $Q$, $U$ and $I$ data into 1~MHz frequency channels, each spanning 12 of the 83.3~kHz-wide channels. At this stage, we also separated the scans into east and west halves, which we then saved as \texttt{SDFITS} files to feed into the S-PASS software. Henceforth we use `scan' to refer to an individual $180\arcdeg$ east or west scan rather than the full $360\arcdeg$ raw observation scans. We applied several initial steps to the full set of scans at each binned frequency to clean up the data prior to inclusion in the maps:
\begin{enumerate}
    \item We removed single-integration outliers to eliminate remaining RFI not flagged in the initial RFI masking (Section~\ref{subsec:rfi}). These outliers were identified as data points having an absolute intensity value (in Stokes $I$, $Q$, or $U$) 3 times higher than the absolute intensity value of either of the two adjacent points in the time series, and were replaced by the mean of the adjacent data points.
    \item We subtracted a daily median from each of Stokes $I$, $Q$ and $U$ at each frequency. This removed any overall frequency- and time-dependent instrumental offset. The calculations of the medians excluded a $5\arcdeg$ latitude range on either side of the Galactic midplane to avoid biasing the median values with strong Galactic emission. For Stokes $Q$ the daily median varied between $\sim$-200 and $\sim$400~Jy~beam$^{-1}$, and for Stokes $U$ the daily median varied between $\sim$-50 and $\sim$100~Jy~beam$^{-1}$ across the band, which corresponds to the Stokes $I$ to $Q$ and $U$ polarization leakage.
    \item The outlier-flagging step was repeated with the scan zero-levels adjusted by the median-subtraction.
    \item We subtracted a constant, fitted base level (average) from each scan in Stokes $I$, $Q$ and $U$ at each frequency to ensure a similar base level between adjacent scans and between east and west scans at the intersection points. For Stokes $Q$ and $U$, the large sinusoidal modulation by the parallactic angle along an east or west scan ensures that the data along each scan should average to zero. This is not the case for Stokes $I$, meaning that the zero-level is lost as a consequence of this step and a correction is required after map-making. This step also removed the constant component of the ground contributions in Stokes $Q$ and $U$ (see Section~\ref{subsec:ground}).
    \item We discarded scans with high rms fluctuations in either $Q$, $U$ or $I$ at each frequency. In the 350--600~MHz range, typically between 30 and 300 of each of the 2812 \highel and 2802 \lowel observed scans\footnote{1406 \highel and 1401 \lowel full $360\arcdeg$ scans yield 2812 \highel and 2802 \lowel $180\arcdeg$ east and west scans.} were discarded, while in the 600--1030~MHz range most channels had less than 30 scans discarded. Channels with more than $\sim$300 discarded scans at the low frequencies, and more than $\sim$100 discarded scans at the high frequencies were excluded from the final maps because of insufficient spatial sampling.
\end{enumerate}

\subsection{Ground Correction}\label{subsec:ground}
\begin{figure*}
    \centering
    \includegraphics[width=1\hsize]{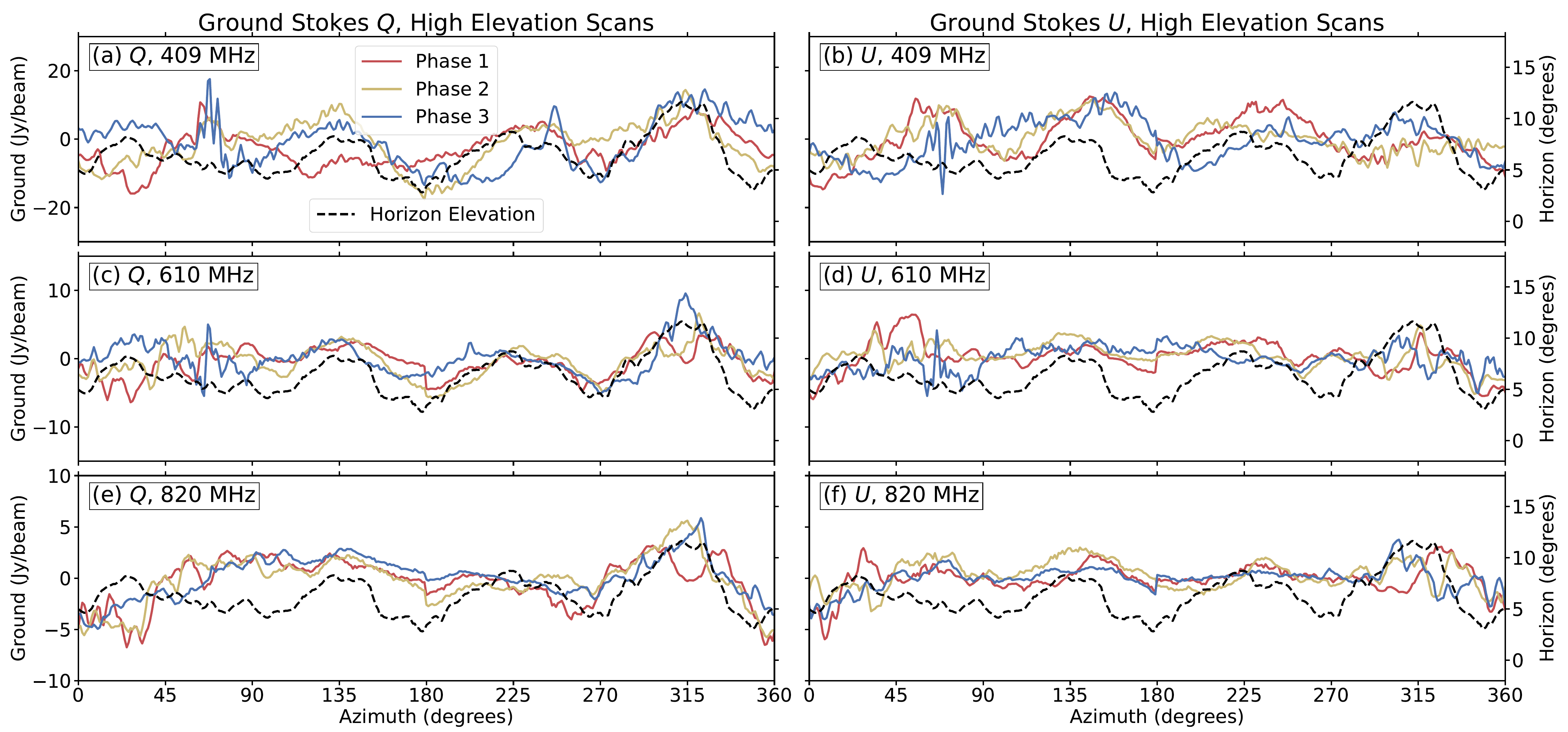}
    \caption{Examples of the seasonally-calculated Stokes $Q$ and $U$ ground profiles for the \highel scans at three example frequencies across the band. The profiles from the survey phases P1, P2, P3 are shown in red, yellow and blue, respectively. The black dashed line in each panel shows the horizon elevation profile around the DRAO (right-hand vertical axis). Here Stokes $Q$ and $U$ are measured in the native azimuth/elevation coordinates of the telescope. 1~Jy~beam$^{-1}$ is approximately equal to 0.05 K.}
    \label{fig:groundhigh}
\end{figure*}

\begin{figure*}
    \centering
    \includegraphics[width=1\hsize]{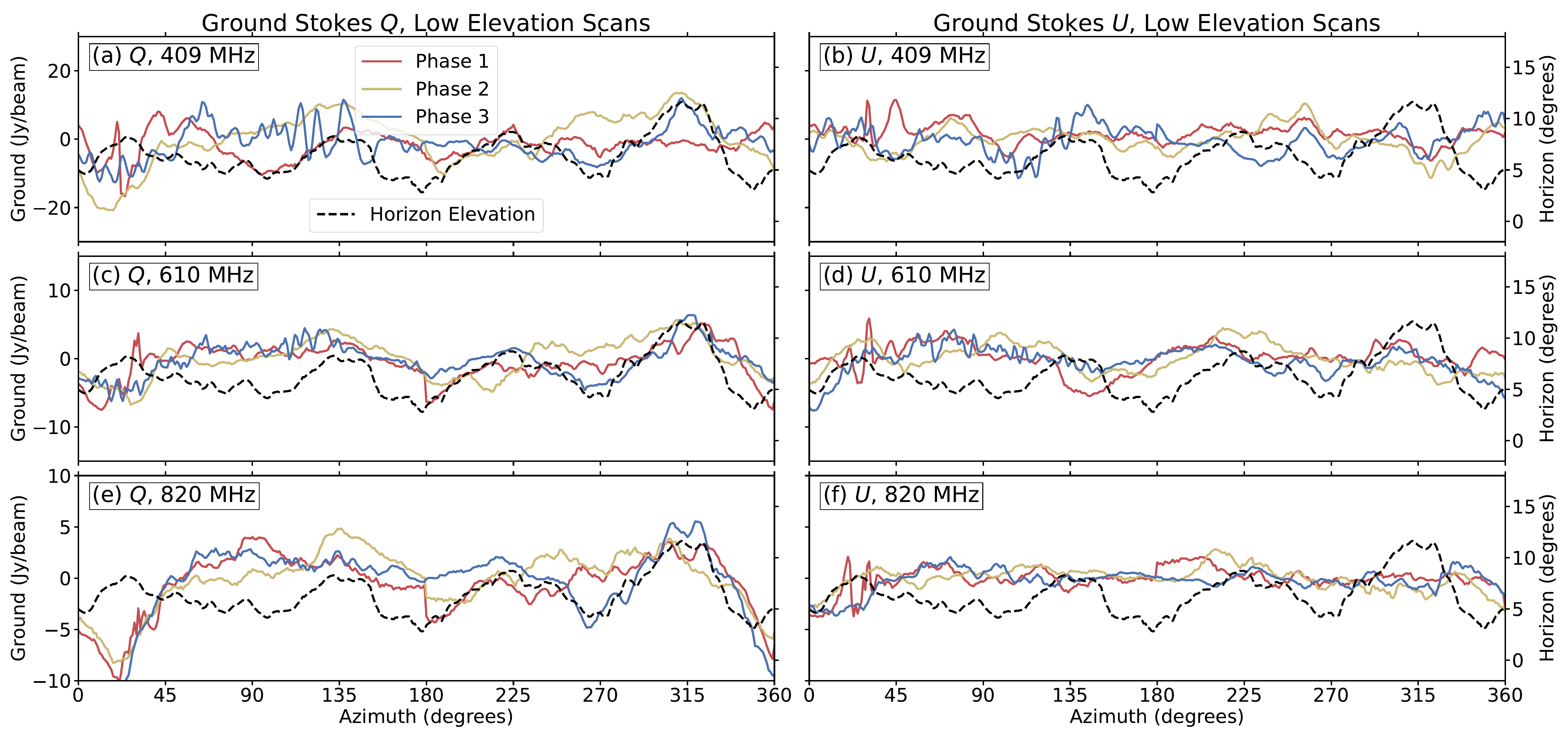}
    \caption{Examples of the seasonally-calculated Stokes $Q$ and $U$ ground profiles for the \lowel scans at three example frequencies across the band. See Figure~\ref{fig:groundhigh} for details.}
    \label{fig:groundlow}
\end{figure*}

With the constant-elevation azimuth scan strategy used, each azimuth in an east or west scan (for each set of \highel and \lowel scans) corresponds to a unique declination (see Figure~\ref{fig:decvsaz}). We begin with the assumption that the sky averages to zero in Stokes $Q$ and $U$ across a large RA range at each declination, and any remaining signal in a narrow declination bin is due to ground contamination from the corresponding azimuth. This is a reasonable assumption, given that we expect Stokes $Q$ and $U$ to fluctuate on scales much smaller than the full RA coverage. We averaged the data in $1\arcdeg$-wide declination bins separately for east, west, \highel and \lowel sets of scans to produce a ground profile in Stokes $Q$ and $U$ at each frequency (examples in Figures~\ref{fig:groundhigh} and \ref{fig:groundlow}). These profiles, which we subtracted from the data, represent the azimuth-dependent component of the ground contribution, while the constant component is already subtracted for each scan in Step 4 described in Section~\ref{subsec:filter}. 

The effectiveness of the ground correction improved when we calculated separate ground profiles for each of the three survey phases to account for seasonal variation, and subtracted these from the data accordingly. This is not unexpected, as ground properties at low frequencies vary rapidly with water content \citep{Hall1979}. The ground profiles for the three phases are shown as the red, yellow, and blue lines in Figures~\ref{fig:groundhigh} and \ref{fig:groundlow}. The dashed black line in each panel depicts the elevation profile of the horizon around the DRAO. Particularly in Stokes $Q$, there is a strong resemblance between the derived ground profiles and the physical features of the horizon. 


We expect the ground emission to be vertically polarized, therefore primarily appearing as positive Stokes $Q$. Since the constant component of the ground emission is already subtracted in Figures~\ref{fig:groundhigh} and \ref{fig:groundlow}, these profiles do not show whether ground Stokes $Q$ is always positive. As well, on rough mountainous terrain the ground emission will be polarized perpendicular to the local ground surface, and there may be considerable Stokes $U$ and negative Stokes $Q$ components. These effects are more pronounced at lower frequencies.

\begin{figure*}[t]
    \centering
    \includegraphics[width=1\hsize]{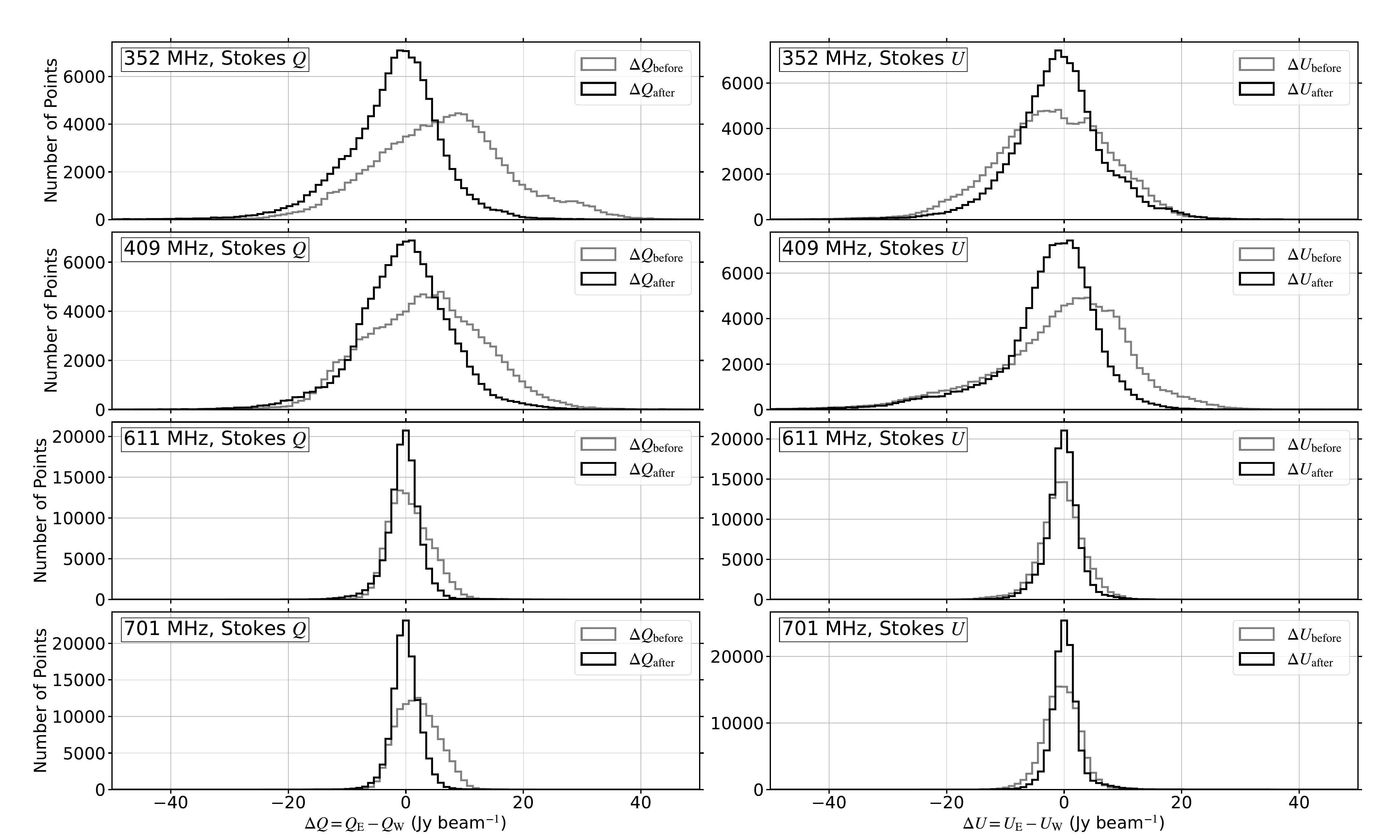}
    \caption{Histograms showing the differences in Stokes~$Q$ (\textit{Left}) and Stokes~$U$ (\textit{Right}) between east and west high-elevation scans at their intersection points before (gray lines) and after (black lines) correcting for ground emission. The distributions generally become narrower and better centered on zero after correcting for the ground, showing the improved agreement between the east and west maps as a result of this step.}
    \label{fig:ground_improve}
\end{figure*}

Since the east and west maps are observed in different ranges of azimuth, the ground contributions in the two maps are not the same. Therefore the east and west maps are not identical prior to ground correction, even though they contain the same sky signal. If our ground correction technique is effective, the difference between east and west maps after correction will be smaller than the difference before correction. Application of this test demonstrated that the differences were indeed reduced, but they were not reduced to zero, indicating that some ground emission remained in the maps. This proved to be a problem at later stages of processing (see Section~\ref{subsec:basket}). At selected frequencies, shown in Figure~\ref{fig:ground_improve}, we compared the distributions of the differences between the east and west values, with and without the ground correction applied, for each of Stokes $Q$ and $U$. The east-west difference distribution typically became 30\% to 50\% narrower with the ground correction applied, indicating substantially improved agreement between the east and west maps. However, at frequencies below $\sim500$~MHz the uncertainties in the ground calculations still produce significant residual differences between the east and west maps. We incorporate these differences into error estimates for the maps as described in Section~\ref{subsec:errors}.

\subsection{Ionospheric Faraday Rotation}\label{subsec:iono}
Faraday rotation in the Earth's ionosphere is indistinguishable from Faraday rotation of astronomical origin in the survey observations alone. The nighttime ionosphere at solar minimum has RM between 0.5 and 2~${\rm{rad}}\thinspace{\rm{m}}^{-2}$, and this translates to a rotation of polarization angle from $21^{\circ}$ to $84^{\circ}$ at our lowest frequency, 350~MHz. Even at solar minimum it would have been essential to correct our survey data for the effects of the ionosphere; since our observations were made on the rising slope of solar cycle 25, it was even more important to make this correction. An ionospheric correction step is not included in the S-PASS software, as it was written for 2.3 GHz, where ionospheric Faraday rotation is negligible.
\begin{figure}[t]
    \centering
    \includegraphics[width=1\hsize]{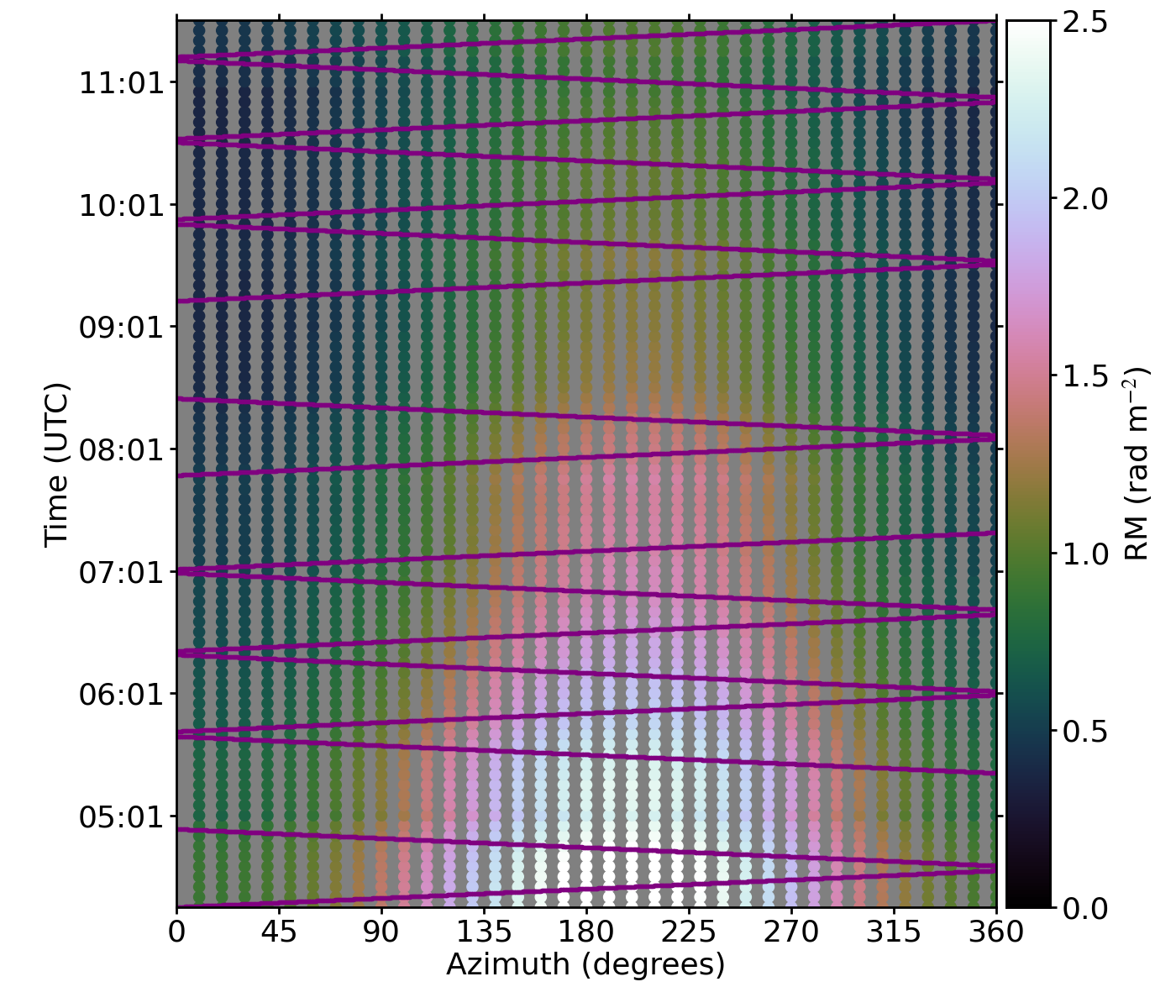}
    \caption{Ionospheric Faraday rotation RM values as a function of azimuth and time, calculated using \texttt{ALBUS}, for the \highel scans on June 7, 2022. The colored points show the RM values at 5-minute cadence with $10\arcdeg$ sampling in azimuth. The purple lines indicate the scans made on this night of the survey, along which the RM values are interpolated to then be applied to the scan polarization data.}
    \label{fig:albustracks}
\end{figure}

Ionospheric modeling is based on the technique developed by \citet{eric01}, using data from Global Positioning System (GPS) receivers. The time delay between two standard GPS frequencies, 1575.42\,MHz and 1227.60\,MHz, measures the total electron content (TEC) along the line of sight from the GPS satellite, through the ionosphere, to the receiver. This technique is the foundation of the widely used routines \texttt{ionFR} \citep{soto13}, \texttt{RMextract} \citep{mevi18} and its more recent upgrade, \texttt{Spinifex}.\footnote{\url{https://git.astron.nl/RD/spinifex}} We used the software package \texttt{ALBUS}{\footnote{\texttt{ALBUS} (Advanced Long Baseline User Software) was originally developed by J.M.\ Anderson to model ionospheric delay as it affects long baseline interferometry. The software is available from: \url{https://github.com/twillis449/ALBUS_ionosphere}.}} whose operating principle is similar, with two key differences. First, other ionospheric modeling routines make use of Vertical TEC calculations from the Center for Orbit Determination in Europe (CODE), which use 240 GPS sites and provide data files of TEC values at a 1-hour cadence. By contrast, \texttt{ALBUS} calculates its own TEC values using additional secondary GPS stations within a selected distance of the observatory, and can do so on a cadence as fast as 30~seconds. Second, most other routines operate by `placing' the electron content at a single fixed height above Earth's surface for calculating the resulting RM. \texttt{ALBUS} instead distributes the electrons into multiple, concentric spherical shells, such that the variations in magnetic field strength and electron density along the LOS are more accurately accounted for. 

Each GPS receiver sees from four to eight satellites (the number above the horizon) at any instant, allowing \texttt{ALBUS} to probe that number of sightlines through the ionosphere. \texttt{ALBUS} fits to the many individual TEC values along different sightlines, deriving a 2D map of ionospheric TEC. This is possible on a 30-second cadence, but we use a 5-minute cadence in all our calculations as there is very little variation over shorter timescales. 

The terrestrial magnetic field is three dimensional, so field intensity and direction vary with height along any LOS. To calculate RM, it is therefore necessary to know how the ionospheric electrons are distributed along the LOS. For this, \texttt{ALBUS} uses the Parameterized Ionospheric Model \citep[PIM;][]{dani95} with three shells (height bins) for the electron distribution, which are equidistant in integrated electron density. 

\texttt{ALBUS} uses the International Geomagnetic Reference Field (IGRF) for the geomagnetic field model \citep{theb15}. The lines of the terrestrial magnetic field slope steeply downwards at the location of the DRAO in Canada, yielding a strong influence of the magnetic field on ionospheric Faraday rotation. \texttt{ALBUS} calculates the ionospheric Faraday rotation by integrating the LOS component of the IGRF magnetic field model weighted by the GPS-derived PIM electron density profile along the LOS path from infinity to the location of the observer for any given telescope pointing.

\begin{figure}
    \centering
    \includegraphics[width=1\hsize]{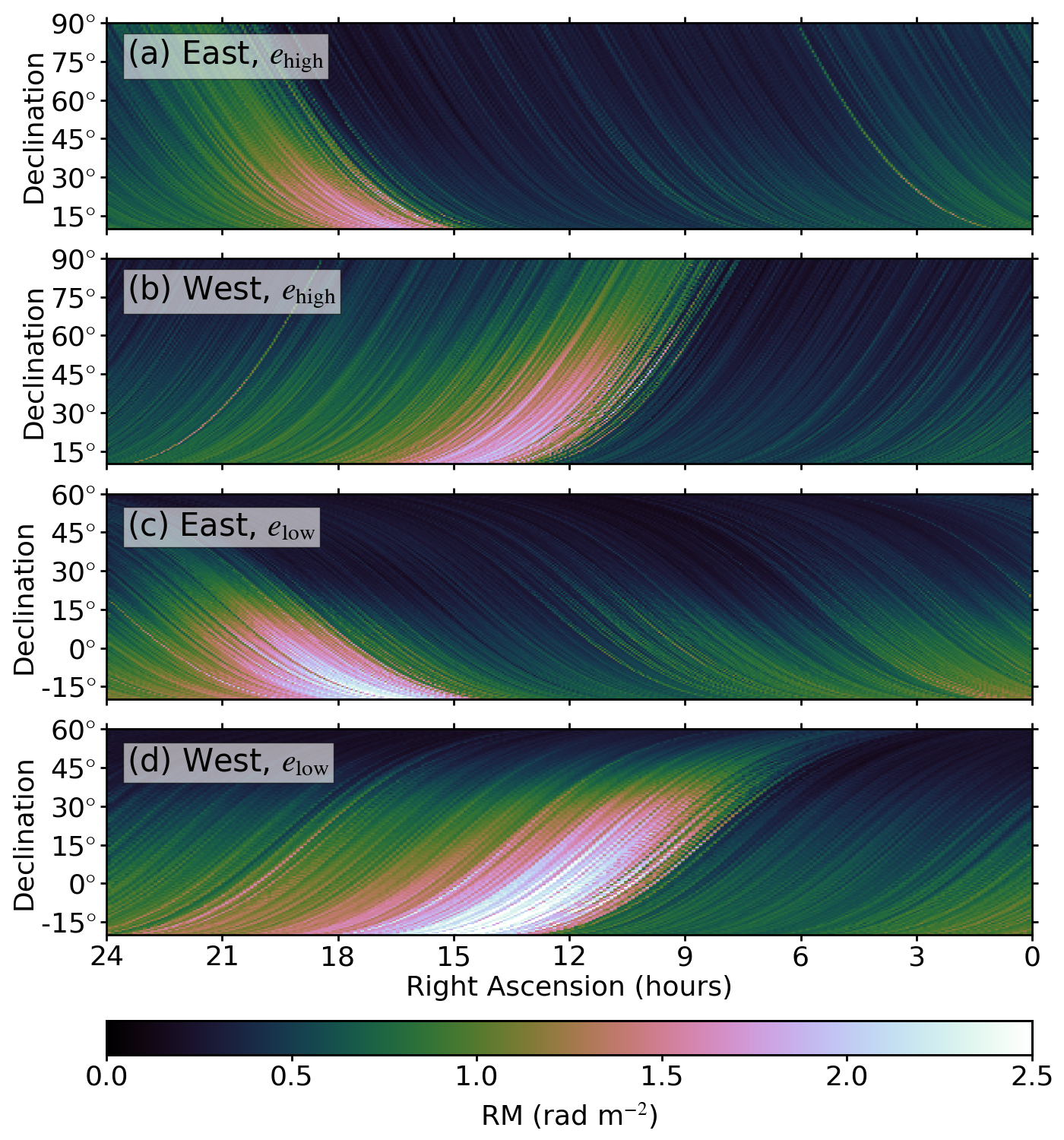}
    \caption{The ionospheric Faraday rotation RM values for all east and west scans in the survey in equatorial coordinates for the two elevations: \highel (\textit{a},\textit{b}) and \lowel (\textit{c},\textit{d}). These are the same scans as shown in Figure~\ref{fig:scans1}. The ionosphere generally produced more Faraday rotation in the summer (P1 and P2) survey phases than in the winter phase (P3).}
    \label{fig:ionomap}
\end{figure}

We used a radius of 350~km around the DRAO 15~m telescope to query available GPS ground station data, yielding 8 to 10 stations for each night of the survey. We implemented a runner script for the \texttt{ALBUS} software, using the GPS data at a 5-minute cadence for every night of the survey and calculating ionospheric RMs from the resulting TEC maps for pointings spaced by $10\arcdeg$ in $0\arcdeg\leq{a}\leq360\arcdeg$ at each of the two survey elevations (\highel and \lowel). This resulted in a grid of ionospheric RM values in time and azimuth for each night of the survey; we show one example in Figure~\ref{fig:albustracks}. We then interpolated these values onto the tracks in time and azimuth traced by the survey scans, thereby producing an ionospheric RM value for each of the 0.6~s integrations and corresponding telescope pointings. The ionospheric RM values for all survey scans are shown in Figure~\ref{fig:ionomap}. 

For each frequency channel, we calculated an ionospheric rotation angle ($\theta_{\text{iono}}$) as
\begin{equation}
    \theta_{\text{iono}}=\lambda^2\text{RM}_{\text{iono}},
\end{equation}
that could then be used to derotate the ionospheric effect from the observed complex polarization, $\tilde{P}_{\text{obs}}=Q+iU$. We applied this rotation along with the parallactic angle ($\theta_{\text{par}}$) correction as
\begin{equation}
    \tilde{P}_{\text{corr}}=\tilde{P}_{\text{obs}}e^{2i(\theta_{\text{par}}-\theta_{\text{iono}})},
\end{equation}
to produce fully calibrated scans with polarization angles referenced to equatorial coordinates.

As shown in Figure~\ref{fig:ionomap}, despite being on the rising slope of the solar cycle, a significant fraction of the survey was observed with relatively low ionospheric RM. This is because the ionospheric Faraday rotation drops rapidly after sunset. For sets of scans conducted during the winter phase (P3; with more than 15 hours of observing possible on many nights), only the scans scheduled close to sunset and sunrise had ionospheric RM values of up to 3~rad~m$^{-2}$, while the majority of scans only suffered $\sim0.5$~rad~m$^{-2}$ of ionospheric rotation. The highest ionospheric RMs that form spatially coherent patches in the maps occurred during the early summer phase, P1 (compare with Figure~\ref{fig:scans1}). In the low elevation scans, these patches coincide with the central regions of the Galaxy, where instrumental polarization is severe, and the data are in any case not trustworthy (see Section~\ref{subsec:errors}). 

\begin{figure*}
    \centering
    \includegraphics[width=1\hsize]{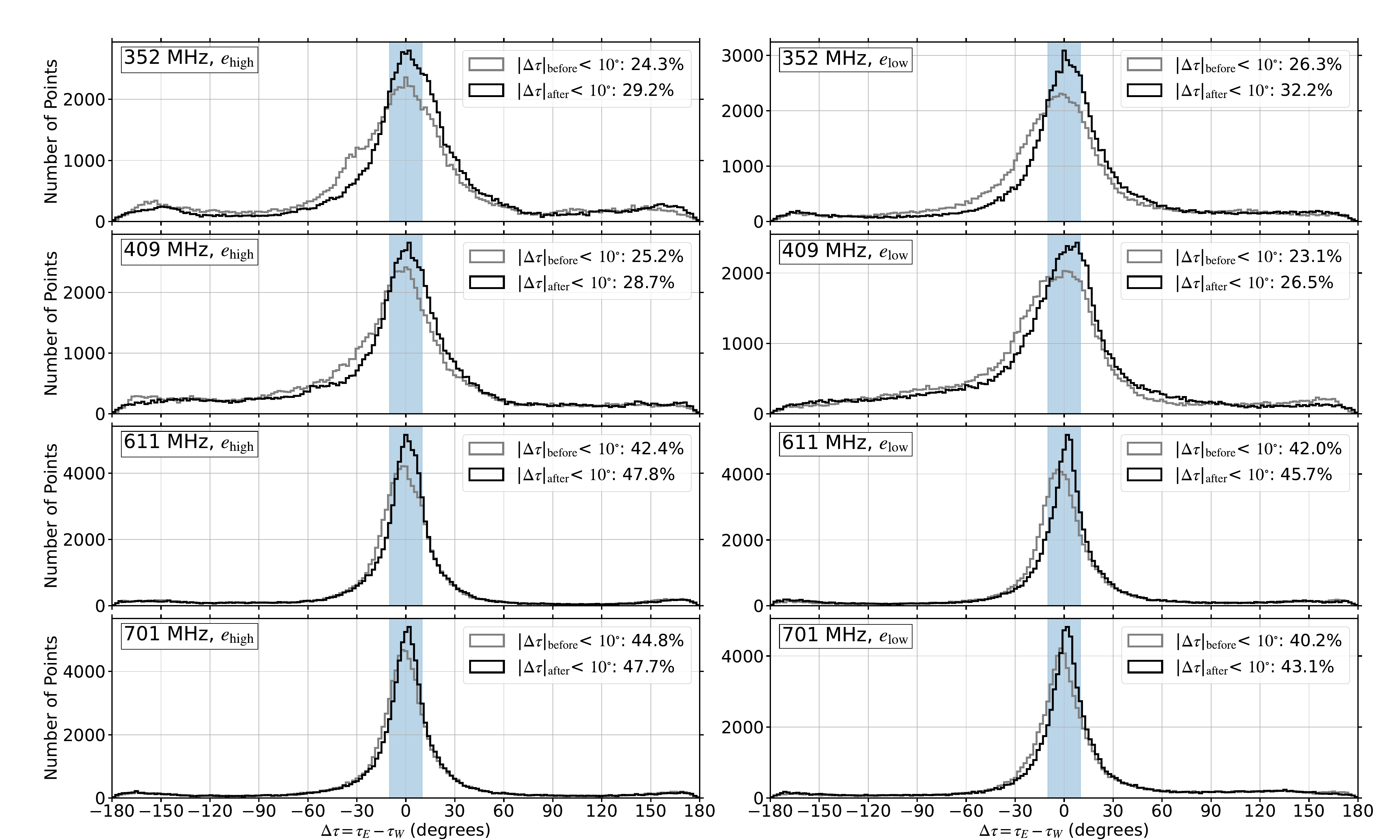}
    \caption{Histograms showing the differences in polarization angle, $\tau$, between east and west high-elevation scans (\textit{Left}) and low-elevation scans (\textit{Right}) at their intersection points before (gray lines) and after (black lines) correcting for ionospheric Faraday rotation. The percentages shown in each panel indicate the fraction of the data set that has less than $10\arcdeg$ difference between east and west, also highlighted by the blue shaded region on each histogram.}
    \label{fig:ionosuccess}
\end{figure*}

The improvement to the maps provided by the ionospheric Faraday rotation correction can be quantified by comparing the polarization angles of the east and west sets of scans at their intersection points before and after derotating by the ionosphere. We illustrate this in Figure~\ref{fig:ionosuccess}, which shows distributions of the differences between the east and west polarization angles, $\Delta \tau=\tau_{\text{E}}-\tau_{W}$ for a sample set of frequencies. For these plots, we excluded $\sim20\%$ of the data using a minimum PI threshold, as low PI data are subject to increased noise in polarization angle. We also masked out the Galactic plane (all data within $|b|<5\arcdeg$), which is dominated by instrumental polarization and is not expected to improve substantially from an ionospheric rotation correction. The resulting difference distributions generally become more peaked at $\tau=0\arcdeg$ after applying the correction. The fraction of the data points with less than $10\arcdeg$ difference between east and west polarization angles is indicated on each histogram. When the ionospheric correction is applied, this fraction increases by 3 to 6\%, with the largest improvement occurring at the lowest frequencies (352~MHz in the examples plotted), as expected.

\subsection{Basket-Weaving}\label{subsec:basket}
The survey was designed in the expectation that basket weaving would be employed to adjust scan base levels, and the S-PASS software package includes routines for that purpose. However, those routines were unsuccessful when applied to our data. We believe that this failure arose from our inability to remove ground emission completely from the data. If there is an error in the ground emission estimate for a west scan characterized by ${\Delta}Q_{\text{ground}}$ and ${\Delta}U_{\text{ground}}$, that error vector is rotated by the ionospheric RM correction, and still remains an error. There may well be a similar error on the east scan, where both the ground emission and ionospheric RM are different. Such errors will vary from point to point along a scan. Basket-weaving calculates one number as a base-level correction for each scan and cannot compensate for errors that vary along a scan. The effect of these discrepancies is most pronounced at the lower frequencies, below $\sim$500~MHz. For a future data release, we plan to improve our estimate of ground emission by using the radiation pattern of the telescope calculated with an electromagnetic simulator to compute the ground contribution. For the current publication and data release, we quantify these errors and include them in uncertainty maps. 

\subsection{Destriping and Combining Maps}\label{subsec:destripe}
In each frequency channel, for each of Stokes $Q$, $U$, and $I$, we binned the calibrated scans into $0.5\arcdeg\times0.5\arcdeg$ pixels on an equatorial grid, separately for east and west scans as well as \highel and \lowel scans, producing four independent maps. For each of \highel and \lowel, the east and west maps are duplicate maps of the same sky, and so their differences can be used to assess the processing and calibration steps. 

We adapted the \cite{emgra88} filtering algorithm to reduce any remaining scanning artifacts in each of the four maps. This algorithm was developed to work with two independent maps produced from sets of orthogonal straight-line scans. The 2D Fourier transform of each map has strong power in a line orthogonal to the scan direction, passing through the center of the transform (at low spatial frequencies). Filtering out power along the two orthogonal lines through the center removes the scanning artifacts. In the case of this survey, the east and west scans in equatorial coordinates are (i) not exactly orthogonal and (ii) deviate from straight lines over a significant declination range (see Figure~\ref{fig:scans1}). Thus we implemented a modification to the \cite{emgra88} algorithm, applying cone-shaped (instead of linear) filters in the Fourier transforms of the east and west maps. The edges of the cones account for the range of slopes encompassed by the curvature of the scan patterns and we tapered these edges using a Hanning window. For each of the \highel and \lowel sets of maps, we combined east and west using weights derived from this filter. Finally, to produce a single map for each frequency and Stokes parameter, the \highel and \lowel maps were combined using a linear weighting function applied in the overlapping declination range.

\subsection{Maps Converted to Brightness Temperature}\label{subsec:abscal}
The conversion to brightness temperature (K) was achieved by assuming a Gaussian main beam and calculating the main beam solid angle, $\Omega$, as a function of frequency, from the measured beamwidths (Figure~\ref{fig:beamw}), $\theta_{\text{HPBW}}$, as
\begin{equation}
    \Omega=\frac{\pi \theta_{\text{HPBW}}^2}{4\ln(2)}.
\end{equation}

\begin{figure}[t]
    \centering
      \includegraphics[width=1\hsize]{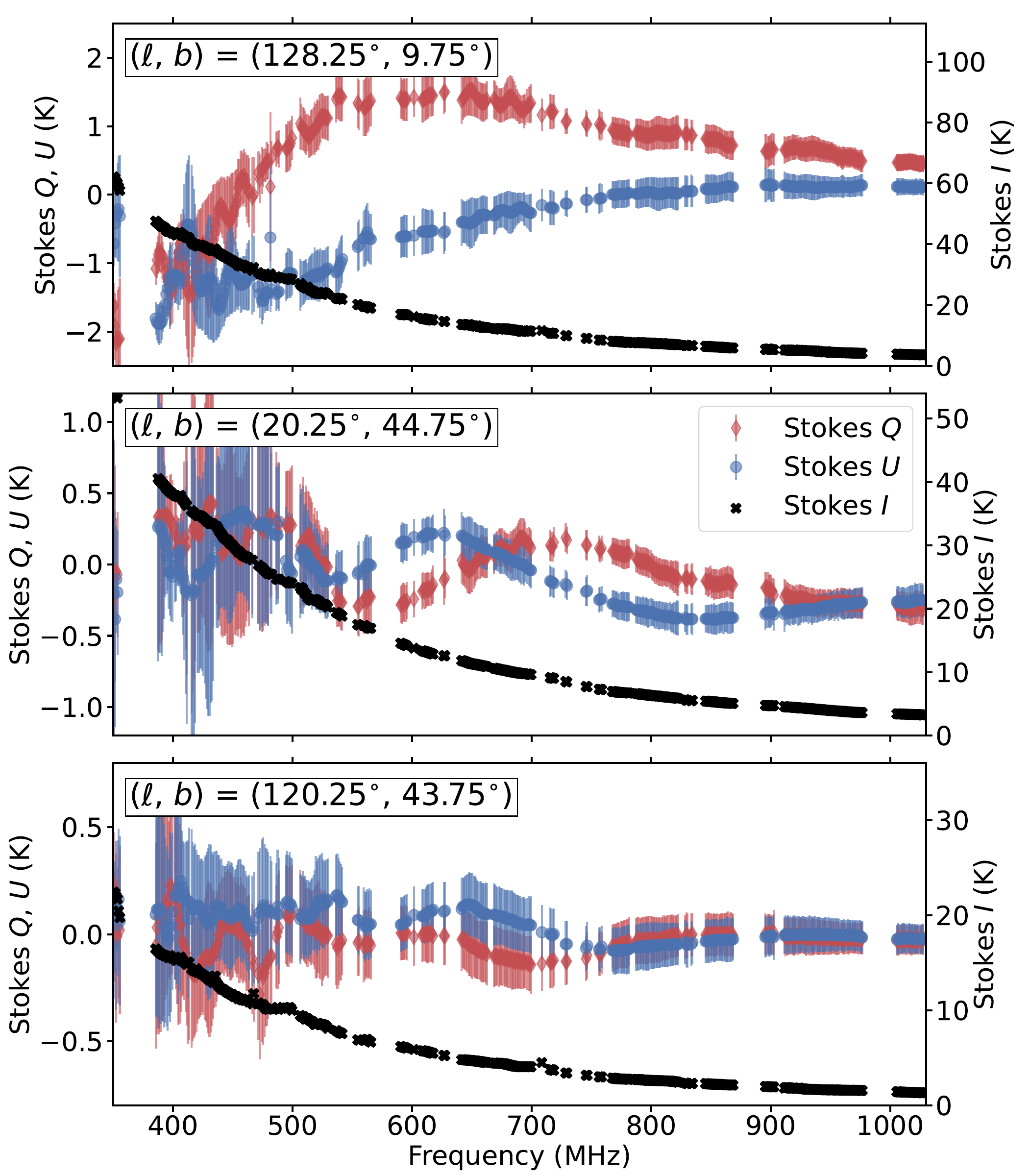} 
    \caption{Frequency spectra for three sample lines of sight. The error bars shown on the Stokes $Q$ and $U$ values are from the error maps described in Section~\ref{subsec:errors}. The values shown here are from the data cubes convolved to a common resolution of $3.6\arcdeg$ for Faraday synthesis.}
    \label{fig:freq_spectra}
\end{figure}

\begin{figure*}
    \centering
    \includegraphics[width=1\hsize]{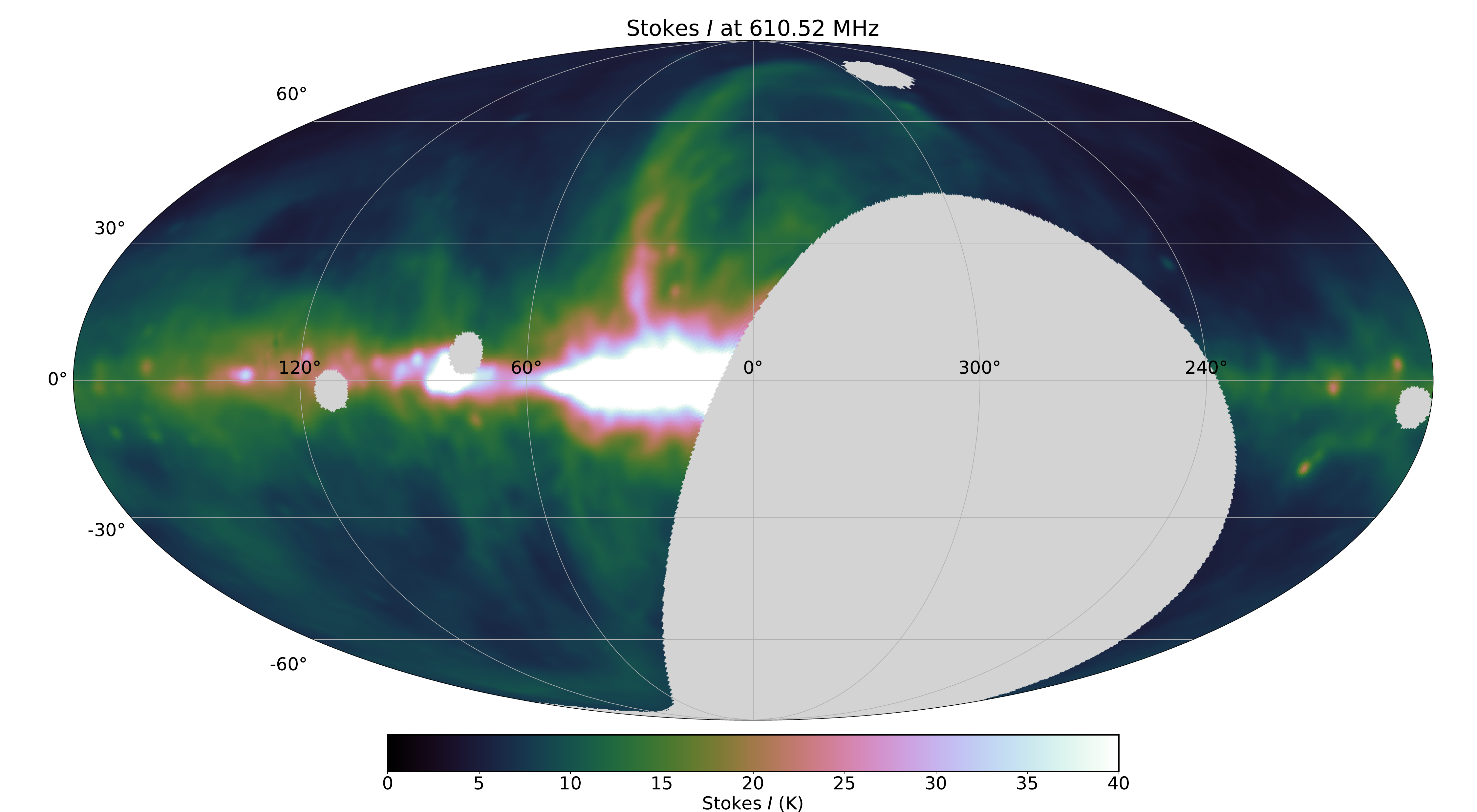}
    \caption{Stokes $I$ map at 610.5~MHz in Galactic coordinates with a Mollweide projection. The four brightest radio sources are masked, as they produce artifacts in the polarization map-making.}
    \label{fig:stokesimap}
\end{figure*}

\begin{figure*}
    \centering
    \includegraphics[width=1\hsize]{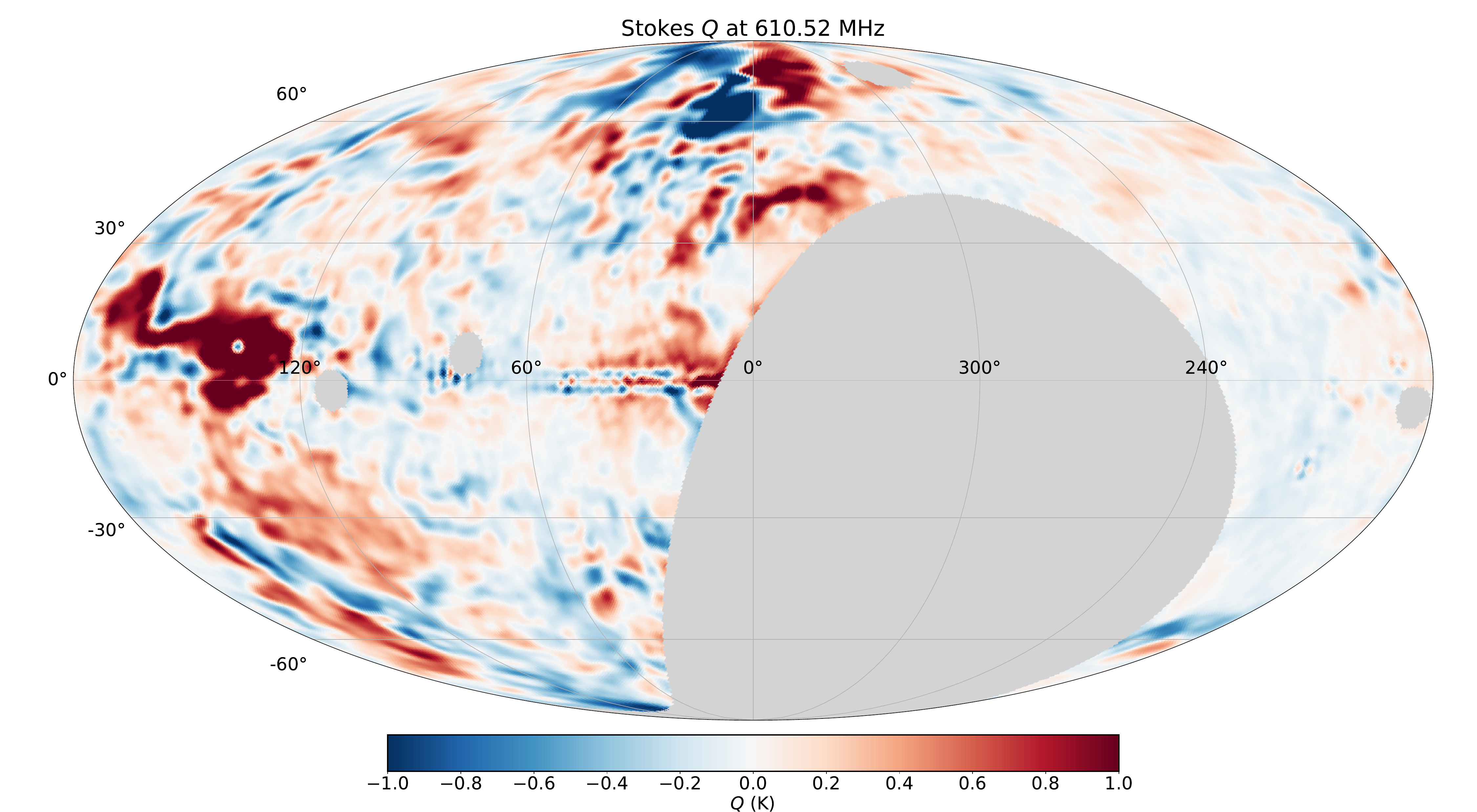}
    \caption{Stokes $Q$ map at 610.5~MHz in Galactic coordinates.}
    \label{fig:qmap}
\end{figure*}

\begin{figure*}
    \centering
    \includegraphics[width=1\hsize]{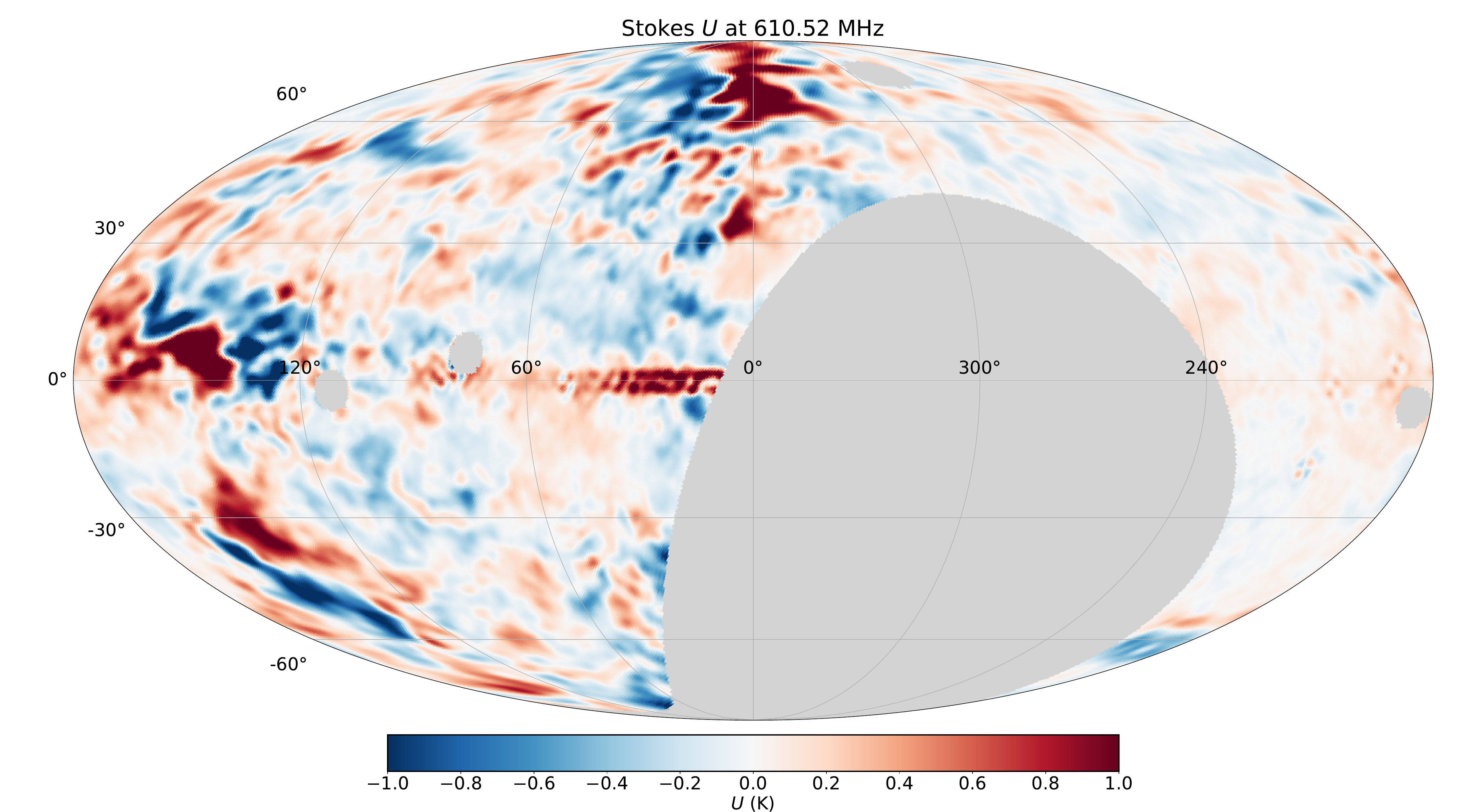}
    \caption{Stokes $U$ map at 610.5~MHz in Galactic coordinates.}
    \label{fig:umap}
\end{figure*}

\begin{figure*}
    \centering
    \includegraphics[width=1\hsize]{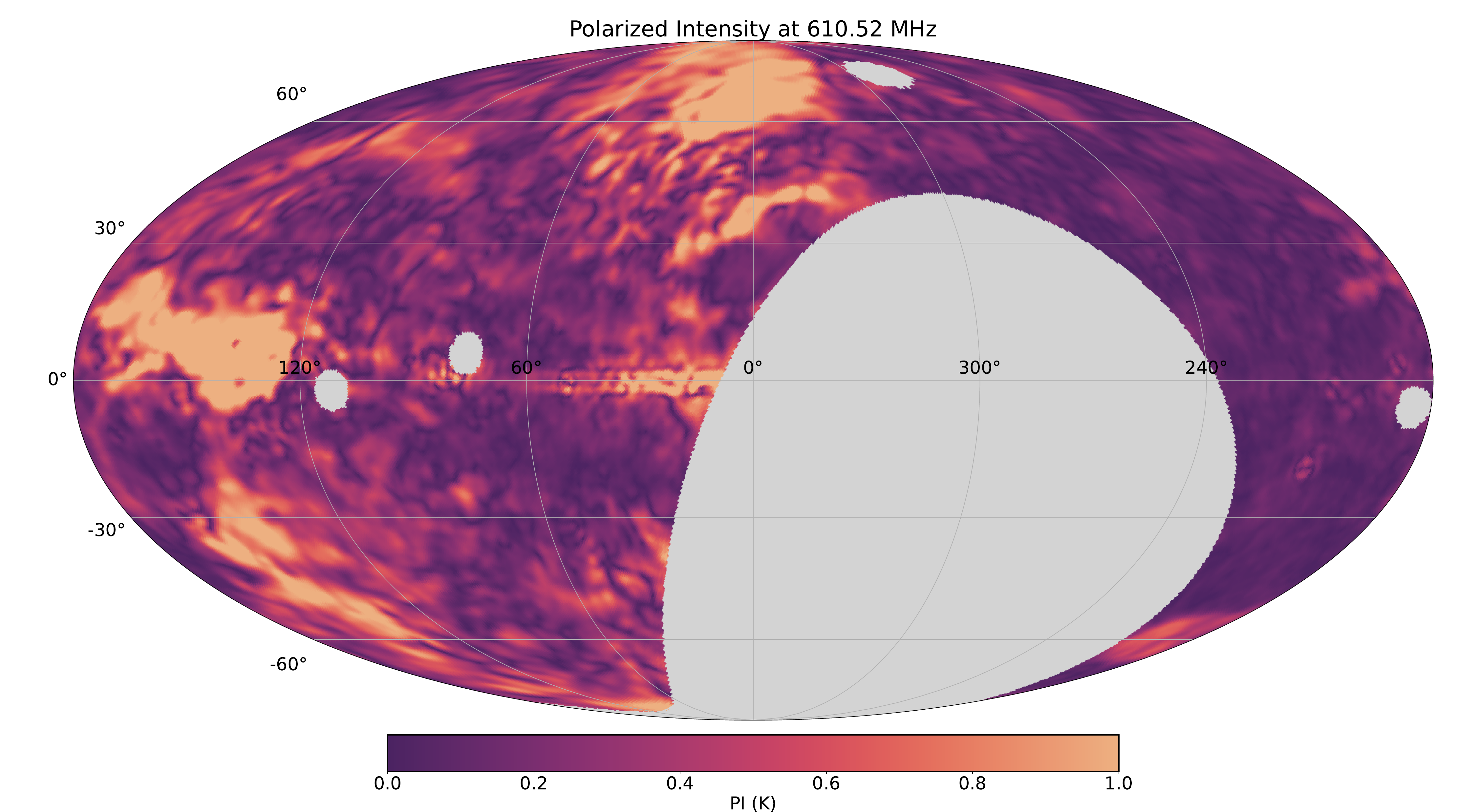}
    \caption{Polarized intensity map at 610.5~MHz in Galactic coordinates.}
    \label{fig:pimap}
\end{figure*}

\begin{figure*}
    \centering
    \includegraphics[width=1\hsize]{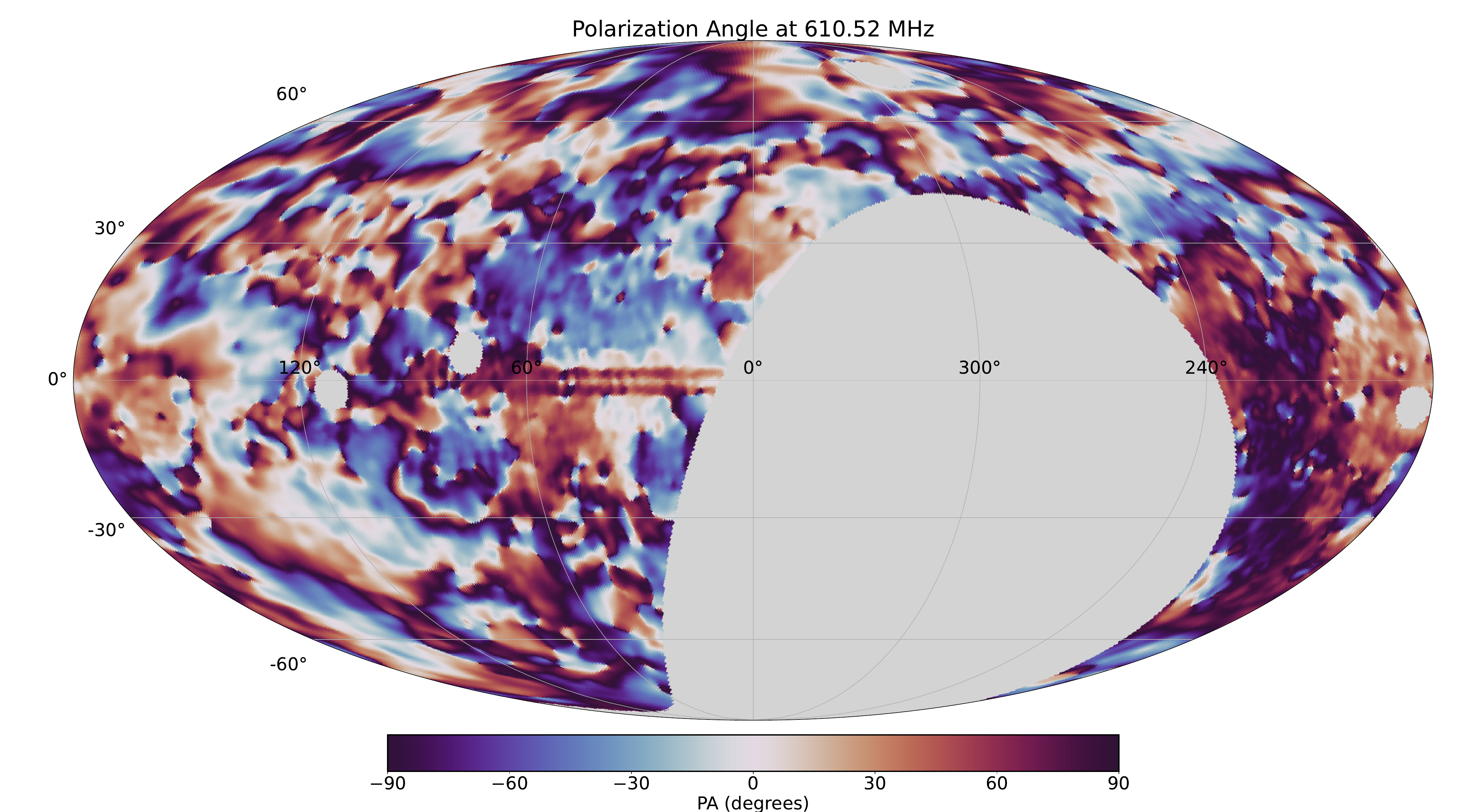}
    \caption{Polarization angle map at 610.5~MHz in Galactic coordinates.}
    \label{fig:pamap}
\end{figure*}

From the Rayleigh-Jeans law, this gives a conversion factor from $I$ in Jy~beam$^{-1}$ to brightness temperature, $T$, in kelvin (K), as
\begin{equation}
    T=\frac{2\ln(2)}{\pi k}\frac{\lambda^2}{\theta_{\text{HPBW}}^2}I,
\end{equation}
where $k$ is the Boltzmann constant and $\lambda$ is the observing wavelength. This conversion factor varies from $\sim$15 to $\sim$19~Jy~beam$^{-1}$~K$^{-1}$, as shown in the bottom panel of Figure~\ref{fig:kconversion}. The variation is due to the fluctuations in aperture efficiency across the band (Figure~\ref{fig:kconversion}, top panel) and variations in beam efficiency (Figure~\ref{fig:kconversion}, middle panel). The frequency-dependent conversion was applied to the maps produced in the final step of map-making. We publish the resulting Stokes $Q$, $U$ and $I$ frequency data cubes along with the Faraday depth cubes that we derive from them (Section~\ref{sec:faraday}). Examples of the spectra are shown in Figure~\ref{fig:freq_spectra} for three lines of sight, in which complex Faraday rotation effects are already evident in the Stokes $Q$ and $U$ fluctuations. Examples of Stokes $I$, $Q$, $U$, PI, and polarization angle maps at a single frequency (610~MHz) are shown in Figures~\ref{fig:stokesimap}--\ref{fig:pamap}.

\section{Data Quality Assessment}\label{sec:quality}
We compare our total intensity and polarization data products to existing data sets, and present an analysis of the errors in our data and their expected effects.

\label{subsec:comp408}
\begin{figure}
    \centering
    \includegraphics[width=1\hsize]{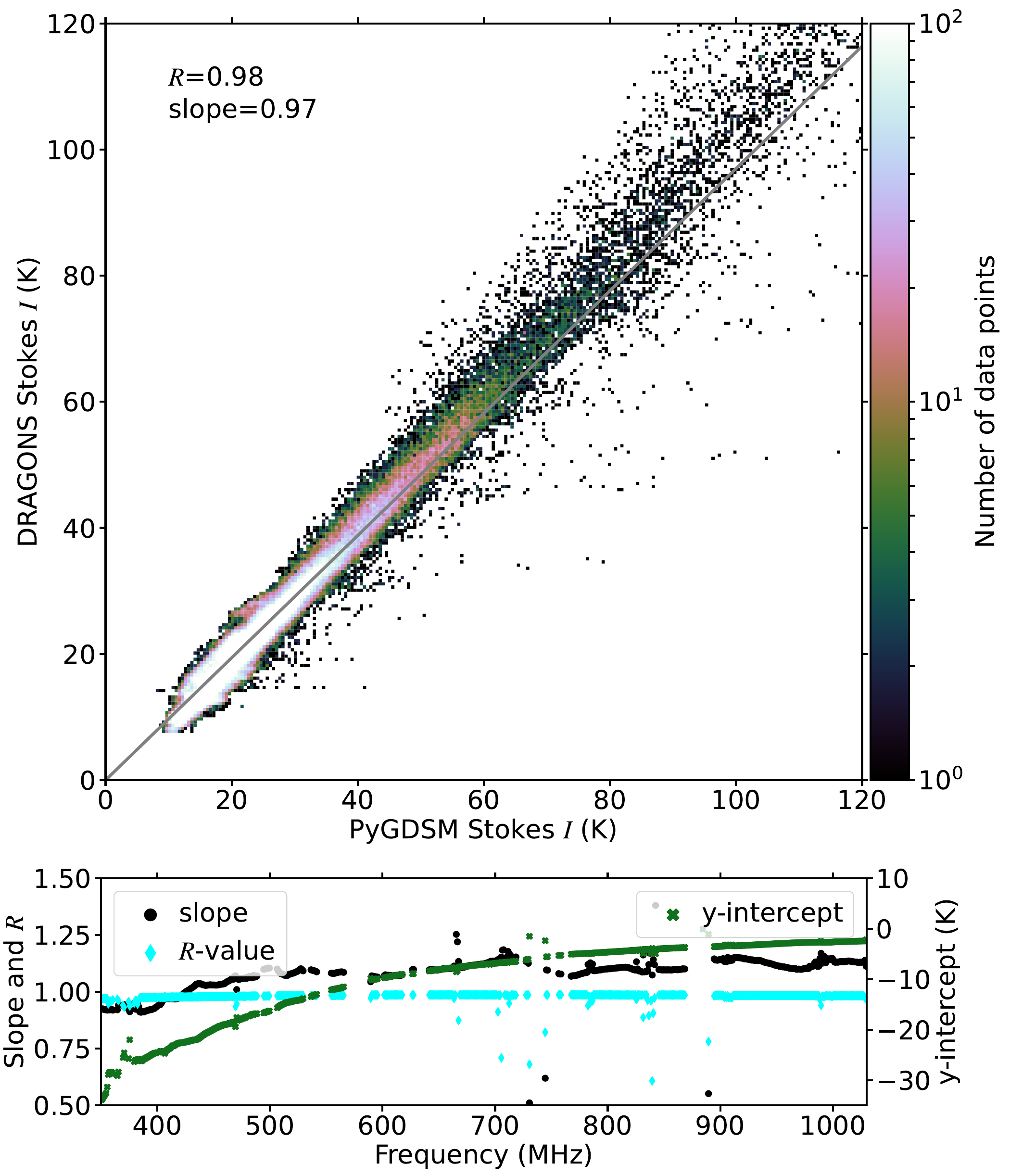}
    \caption{DRAGONS total intensity (Stokes $I$) compared to the \texttt{PyGDSM} sky model. \textit{Top panel}: $T$-$T$ plot at 408~MHz, with the DRAGONS zero-offset corrected using \texttt{PyGDSM}. \textit{Bottom panel}: the slopes, $y$-intercepts and Pearson correlation coefficients from the DRAGONS-\texttt{PyGDSM} comparison across the band. The slopes and Pearson $R$ values are consistently close to one, indicating good agreement with \texttt{PyGDSM} across the band. The $y$-intercepts provide the zero-offset corrections (already applied to the $T$-$T$ plot in the top panel).}
    \label{fig:TTstokeI}
\end{figure}

\subsection{Total Intensity Assessment}\label{sec:stokesi}
To assess the accuracy of our Stokes $I$ maps, we compared our data with the \texttt{PyGDSM} sky model \citep{price16,PGSM17} across all frequencies. An example of a $T$-$T$ plot, in the form of a 2D histogram, is shown in the top panel of Figure~\ref{fig:TTstokeI} for 408~MHz, where \texttt{PyGDSM} is dominated by the \cite{haslam82} map. There is a strong correlation between DRAGONS and the \texttt{PyGDSM} map (Pearson correlation of 0.97 with a DRAGONS versus \texttt{PyGDSM} slope of 0.98). This correlation is consistent across all frequencies as shown in the bottom panel of Figure~\ref{fig:TTstokeI}, and the slopes also remain close to unity. We have corrected the missing zero-level in the Stokes $I$ maps by subtracting the $y$-intercepts (which are negative) of the $T$-$T$ plots. This correction is already applied in the top panel of Figure~\ref{fig:TTstokeI}. The apparent excess of Stokes $I$ in DRAGONS above $\sim60$~K compared to \texttt{PyGDSM} is the result of a slight mismatch in the spatial profile of the Galactic disk between DRAGONS and \texttt{PyGDSM}. A spatial cut across the Galactic disk peaks more sharply toward the midplane in \texttt{PyGDSM} than in DRAGONS, with the shoulders of this profile having higher intensity in DRAGONS than in \texttt{PyGDSM}, resulting in a bias toward higher Stokes $I$ values in DRAGONS.

\subsection{Polarization Comparison with Dwingeloo}
\label{subsec:dwingeloo}
This paper describes a survey of polarized radio emission over a large fraction of the sky, thus it is instructive to compare our survey with the previous best knowledge of the polarized radio sky in this frequency range. \citet{brou76} published maps of the polarized emission at 408, 465, 610, 820, and 1411\,MHz. The data were obtained with the Dwingeloo 25\,m Telescope in the 1960s, and carefully calibrated in amplitude and corrected for instrumental polarization, for spurious polarized emission generated by radiation from the ground, and for Faraday rotation in the ionosphere. We refer to this dataset as the `Dwingeloo data'. The Dwingeloo data set was a good representation of the polarized sky in its frequency range, but its spatial sampling was very far from complete and its frequency sampling very sparse. Our work presents a survey that is fully Nyquist sampled on the sky, is fully sampled in frequency from 350 to 1030\,MHz (to the extent permitted by RFI at the DRAO at the time the observations were made), and covers a much larger area of the sky. This makes mapping a range of scales of ISM structures in this frequency range now possible, in both the spatial and Faraday depth domains.

\begin{figure*}
    \centering
    \includegraphics[width=1\hsize]{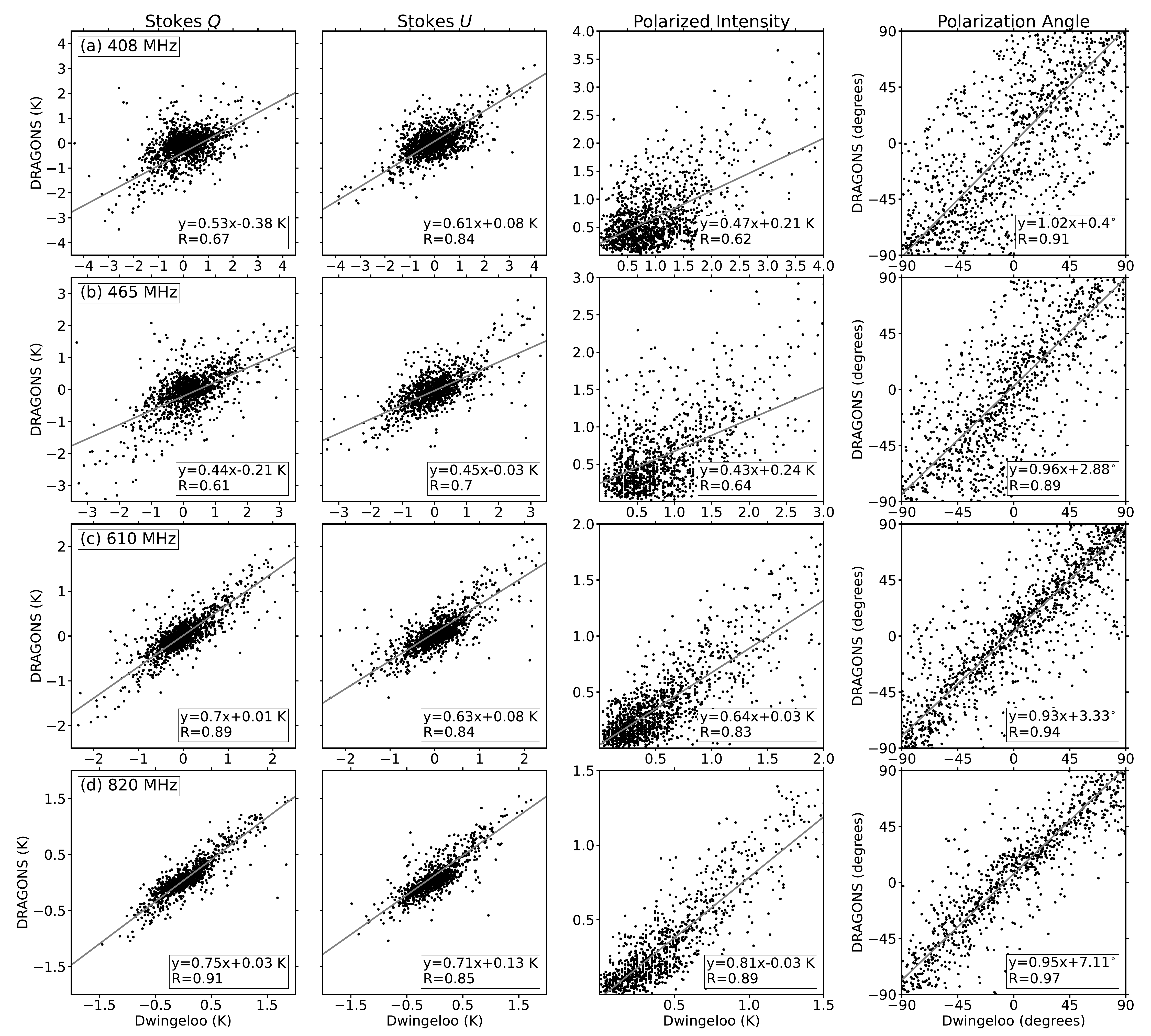}
    \caption{Comparisons between DRAGONS and Dwingeloo polarization data for the four frequencies common to both surveys (rows \textit{a}--\textit{d}). The first two columns show Stokes $Q$ and $U$ comparisons, the third column shows PI, and the fourth column shows polarization angles. Slopes, intercepts, and correlation coefficients are shown for all four quantities. For Stokes $Q$ and $U$ we exclude data points with PI below 3 times the noise level from the linear fit calculations to avoid biasing the fits with low signal values. The generally lower polarized intensities for DRAGONS compared to Dwingeloo can be attributed to the lower angular resolution of DRAGONS resulting in more beam depolarization.}
    \label{fig:comparedwingeloo}
\end{figure*}

A comparison of the DRAGONS and Dwingeloo survey parameters is shown in Table~\ref{tab:comp_dwing} and a comparison between the two data sets is shown in Figure~\ref{fig:comparedwingeloo}. To produce the comparison plots we calculated the mean values of DRAGONS Stokes $Q$ and $U$ within the DRAGONS beam at the location of each Dwingeloo pointing. We then calculated PI and polarization angle from these mean $Q$ and $U$ values. The polarization angles are `un-wrapped' by constraining the differences between the DRAGONS and Dwingeloo angles to be less than $90\arcdeg$ and adjusting the angles by $180\arcdeg$ accordingly. We fitted a linear equation to each scatter plot, using data points for which DRAGONS PI is above 3 times the noise level, and calculated the Pearson correlation coefficient, $R$. There is overall good correlation between DRAGONS and Dwingeloo both in Stokes $Q$ and $U$, as well as in PI ($R$ values of 0.6 to 0.9). The polarization angle agreement is even stronger ($R$ mostly above 0.9).  The poorer correlations at 408 and 465~MHz compared to 610 and 820~MHz are likely in part caused by the uncertainties in the DRAGONS maps, which also tend to be higher at the lower frequencies (see Section~\ref{subsec:errors}).

\begin{table}[!h]
\caption{Resolution and polarization sensitivity comparison with Dwingeloo surveys. }
\label{tab:comp_dwing}
\begin{center}
\begin{tabular}{ccccc}
\hline
Frequency & Angular   & Angular    & Noise (mK) & Noise (mK)\\
(MHz)     & resolution & resolution & rms error    & mean error    \\
          & DRAGONS & Dwingeloo & DRAGONS & Dwingeloo \\
\hline
408       & $3.0^{\circ}$ & $2.3^{\circ}$ &  160   & 340 \\
465       & $2.7^{\circ}$ & $2.0^{\circ}$ &  180   & 330 \\
610       & $2.0^{\circ}$ & $1.5^{\circ}$ &  100   & 160 \\
820       & $1.5^{\circ}$ & $1.0^{\circ}$ &  80   &  110 \\
\hline
\end{tabular}
\end{center}
\end{table}

The PI in DRAGONS is consistently lower than in Dwingeloo, and the discrepancy is more pronounced at the lower frequencies. This is likely caused by more beam depolarization present in DRAGONS due to its beam having approximately twice the area of the Dwingeloo beam. Since the amount of beam depolarization depends not only on the size of the beam, but also on the angular scale of the fluctuations in Stokes $Q$ and $U$, it is not possible to determine the expected difference in beam depolarization between the two surveys without comparing to a higher resolution dataset. Once the full-sky Galactic foreground polarization data from CHIME are calibrated, the CHIME Stokes $Q$ and $U$ maps, at $\sim20\arcmin$ resolution, can provide the necessary small-scale information for this calculation.

\subsection{Error Analysis}\label{subsec:errors}
We account for two types of errors present in the maps. The first is a position-dependent, systematic error characterized by the residual differences between the east and west maps prior to filtering and combining, and the second is noise characterized by random spatial variability in the combined maps. Both types of error are frequency-dependent, while the systematic error also has a structured, spatial dependence.

\begin{figure*}
    \centering
    \includegraphics[width=1\hsize]{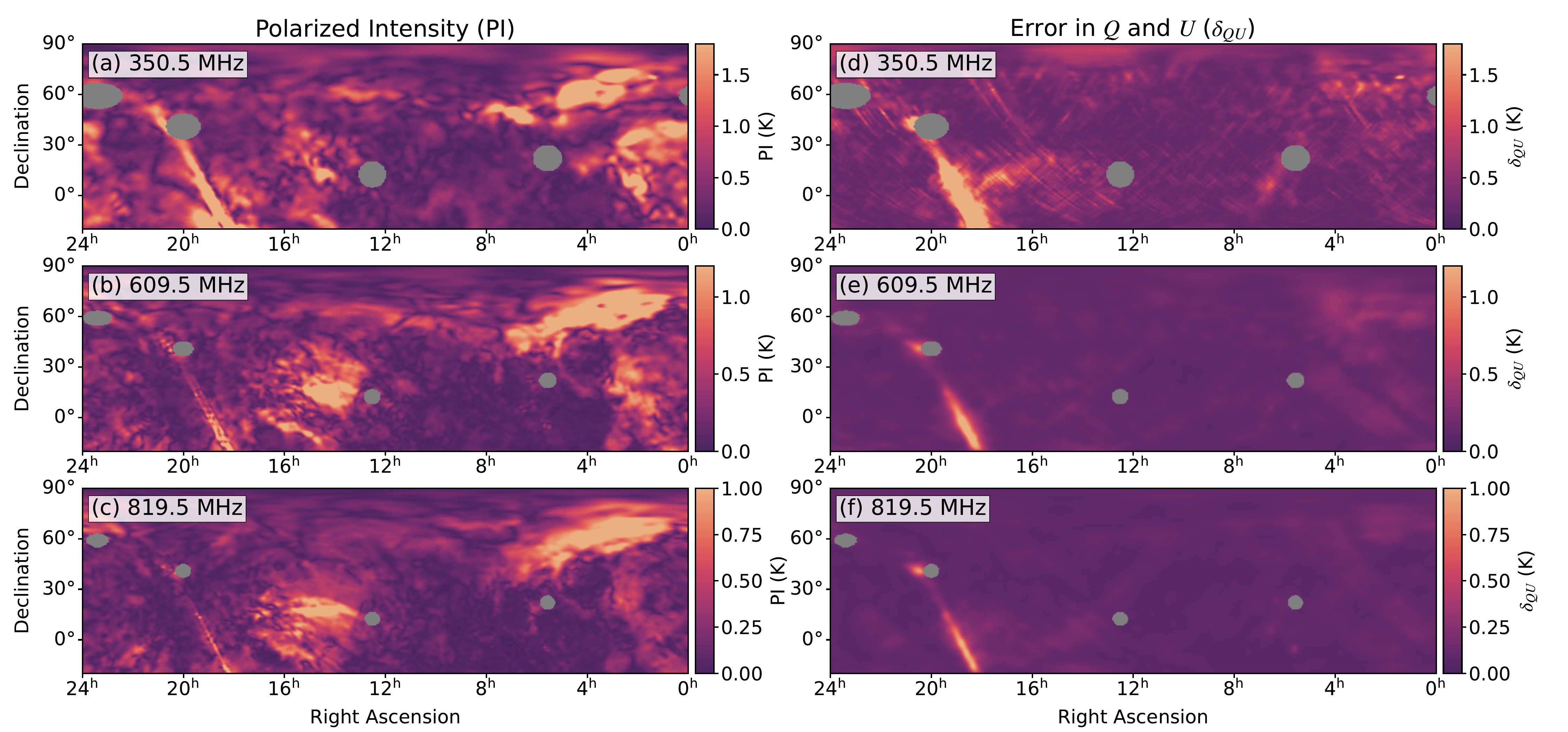}
    \caption{Polarized intensity (\textit{a}--\textit{c}) and systematic error in Stokes $Q$ and $U$ (\textit{d}--\textit{f}) for three sample frequency channels. The spatially-dependent, systematic error, $\delta_{QU_{\text{sys}}}$, is calculated as the average of the errors in $Q$ and $U$, which are in turn calculated as $\delta Q=|Q_E-Q_W|$ and $\delta U=|U_E-U_W|$. These error estimates reflect both spatial and spectral variation in the reliability of the polarization maps based on the differences between the independent east and west maps that are ultimately combined. We combine these systematic errors in quadrature with the measured random spatial variability to provide uncertainty maps for Faraday synthesis. The gray regions correspond to the four brightest radio sources, which were masked out in the map-making process.}
    \label{fig:QUerrors}
\end{figure*}

Particularly at frequencies below $\sim500$~MHz, residual errors following the ground and ionosphere correction steps (see Section~\ref{subsec:basket}) lead to the east and west maps differing over some regions of the sky. At each frequency, we use the differences between the east and west maps as a proxy for the resulting position-dependent, systematic error in Stokes $Q$ and $U$, calculated as
\begin{equation}
    \delta_{QU_{\text{sys}}}=\frac{1}{2}\Big(|Q_{\text{E}}-Q_{\text{W}}|+|U_{\text{E}}-U_{\text{W}}|\Big),
\end{equation}
where $Q_{\text{E}}$ ($U_{\text{E}}$) is the Stokes $Q$ ($U$) east map and $Q_{\text{W}}$ ($U_{\text{W}}$) is the Stokes $Q$ ($U$) west map. Examples of the resulting error maps are shown in Figure~\ref{fig:QUerrors}, compared to the PI maps. Instrumental polarization leakage leads to enhanced errors along the brightest part of the Galactic disk (RA=18 to 22 hours). This is present across the full band, generally making the area within approximately two beamwidths of the Galactic plane toward the inner Galaxy unreliable. Enhanced errors resulting from the residual systematics (inadequately corrected ground and ionosphere effects) occur mostly at low frequencies. In particular, the errors are enhanced in patches of the sky that were observed under significantly different ionospheric conditions between east and west maps.

To quantify the random spatial variability (noise) in the maps, we calculated the standard deviation in Stokes $Q$ and $U$ in a region of low polarized intensity in the longitude range $180\arcdeg\leq\ell\leq240\arcdeg$. These values range from 0.5 to 0.05~K across 350 to 1030~MHz. We combine the two types of errors described here in quadrature, and use the resulting error cubes (with both spatial and spectral dependence) as inputs to Faraday synthesis (Section~\ref{sec:faraday}) for determining position-dependent channel weighting.

In the Faraday depth cubes, residual contamination from inadequately removed ground contribution will appear as a feature in the spectra at a Faraday depth of 0~rad~m$^{-2}$. This is because even though the Stokes $Q$ and $U$ values of the ground emission change with frequency, we do not expect the effective `polarization angle' of the ground to change with frequency and mimic Faraday rotation. The effect of remaining contamination is likely negligible in regions of the sky having low error in polarization compared to PI (see Figure~\ref{fig:QUerrors}). Therefore it is only in the regions with high polarization error that Faraday depth features near 0~rad~m$^{-2}$ should be interpreted with caution.

\subsection{Polarization Leakage}\label{subsec:leakage}
For circular feeds, imperfections in the correlated outputs, RL* and LR*, can lead to on-axis leakage of Stokes $I$ into $Q$ and $U$. This leakage is almost completely removed by the daily median-subtraction (Step 2 in Section~\ref{subsec:filter}). We examine the final Stokes $I$, $Q$, and $U$ maps to quantify the amount of residual leakage. The Orion A $\HII$ region ($\ell=209\arcdeg$, $b=-19\arcdeg$), at $\sim1\arcdeg$ in angular size, is unresolved at most DRAGONS frequencies and produces the expected cloverleaf pattern characterizing the DRAO-15 beam pattern in Stokes $Q$ and $U$ (Figure~\ref{fig:qmap} and \ref{fig:umap}). At its center, we measure less than 1\% fractional polarization across most of band, increasing to 2\% at the higher frequencies where the source may be partially resolved. This allows us to identify an upper limit of 1--2\% on-axis polarization leakage across the DRAGONS band.

The Galactic midplane stands out prominently as a band of instrumental polarization in the Stokes $Q$, $U$, PI, and polarization angle maps (Figures~\ref{fig:qmap}--\ref{fig:pamap}) at Galactic longitudes $\ell<90\arcdeg$. The midplane should be depolarized in the DRAGONS data due to polarization structures in the Galactic disk on scales much smaller than the beam. Although the \textit{fractional} polarization in the midplane does not differ significantly from the surroundings in our maps, there is a significant increase in PI within two beamwidths on either side of the Galactic midplane for $\ell<90\arcdeg$. Therefore, $\pm7\arcdeg$ in latitude from the midplane at 350~MHz, and $\pm2\arcdeg$ at 1030~MHz should be excluded from analysis. For interpreting Faraday synthesis results we recommend masking $\pm7\arcdeg$ since the beam is convolved to the common resolution of the beam at 350~MHz and all frequencies are included in calculating the spectra.

\begin{table}\label{tab:fdspecs}
\caption{Characteristics of published survey data.}
\begin{center}
\begin{tabular}{ll}
\hline
Frequency range, $I$, $Q$ and $U$ & 350 to 1030\,MHz \\
Channel width                     & 1.0\,MHz \\ 
Noise, $Q$ and $U$ images & 500 to 60\,mK \\
(single channel) &  \\
Available data formats & Galactic coordinates,  \\
 & FITS and HEALPix \\
Coverage of Faraday cube   & ${\pm}200$~rad~m$^{-2}$\\
Channel width in Faraday cube  & 0.5~rad~m$^{-2}$ \\
Largest detectable Faraday depth & 450~rad~m$^{-2}$\\
Resolution in Faraday depth  & 6~rad~m$^{-2}$ \\
Largest measurable& 38~rad~m$^{-2}$ \\ 
Faraday depth structure & \\ 
Sensitivity in Faraday depth cube & 11~mK~RMSF$^{-1}$ \\
\hline
\hline
\end{tabular}
\end{center}
\end{table}

\begin{table*}
\caption{Data available on CADC.}
\label{tab:data}
\begin{center}
\begin{tabular}{lll}
\hline
Type & Angular resolution & Data file \\
\hline
Frequency cube (3D) & Native channel resolution & Stokes $Q$ \\
                    &                           & Stokes $U$ \\
                    &                           & Stokes $I$ \\
                    &                           & Stokes $Q$, $U$ errors \\
                    &                           & Mask for bad data \\
                    &                           & Mask for bright sources \\
\hline
Faraday synthesis cube (3D) & 3.6\arcdeg (beam at 350~MHz) & dirty PI \\
                            &                              & CLEANed PI \\
                            &                              & RMSF \\
\hline
Faraday synthesis map (2D) & 3.6\arcdeg (beam at 350~MHz) & Rayleigh PI noise \\
                           &                              & Zeroth moment, M0 \\
                           &                              & First moment, M1 \\
                           &                              & Square root of second moment, m2 \\
                           &                              & Number of Faraday depth peaks \\
\hline
\hline
\end{tabular}
\end{center}
\end{table*}

\section{Faraday Depth Cubes}\label{sec:faraday}
The primary data product from the DRAGONS survey is a diffuse-emission Faraday depth cube, which can be studied spectrally along individual lines of sight, or in 2D maps, either as Faraday depth slices or as moments of the Faraday spectra \citep{dickey19}. The strength of the DRAGONS survey is that the frequency range allows for high resolution in Faraday depth with simultaneous sensitivity to broad Faraday depth structures. The Faraday depth resolution, $\delta \phi$, is inversely proportional to the $\lambda^2$ coverage \citep{brentjens_de_bruyn_2005}, and is 6~rad~m$^{-2}$ for DRAGONS. The widest detectable (not completely depolarized) feature in Faraday depth space, $\phi_{\mathrm{max-scale}}$, is inversely proportional to the smallest $\lambda^2$ in the survey. For DRAGONS this is 38~rad~m$^{-2}$ based on \cite{brentjens_de_bruyn_2005}\footnote{A more conservative estimate of this quantity is presented in \cite{Rudnick2023}}. Diffuse emission Faraday synthesis has the potential to detect the widest range of scales of Faraday depth structures when $\phi_{\mathrm{max-scale}} > \delta \phi$. DRAGONS is the only GMIMS component survey to date to achieve this criterion by a significant margin, resulting in DRAGONS having unprecedented sensitivity to diffuse emission Faraday complexity. The published data parameters are summarized in Table~\ref{tab:fdspecs} and the available data products are listed in Table~\ref{tab:data}.

\subsection{Faraday Synthesis}\label{subsec:rmsynth}
For applying Faraday synthesis, we convolved the 1~MHz binned frequency cubes of Stokes $Q$, $U$, and $I$, with persistent RFI channels flagged, to a common angular resolution of 3.6\arcdeg (the survey resolution at 350~MHz). In addition to the flagged RFI channels, we also conducted a visual inspection to mask out either entire maps or portions of maps in channels heavily contaminated by other artifacts. We then fed the remaining set of 393 full maps and 11 partial maps (out of a total of 680) into the CIRADA \texttt{RMTools} package \citep{Purcell:2020}\footnote{\url{https://github.com/CIRADA-Tools/RM-Tools}} for Faraday synthesis and RM CLEAN calculations. 
\begin{figure}
    \centering
      \includegraphics[width=1\hsize]{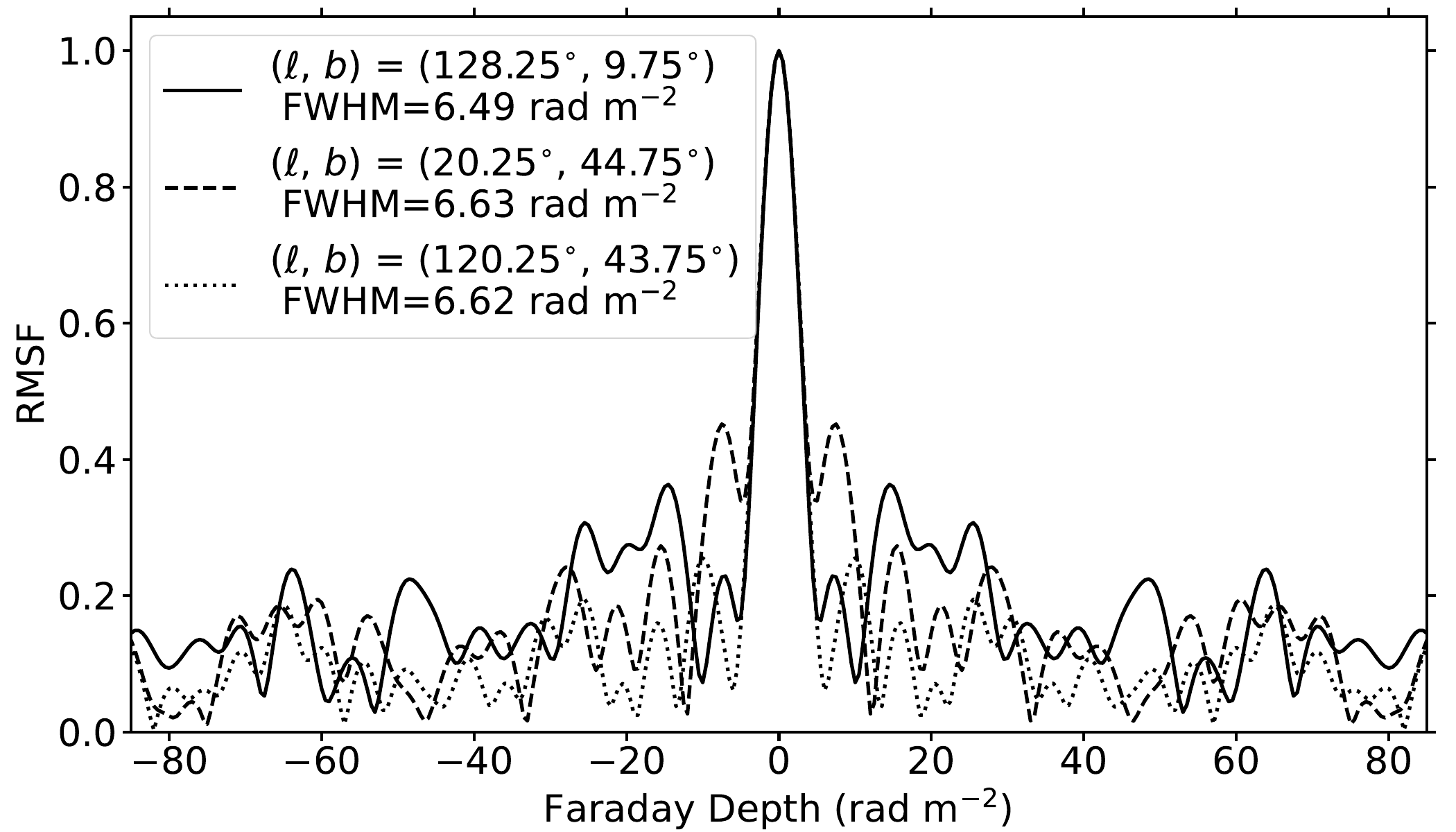} 
    \caption{The RMSF for three sample lines of sight (spectra shown in Figure~\ref{fig:FD_spectra}), along with the FWHM values of the fitted Gaussian clean beams.}
    \label{fig:rmsf}
\end{figure}
\begin{figure}
    \centering
      \includegraphics[width=1\hsize]{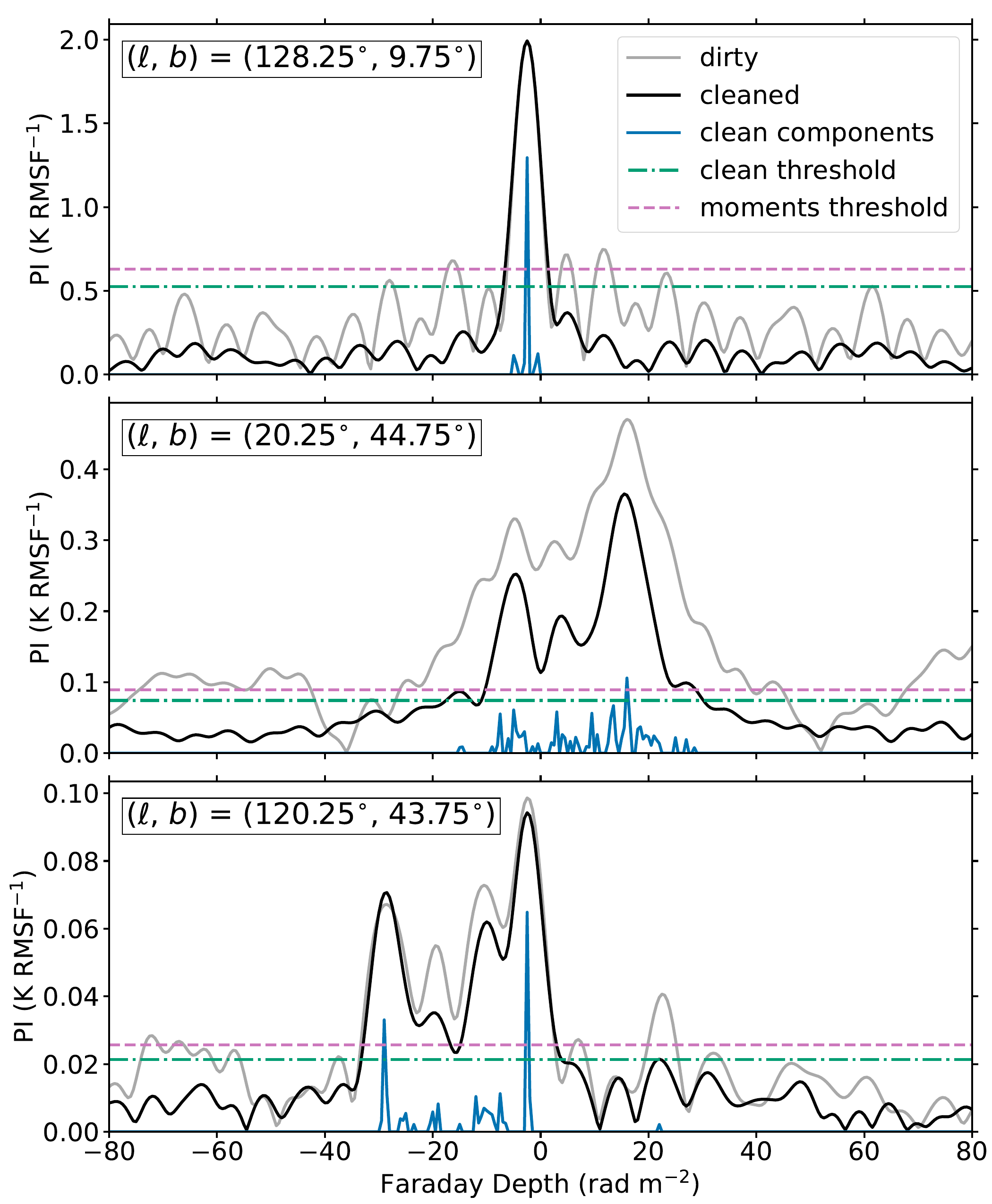} 
    \caption{Sample Faraday depth spectra, corresponding to the frequency spectra shown in Figure~\ref{fig:freq_spectra}. The dirty (cleaned) spectra are shown in gray (black), the CLEAN threshold is shown by the green dot-dashed line, the minimum PI threshold for the moment calculations is the pink dashed line, and the CLEAN components are in blue.}
    \label{fig:FD_spectra}
\end{figure}

For Faraday synthesis, we used inverse variance weighting of the channels to account for the higher noise and systematic errors at lower frequencies. This was based on the error maps described in Section~\ref{subsec:errors}. We incorporated Stokes $I$ data (corrected for the zero-offset as described in Section~\ref{sec:stokesi}) as a fitted model along each LOS to account for the spectral index variation across the 680~MHz-wide band. The effect of dividing by the Stokes $I$ model (and thus using fractional polarization as the input to Faraday synthesis) was to reduce the uncertainties at the lower frequencies and thus partially offset the down-weighting of the low-frequency channels introduced by the inverse variance weighting. Examples of the resulting rotation measure spread function (RMSF) are shown in Figure~\ref{fig:rmsf}. The investigation of alternative weighting schemes will be the subject of future studies.

We used the method introduced by \cite{raycheva25} to determine a position-dependent threshold for RM CLEAN. For each pixel in the map the noise was calculated from the Faraday depth channels covering $-1500$ to $-500$~rad~m$^{-2}$ and $500$ to $1500$~rad~m$^{-2}$. This range was selected because it contains only noise in the spectra, and is free of real Faraday depth features and their sidelobes. The resulting noise estimate ranges from 0.5 to 200~mK~RMSF$^{-1}$ across the map, with a median value of 11~mK~RMSF$^{-1}$. We used a $5\sigma$ CLEAN threshold and the cleaned spectra were restored with a Gaussian beam fitted to each RMSF. Examples of dirty and cleaned spectra are shown in Figure~\ref{fig:FD_spectra}. The narrow Faraday depth resolution of $\sim6$~rad~m$^{-2}$ allows multiple peaks to be resolved along many lines of sight with high signal-to-noise ratio.

\subsection{Faraday Moment Maps}\label{subsec:moment}
\begin{figure*}[t!]
    \centering
    \includegraphics[width=1\hsize]{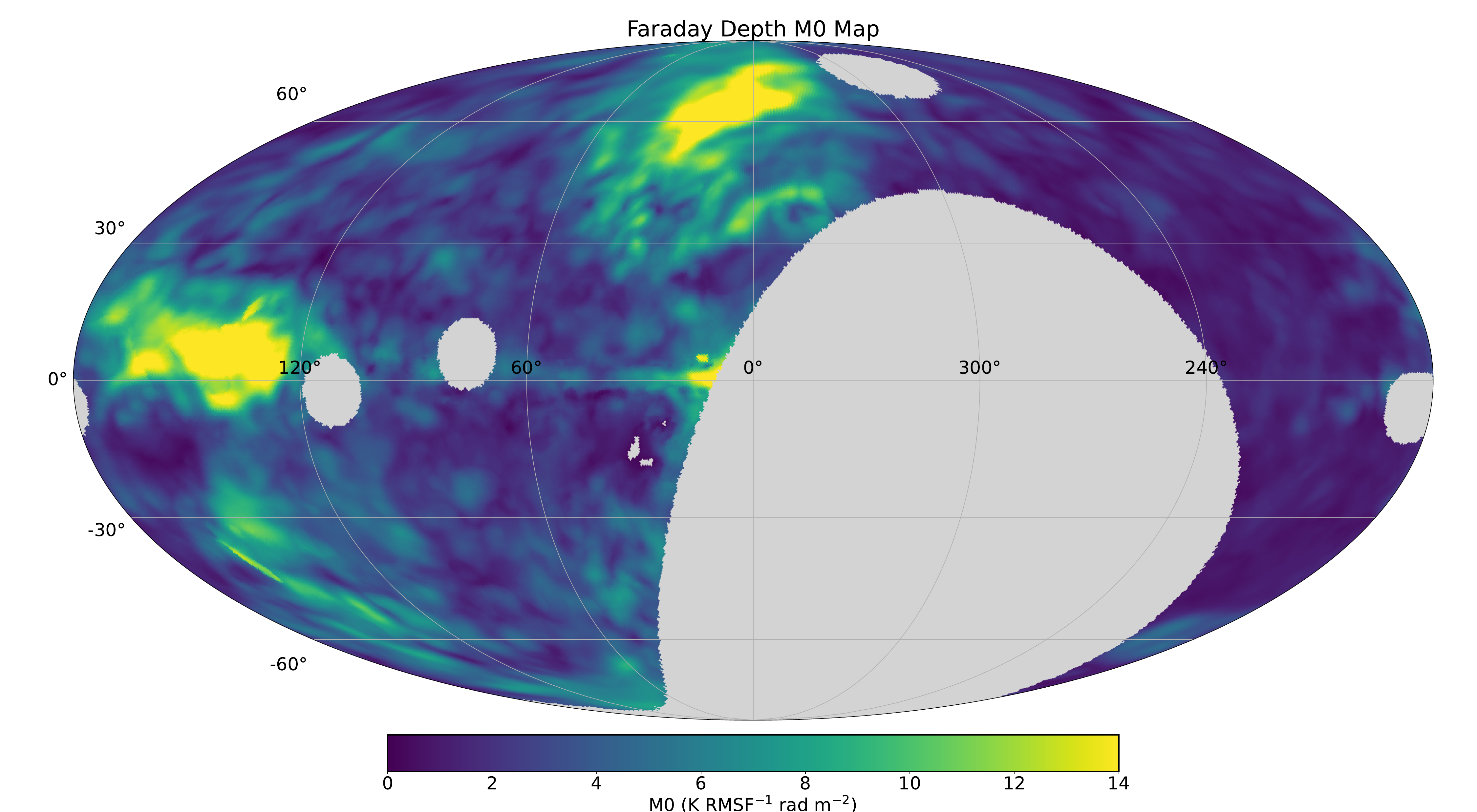}
    \caption{Map of zeroth moments (M0) of the cleaned Faraday depth spectra in Galactic coordinates with a Mollweide projection. The four brightest radio sources are masked, as the regions within two beamwidths of these are contaminated by artifacts. We do not mask the Galactic plane in the maps displayed here, but for Galactic longitudes $\ell\leq90\arcdeg$ the region within a $\pm7\arcdeg$ latitude range from the mid-plane is also unreliable. The small patches of missing data near $\ell=30\arcdeg,b=-15\arcdeg$ are masked for insufficient signal to noise.}
    \label{fig:M0map}
\end{figure*}

\begin{figure*}[t!]
    \centering
    \includegraphics[width=1\hsize]{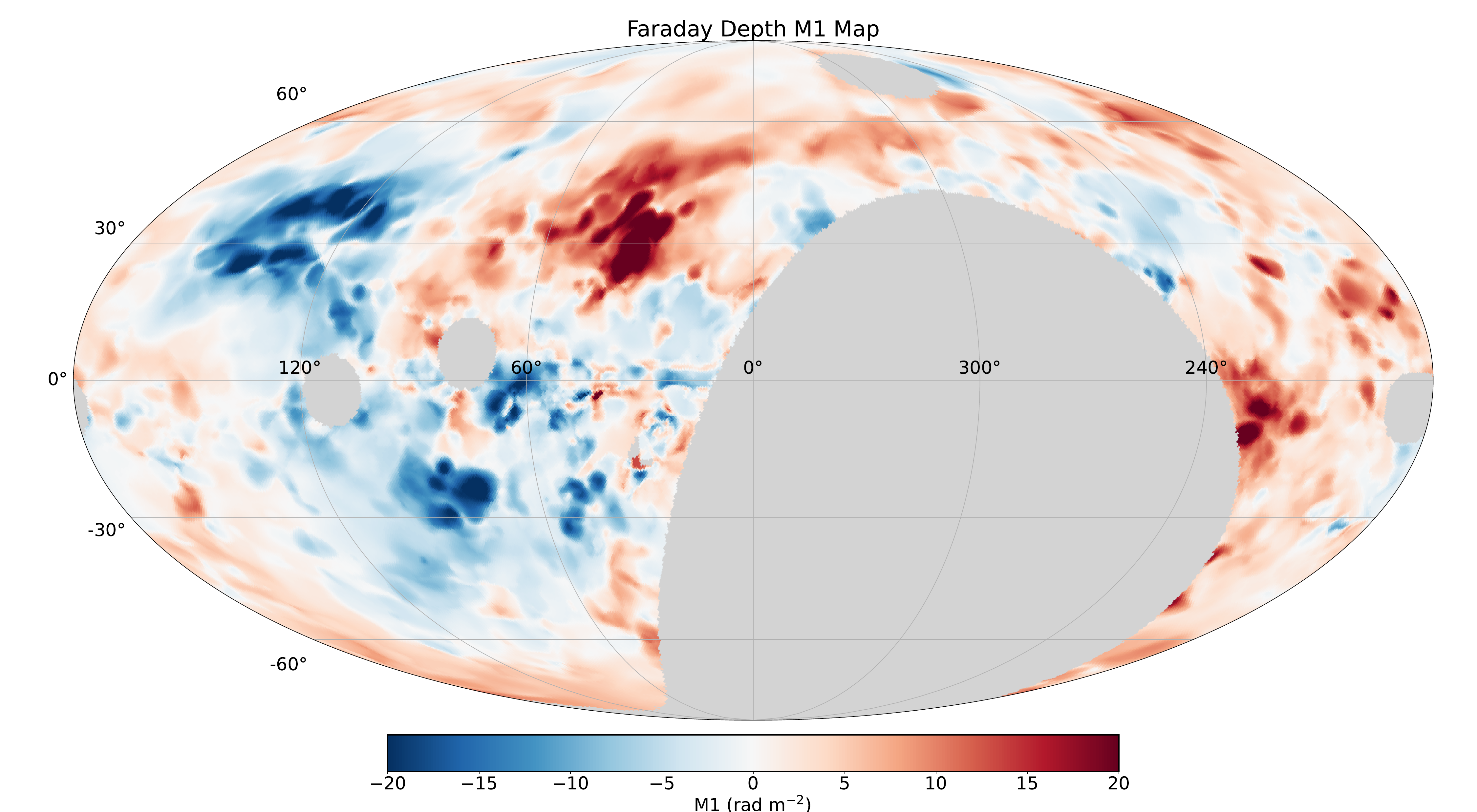}
    \caption{Map of first moments (M1) of the cleaned Faraday depth spectra. Projection and masking as in Figure~\ref{fig:M0map}.}
    \label{fig:M1map}
\end{figure*}

Following previous GMIMS work, such as \cite{dickey19} and \cite{raycheva25}, we produce Faraday depth moment maps in order to collapse the three-dimensional data cubes into 2D maps for ease of visualization and for studying the large-scale plane-of-sky structures in the Faraday rotation. The zeroth moment (M0; Figure~\ref{fig:M0map}) is the total polarized intensity integrated over a selected Faraday depth range, the first moment (M1; Figure~\ref{fig:M1map}) is the PI-weighted average of the Faraday depths, and the square root of the second moment (m2; Figure~\ref{fig:M2map})\footnote{We use `m2' (units of rad~m$^{-2}$) to refer to the square root of the second moment, M2.} represents the width of the distribution of Faraday depths. The equations defining these metrics are given in \cite{dickey19}. For the maps presented in Figures~\ref{fig:M0map}--\ref{fig:M2map}, we calculated the moments on the cleaned Faraday depth cube out to $\pm100$~rad~m$^{-2}$, using a threshold of $6\sigma$ (derived from the dirty Faraday depth cubes) as the minimum polarized intensity to include in the calculations. This is the first diffuse emission polarization survey that has the sensitivity to compute Faraday moments almost everywhere, with only a small blank patch (near $\ell=30\arcdeg,b=-15\arcdeg$) where the PI is too low to give a clear detection in the Faraday spectrum.

The overall, large-scale patterns seen in the M0 and M1 maps are similar to their counterparts in GMIMS-HBN \citep{woll21}, indicating that the same large-scale magnetic field structures dominate the depths probed by the two surveys covering different frequency ranges. However, the degree of Faraday complexity indicated by the m2 map (discussed further in Section~\ref{subsec:complexity}) warrants exploration of the spatial structure in the Faraday depth slices. Although the M1 structures are similar, the magnitudes of the Faraday depths are generally smaller, as expected for the larger beam of DRAGONS compared to GMIMS-HBN. This is more likely a consequence of beam depolarization than bandwidth depolarization, as a Faraday depth of 50~rad~m$^{-2}$ (the maximum seen in DRAGONS) produces only 12$\arcdeg$ of polarization angle rotation across 1~MHz at the 350~MHz end of the DRAGONS band.

\section{Results}\label{sec:results}
We present a small set of preliminary results and analyses based on the DRAGONS frequency and Faraday depth cubes, highlighting the strengths of the survey and potential for exciting new polarization science.

\subsection{Faraday Complexity}\label{subsec:complexity}
\begin{figure*}[t!]
    \centering
    \includegraphics[width=1\hsize]{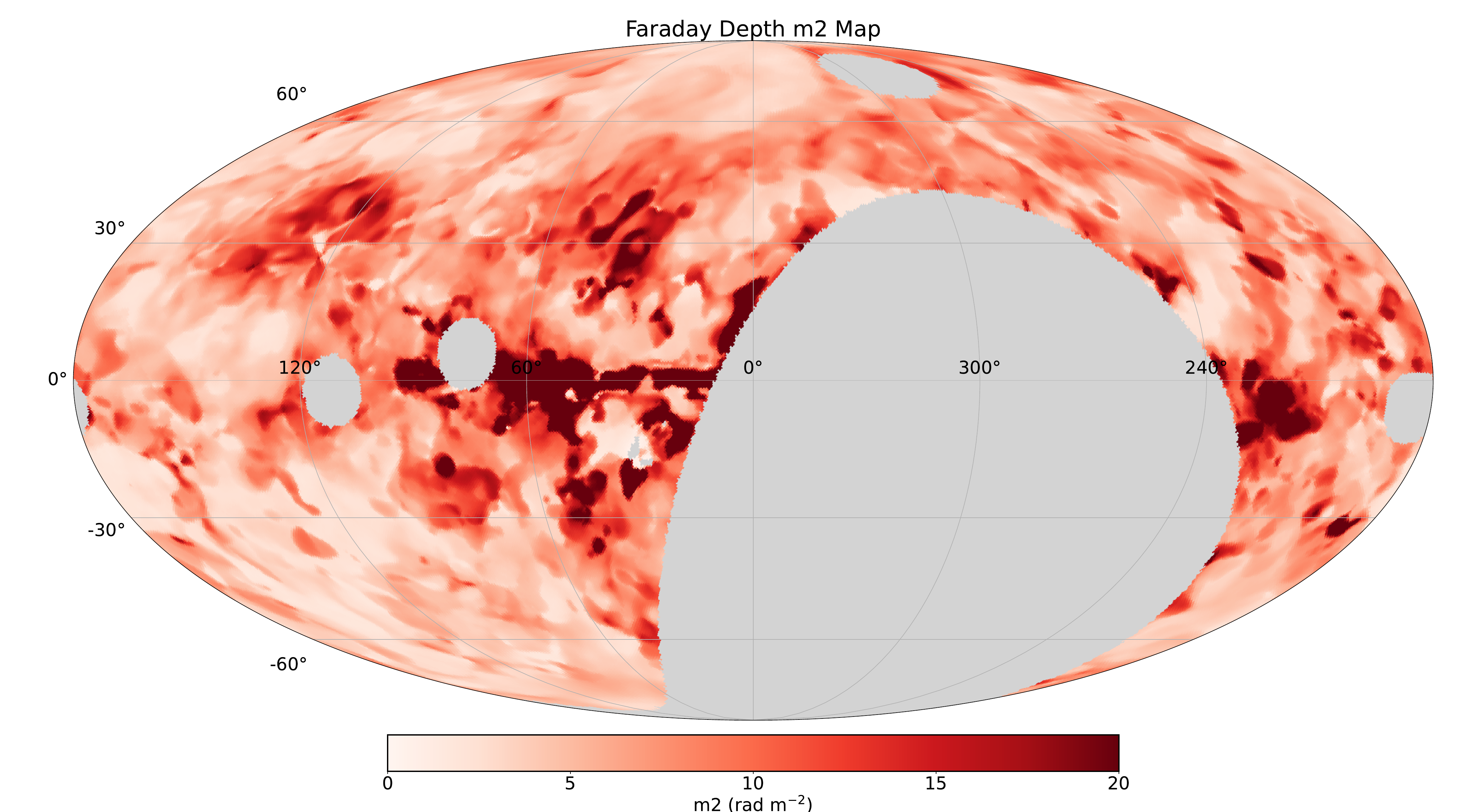}
    \caption{Map of the square root of the second moments of the cleaned Faraday depth spectra (m2). Projection and masking as in Figure~\ref{fig:M0map}.}
    \label{fig:M2map}
\end{figure*}

Approximately 55\% of the sky covered by DRAGONS reveals Faraday complexity in the form of multiple distinct Faraday depth peaks, or blended, broadened features. The second moment map (m2; Figure~\ref{fig:M2map}) illustrates this, with m2 values larger than the RMSF width ($\sim6$~rad~m$^{-2}$) indicating Faraday complexity. 

\begin{figure*}[t!]
    \centering
    \includegraphics[width=1\hsize]{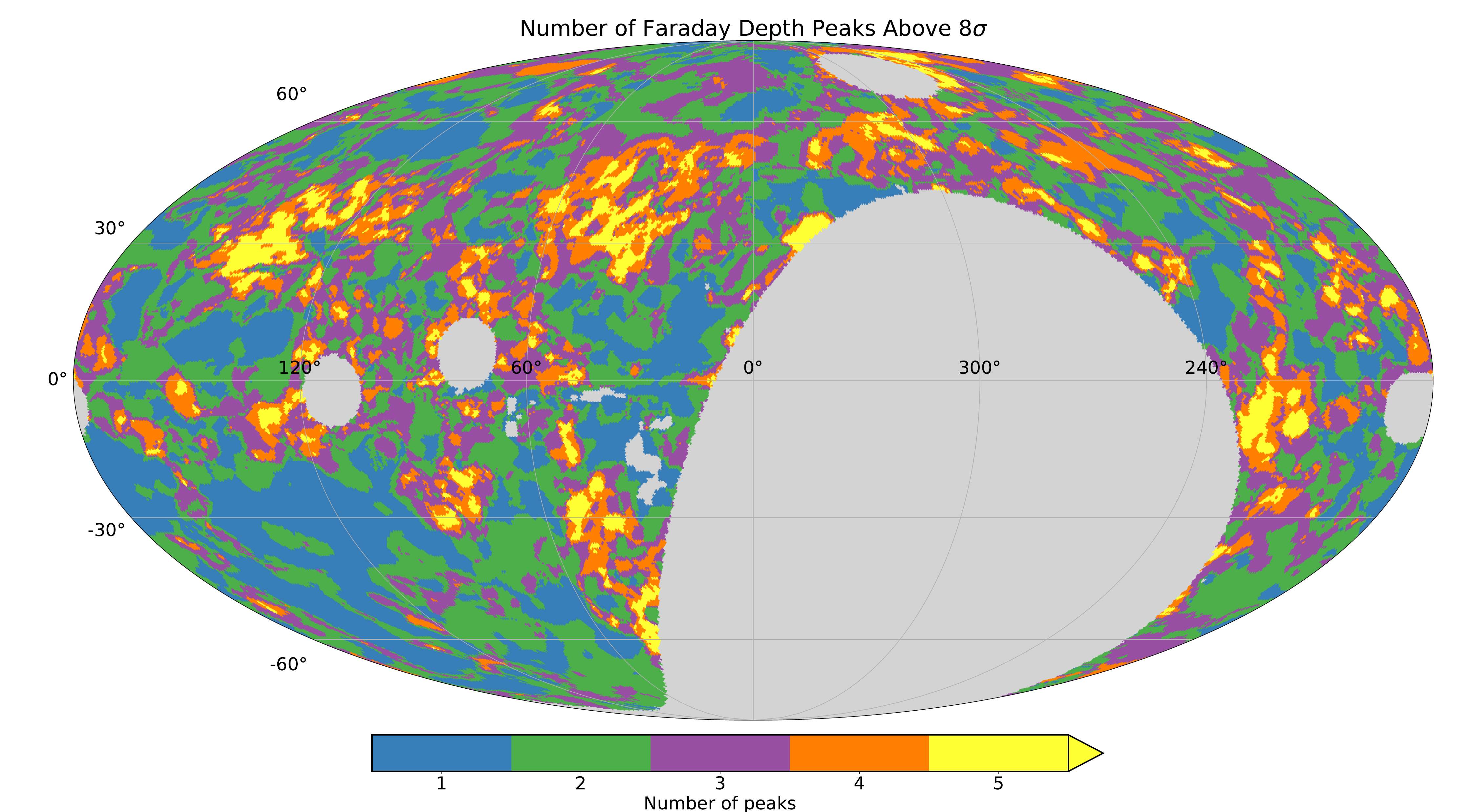}
    \caption{Number of Faraday depth peaks in the cleaned cube above a polarized intensity threshold of 8$\sigma$. 55\% of the sky has more than one Faraday depth peak. Note that yellow indicates five Faraday depth peaks or more. Masking as in Figure~\ref{fig:M0map}.}
    \label{fig:numpeaksmap}
\end{figure*}

We study the effect further by counting the number of distinct restored Faraday depth peaks above $8\sigma$ along each LOS and mapping the number of peak, as shown in Figure~\ref{fig:numpeaksmap}. We note that the Fan Region, ($\ell\approx130\arcdeg$, $b\approx5\arcdeg$) is one of the few large-scale Faraday simple regions (having a single Faraday depth peak) in the northern Galactic hemisphere, while in the southern hemisphere there is a larger area of Faraday simple lines of sight. The North Polar Spur \citep[NPS;][]{sun2015}, near $\ell\approx20\arcdeg$, $b>25\arcdeg$, and the North Celestial Pole loop \citep[NCPL;][]{Meyerdierks1991,Marchal2023}, near $\ell\approx30\arcdeg$, $120\arcdeg<b<150\arcdeg$, are two of the regions exhibiting the most Faraday complexity in the DRAGONS data, likely indicating multiple emission and rotation regions along the lines of sight. The second and third panels of Figure~\ref{fig:FD_spectra} show sample Faraday depth spectra from the NPS and NCPL respectively. The spread of the features in Faraday depth space being comparable to the widest detectable Faraday depth scale of $\phi_{\mathrm{max-scale}}=38$~rad~m$^{-2}$ is also consistent with partially depolarized extended features in Faraday depth resulting from mixed emission and rotation. Modeling these lines of sight using methods like `$QU$-fitting' \citep{Farnsworth:2011,OSullivan:2012,Ideguchi:2014,SunRudnick:2015} will be enlightening in revealing the ISM geometry that produces the complex structures.

The only other diffuse emission dataset to date having a comparable degree of Faraday complexity is LoTSS-DR2 \citep{Erceg_2022,Erceg_2024}. While LoTSS-DR2 has better Faraday depth resolution than DRAGONS, with $\delta\phi=1$~rad~m$^{-2}$, its $\phi_{\mathrm{max-scale}}$ is also limited to 1~rad~m$^{-2}$. The complementarity of these two datasets in terms of the Faraday depth regimes they are sensitive to makes them well suited to Faraday tomography studies probing the 3D nature of ISM magnetic fields.

\begin{figure}
    \centering
      \includegraphics[width=0.9\hsize]{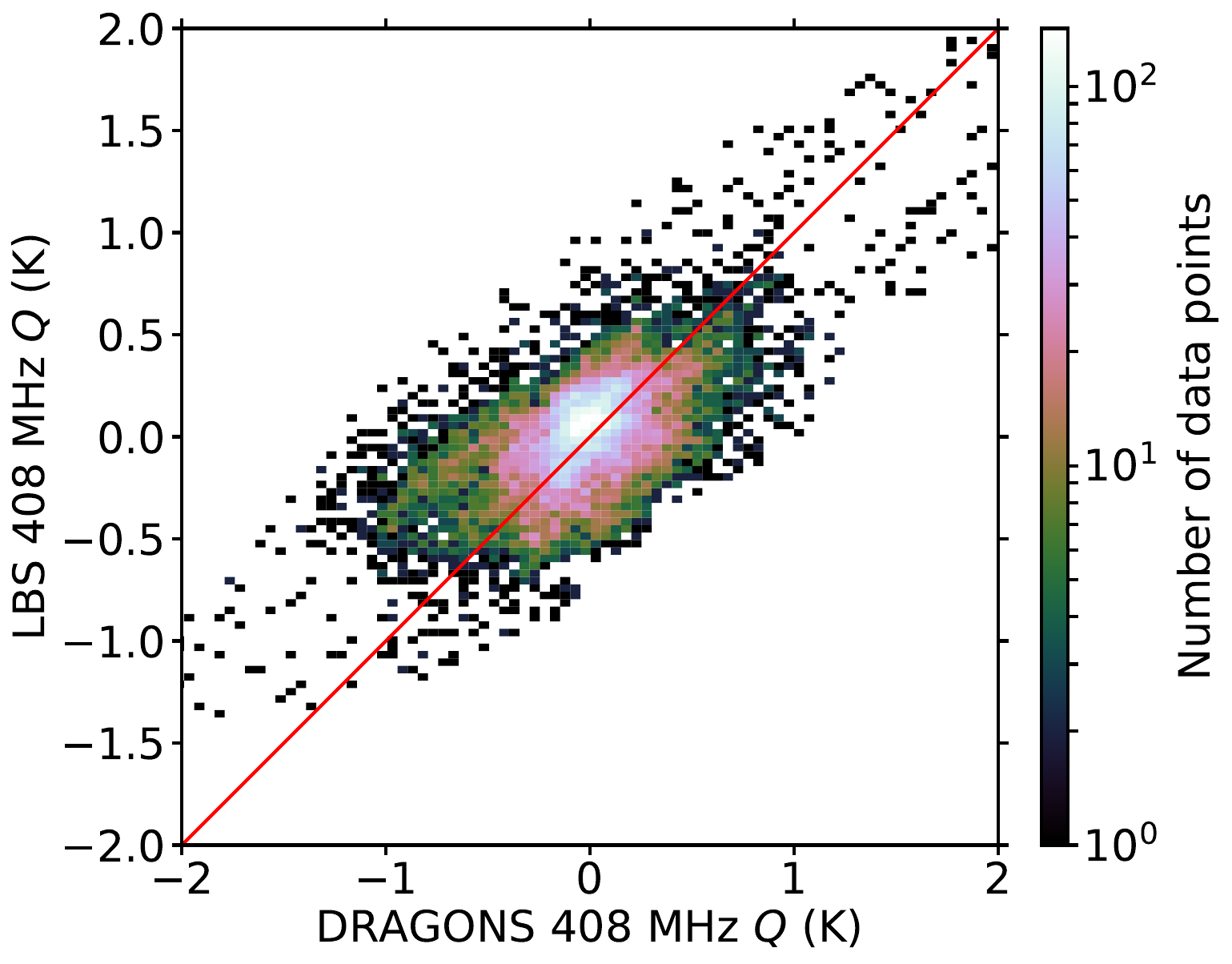} \\ 
      \includegraphics[width=0.9\hsize]{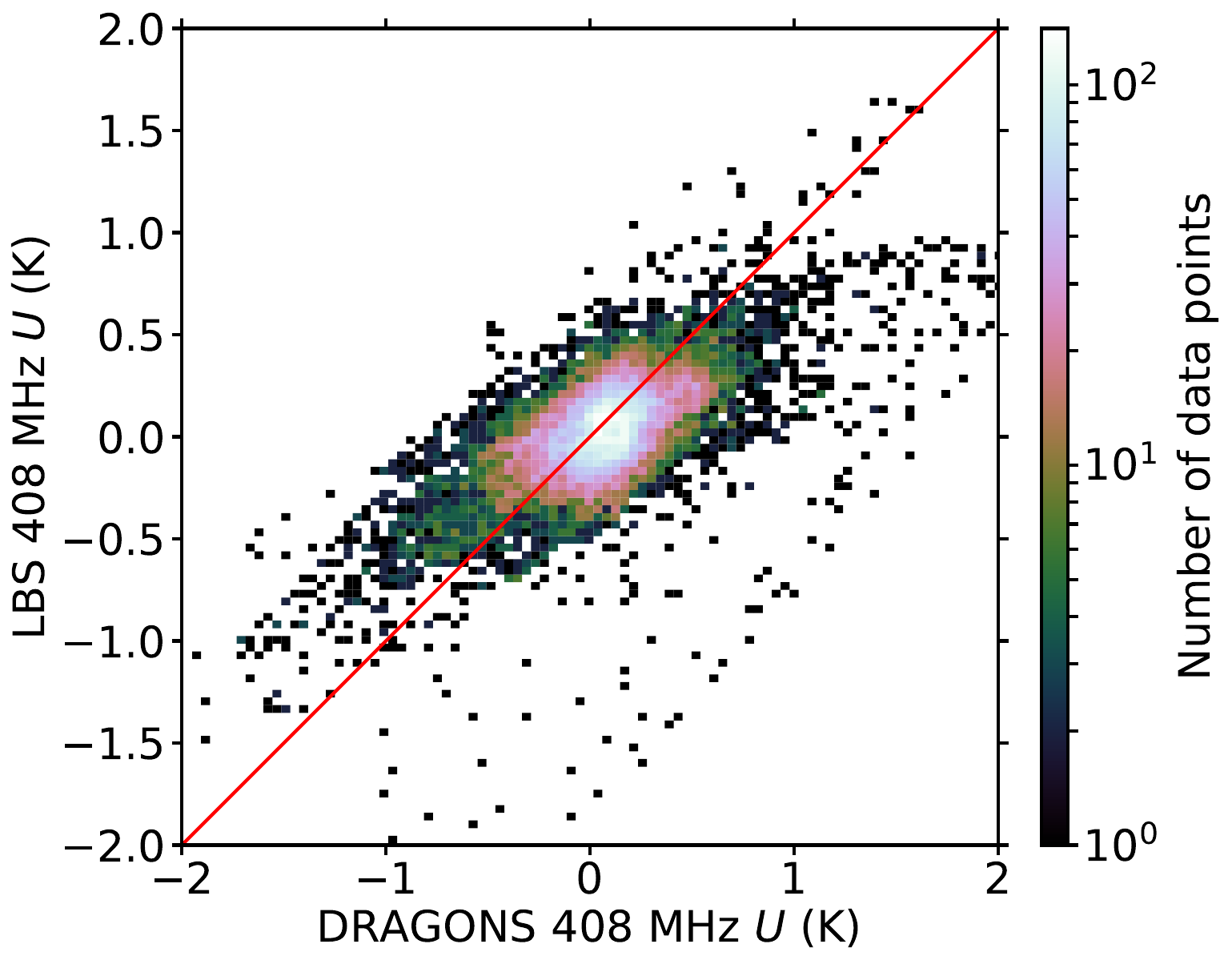} 
    \caption{The 408 MHz GMIMS-LBS polarization data compared to the DRAGONS data in the same channel. The \textit{top} shows Stokes $Q$ and the \textit{bottom} shows Stokes $U$. The Pearson correlation coefficient for both is 0.63.}
    \label{fig:LBS_QQ_UU}
\end{figure}

\subsection{Comparison with GMIMS-LBS}\label{subsec:compGMIMS}
DRAGONS and its southern-hemisphere counterpart, GMIMS-LBS, overlap in frequency between 350~MHz and 480~MHz, and in declination coverage from $\delta=-20\arcdeg$ to $\delta=20\arcdeg$, corresponding to 34\% of the sky in solid angle. This makes the surveys ideally suited for comparison and GMIMS data-product validation. Given the 64~m diameter of the Murriyang telescope (used for all southern-hemisphere GMIMS surveys), DRAGONS has approximately 4 times the beamwidth of GMIMS-LBS. Thus, prior to comparing, we convolve the GMIMS-LBS maps to the DRAGONS beam. We show a point-by-point comparison on a HEALPix grid with \texttt{nside}=64 for Stokes $Q$ and $U$ at 408~MHz in Figure~\ref{fig:LBS_QQ_UU}, masking out a region $\pm10\arcdeg$ around the Galactic plane, where both data sets suffer from instrumental polarization leakage. Both Stokes $Q$ and $U$ are strongly correlated between the two surveys with Pearson correlation coefficients of 0.63, and have ratios close to unity, indicating successful calibration of both surveys.

\begin{figure}
    \centering
      \vspace{0.5cm}
      \includegraphics[width=\hsize]{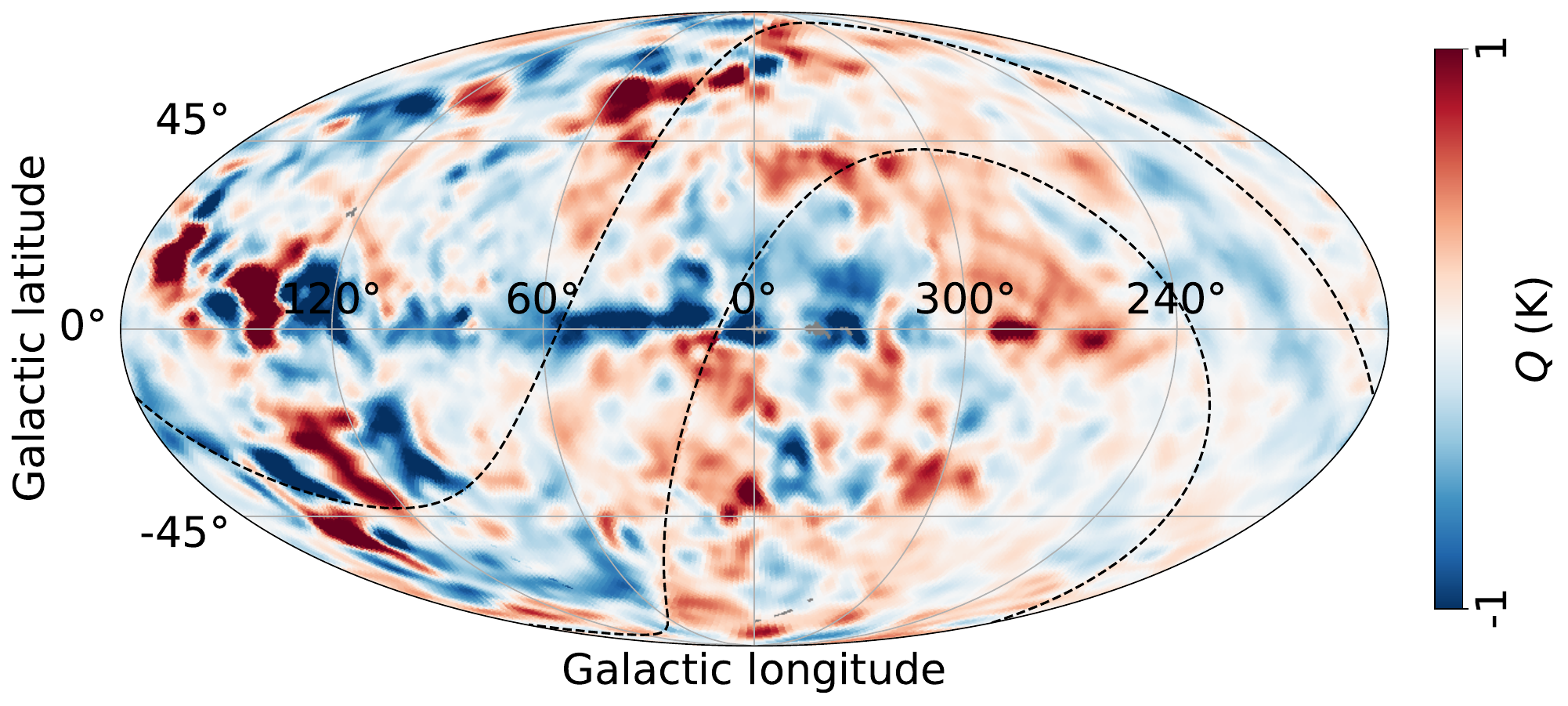} \\ 
      \includegraphics[width=\hsize]{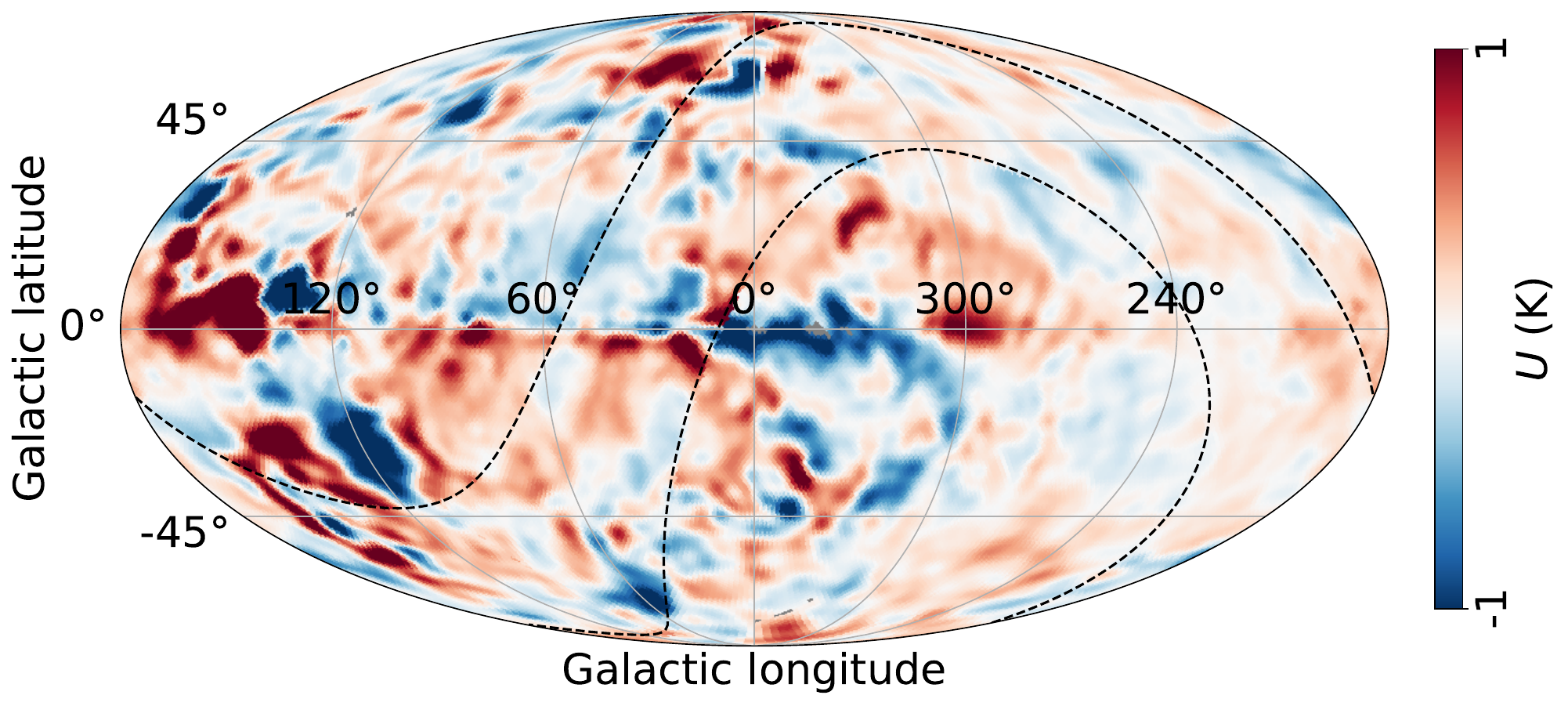} 
    \caption{DRAGONS and LBS 408 MHz merged maps. \textit{Top:} Stokes $Q$; \textit{bottom:} Stokes $U$.}
    \label{fig:LBS_maps}
\end{figure}

The strong agreement between GMIMS-LBS and DRAGONS allows for the maps to be combined into a single, full-sky image across the overlapping frequency range. We show examples of such a combined map in Figure~\ref{fig:LBS_maps}, where we have averaged the maps in the overlapping declination range. This highlights the potential of the full GMIMS initiative in allowing for polarization and Faraday depth mapping of large-scale patterns simultaneously covering both northern and southern Galactic hemispheres. Examples of existing and forthcoming studies that take advantage of this are: (i) a follow-up to the analysis of the asymmetry between the northern and southern Galactic latitudes explored in \cite{dickey2022} using GMIMS-HBN and STAPS (C.~van~Bergen et al., 2026, in preparation) and (ii) an analysis of the large-scale magnetic field reversal using STAPS and the higher frequencies ($>500$~MHz) of DRAGONS \citep{Booth2026}. Other future work will be able to benefit from the high Faraday depth resolution provided by GMIMS-LBS and the lower frequencies of DRAGONS to investigate Faraday complexity in the more nearby Galactic volume by using only the lower end of the DRAGONS band ($<500$~MHz).

\subsection{Perseus-Taurus Bubble}\label{subsec:pertau}
Faraday depth maps and studies can help identify unique Galactic structures \citep[e.g.,][]{Mohammed_2024}. One Galactic structure clearly revealed in DRAGONS Faraday depth maps is the Per-Tau bubble, situated between the Perseus, Taurus, and California molecular clouds. Previously, \cite{Shimajiri_2019} identified an \HI\ bubble in this region (see their Figure 10). Subsequently, \cite{Doi_2021} determined the 3D shape of this bubble, and \cite{Bialy_2021} presented 3D dust maps of this feature, naming it Per-Tau and suggesting it formed through multiple supernova events. This formation mechanism can make the bubble distinctly identifiable in Faraday depth maps. 

\begin{figure}
    \centering
    \includegraphics[width=1\linewidth]{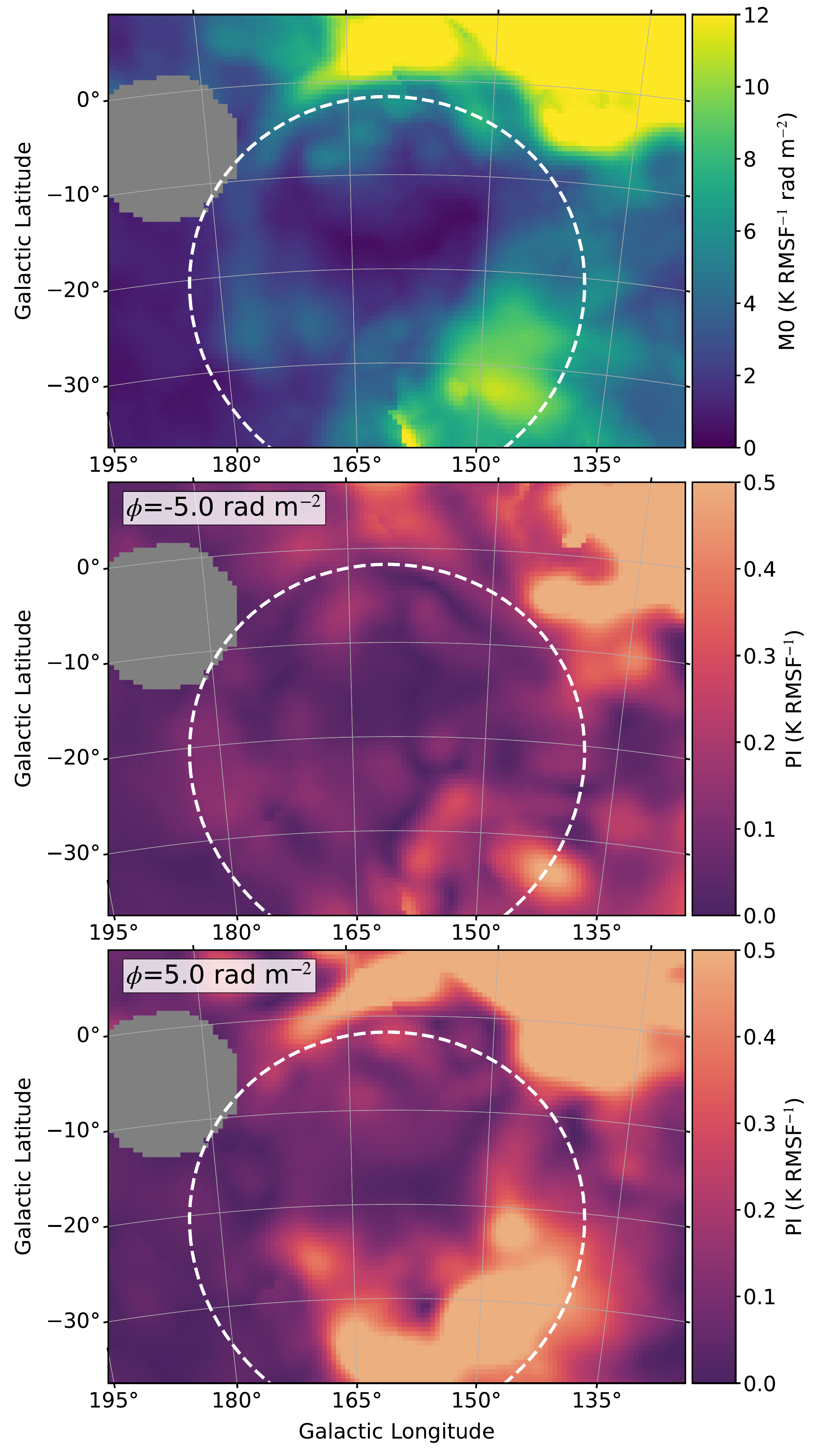}
    \caption{The Per-Tau bubble in DRAGONS M0 (\textit{top}), PI at Faraday depth -5.0~rad~m$^{-2}$ (\textit{middle}), and PI at 5.0~rad~m$^{-2}$ (\textit{bottom}). The white dashed circle is the boundary of the bubble given by \cite{Bialy_2021}. The masked (gray) region is Tau A, with the same masking as in Figure~\ref{fig:M0map}--\ref{fig:numpeaksmap}.}
    \label{fig:pertau}
\end{figure}

\citet[][]{Tahani_2022a} further reconstructed the 3D magnetic field vector of the Perseus molecular cloud, revealing that another structure beyond the Per-Tau bubble must have interacted with the Perseus cloud and shaped its field lines. They called this feature Per2, a finding subsequently confirmed by the kinematic studies of \citep{Kounkel2022}. Faraday depth maps can be used not only to study the magnetic field morphology of the Per-Tau bubble and its surrounding environment, but also to help identify the Per2 structure, its shape, and the extent of its impact (M.~Tahani et al., 2026, in preparation). The bubble stands out as a depolarized region in the DRAGONS M0 map (top panel of Figure~\ref{fig:pertau}) and its outline, along with features within the bubble, appears across multiple Faraday depth values (examples shown the bottom two panels of Figure~\ref{fig:pertau}). 

\section{Conclusions and Future Work}
We have used the DRAO-15 telescope to produce the DRAGONS polarization survey, which comprises the  single-antenna, low-frequency, northern-hemisphere component of GMIMS. This survey included a science commissioning phase for the DRAO-15, since this was the first science initiative with this instrument. The DRAGONS survey is spatially Nyquist sampled, covering a declination range of $-20\arcdeg$ to $90\arcdeg$, and a frequency range of 350--1030~MHz. The 83~kHz sampling allowed for effective RFI excision prior to binning into 1~MHz channelization for the final data cubes. We used the GPS-based \texttt{ALBUS} software to correct for ionospheric Faraday rotation effects in the Stokes $Q$ and $U$ data, which can be significant at the lower frequencies of DRAGONS, particularly near solar maximum. The observing strategy of $360\arcdeg$ azimuth scans over the course of seven months allowed for adequate baseline correction and ground subtraction, while the residual differences between the resulting duplicated maps of the sky (east and west) provided a position-dependent estimate of the errors in the maps. 

The published data products (available in both plate-carr\'ee and HEALPix FITS files in Galactic coordinates) are: Stokes $I$, $Q$, and $U$ frequency cubes, and error cube for $Q$ and $U$, the derived complex Faraday depth cubes, and the Faraday depth moment maps (Table~\ref{tab:data}). The frequency cubes have a resolution between $3.6\arcdeg$ at 350~MHz and $1.3\arcdeg$ at 1030~MHz. For the Faraday depth data products, all frequency channels were convolved to $3.6\arcdeg$ resolution prior to applying Faraday synthesis. A position-dependent RM CLEAN threshold allowed for Faraday depth spectra with a wide range of polarized intensities to be accurately recovered.

The DRAGONS cubes provide an immense improvement over the Dwingeloo data \citep{brou76}, the best previously available polarization data in this frequency range, in terms of both spatial and spectral sampling and coverage. This allows for detailed studies of Faraday complexity and large-scale patterns in Faraday rotation across the entire northern sky. We observe Faraday complexity in the form of multiple Faraday depth peaks and broadened structures in over half of the observed region. While the Faraday depth first moment, M1, map reveals many similarities with the M1 maps of other GMIMS component surveys, the structures along the Faraday depth axis will undoubtedly provide critical information for probing the LOS component of the Galactic magnetic field.

While the current data release is already of high quality and ready for use in Faraday rotation analyses and magnetized ISM studies, future work on the processing pipeline will provide further refinements. This will include: (i) improved ground emission estimation using the simulated radiation patterns of the antenna convolved with the physical ground profile at the DRAO site, (ii) an implementation of basket-weaving to fully correct the base levels of the scans that go into the map-making, and (iii) a correction to the Stokes $I$ base levels that is independent of previous datasets (Haslam 408~MHz and \texttt{PyGDSM}).

There are already several projects completed or in progress that make use of the DRAGONS dataset, including modeling of the large-scale Galactic magnetic field reversal \citep{Booth2026}, a comparison of Faraday depth spectra and moments maps between DRAGONS and LoTSS DR2, and efforts to produce $20\arcmin$ angular resolution polarization maps by combining DRAGONS with CHIME data in 400--800~MHz. In combination with other GMIMS datasets, DRAGONS provides an exciting opportunity to investigate the effects of depolarization and the polarization horizon by studying Faraday depth spectra produced using different frequency sub-bands. DRAGONS will also be useful for total intensity, Stokes $I$ studies. In the frequency range covered by DRAGONS, the \texttt{PyGDSM} total intensity model is constrained by only the \citet{haslam82} 408~MHz data and the Dwingeloo 820~MHz data \citep{Berkhuijsen1971}. The near-complete frequency sampling of DRAGONS may allow for more detailed spectral index mapping in total intensity. The DRAGONS polarization data cube in conjunction with this could also be used to separate out the synchrotron component.

The DRAGONS survey is part of the next generation of wideband spectropolarimetric radio surveys, critical for disentangling the 3D magnetic field geometry of the ISM.

\label{sec:concl}

%\begin{acknowledgments}
\section*{Acknowledgments}
This paper relies on observations obtained using a telescope at the Dominion Radio Astrophysical Observatory, which is located on the traditional, ancestral, and unceded territory of the syilx people. We benefit enormously from the stewardship of the land by the syilx Okanagan Nation and the radio frequency interference environment protection work by the syilx Okanagan Nation and NRC. DRAO is a national facility operated by the National Research Council Canada.

The telescope is a prototype developed through an international collaboration for the SKA project. We acknowledge the crucial contributions of the US SKA Technology Development Project (TDP), supported by the National Science Foundation (NSF), for the antenna's advanced optical design, high-speed mechanical drive system, and stable tower structure. The instrument's excellent performance also relies on the Band 1 feed, developed by Chalmers University of Technology at the Onsala Space Observatory (OSO) as part of their contribution to the SKA.

We express our gratitude to DRAO staff for their work on the site and the telescope used in this project. In particular, we thank Richard Hellyer for coordinating the technical team and ensuring smooth operation of the telescope, Geordie Goodall and Mohammad Islam for the necessary corrections to the telescope encoder system, Mohammad Islam and Elena Bergeron for their work on the front-end electronics enclosure, Gordon Lacy for his expertise on the reflector system, Nicholas Bruce and Tayron Dueck for being on snow-removal duty during the winter observing season, and Andrew Gray for his role as DRAO telescope manager. 

We also thank Bryan Gaensler, Gary Hinshaw, Lewis Knee, Roland Kothes, Josh MacEachern, Wasim Raja, Michael Rupen, Ken Tapping, Alec Thomson, and Cameron Van Eck for their constructive feedback over the course of the project, and we thank Paeton Hughes for helping us with data quality assessment. We appreciate the comments from the anonymous referee, which helped clarify some details of the data processing.

This work benefited from presentations and discussions during the program ``Towards a Comprehensive Model of the Galactic Magnetic Field'' at Nordita in April 2023, which is partly supported by NordForsk, and during the Interstellar Institute’s programs ``II6'' and ``II7’’ at the Paris-Saclay University’s Institut Pascal in July 2023 and 2025.

A.O. was partly supported by the Dunlap Institute at the University of Toronto. The Dunlap Institute is funded through an endowment established by the David Dunlap family and the University of Toronto. R.A.B. was supported by the Natural Sciences and Engineering Research Council of Canada (NSERC) Vanier scholarship and the University of Calgary Izaak Walton Killam Doctoral Scholarship. This research has been supported by NSERC Discovery Grants (PIs: J.C.B., T.L.L., and A.S.H). G.M. was supported by an NSERC Undergraduate Student Research Award, and L.M.C. was supported by an International Undergraduate Research Award funded by UBC Okanagan's International Student Initiative and delivered in partnership with the Irving K. Barber Faculty of Science.

We are indebted to Gavin Miller of Disman Bakner Northwest and Mandana Amiri of the University of British Columbia for help in obtaining the low-noise amplifiers used in this project, Mini-Circuits prototypes ZKL-82ULN-1+.

%\end{acknowledgments}

\facilities{DRAO:15m}
\software{astropy \citep{Astropy_2022}; matplotlib \citep{Hunter_2007}; numpy; RM-tools \citep{Purcell:2020}; healpy \citep{Zonca2019,Healpix2005}; ALBUS.}

\section*{Data Availability}
FITS files containing the data summarized in Table~\ref{tab:data} are
available through the Canadian Advanced Network for Astronomical Research (CANFAR), doi: \href{https://www.canfar.net/citation/landing?doi=25.0104}{10.11570/25.0104}.


%\appendix

%\section{Appendix}\label{sec:app}


\bibliography{drao15_refs}{}
\bibliographystyle{aasjournal}



\end{document}